\documentclass[apj,twocolumn]{emulatearxiv}
\usepackage{amsmath}
\usepackage{longtable}

\begin{document}

\title{Explosive Nucleosynthesis in Sub-Chandrasekhar Mass White Dwarf Models for Type Ia Supernovae:
Dependence on Model Parameters}

\author{Shing-Chi Leung\thanks{Email address: shingchi.leung@ipmu.jp}}

\affiliation{Kavli Institute for the Physics and 
Mathematics of the Universe (WPI),The University 
of Tokyo Institutes for Advanced Study, The 
University of Tokyo, Kashiwa, Chiba 277-8583, Japan}

\affiliation{TAPIR, Walter Burke Institute for Theoretical Physics, 
Mailcode 350-17, Caltech, Pasadena, CA 91125, USA} 

\author{Ken'ichi Nomoto\thanks{Email address: nomoto@astron.s.u-tokyo.ac.jp}}

\affiliation{Kavli Institute for the Physics and 
Mathematics of the Universe (WPI),The University 
of Tokyo Institutes for Advanced Study, The 
University of Tokyo, Kashiwa, Chiba 277-8583, Japan}

\date{\today}

\begin{abstract}

Recent observations of Type Ia supernovae (SNe Ia) have shown
diversified properties of the explosion strength, 
light curves and chemical composition. To investigate possible origins of
such diversities in SNe Ia, we have presented multi-dimensional hydrodynamical study 
of explosions and associated nucleosynthesis in the near Chandrasekhar
mass carbon-oxygen (CO) white dwarfs (WDs) for a wide range of parameters (Leung and Nomoto 2018 ApJ).
In the present paper, we extend our wide parameter survey of models to the explosions of
sub-Chandrasekhar mass CO WDs.
We take the double
detonation model for the explosion mechanism.
The model parameters of the survey include the metallicity of $Z = 0 - 5~Z_\odot$, 
the CO WD mass of $M = 0.90 - 1.20~M_\odot$, and
the He envelope mass of $M_{\rm He} = 0.05 - 0.20~M_\odot$.
We also study how the initial He detonation configuration, 
such as spherical, bubble, and ring shapes, 
triggers the C detonation. 
For these parameters, we derive the minimum He envelope
mass necessary to trigger the C detonation. We then
examine how the explosion dynamics and associated nucleosynthesis depend 
on these parameters, and compare our results with the previous representative models.  
We compare our nucleosynthesis yields with the unusual abundance patterns of Fe-peak elements and isotopes
observed in SNe Ia 2011fe, 2012cg and 2014J, as well as SN Ia remnant 3C 397
to provide constraints on their progenitors and environments.
We provide the nucleosynthesis yields table of the sub-Chandrasekhar mass explosions,
to discuss their roles in the galactic chemical evolution and archaeology.

\end{abstract}


%\maketitle


\section{Introduction}
\label{sec:intro}

Type Ia supernovae (SNe Ia) are known to have almost homogenized light
curves and spectra, thus being used as a standard candle for studying
the cosmic acceleration that led to the discovery of dark energy
\citep[e.g.,][]{Bergstroem2004,Branch2017}.
%******************************             

The basic properties of SNe Ia have been well-modeled as the
explosions of CO white dwarfs (WDs) which have both near-Chandrasekhar
mass and sub-Chandrasekhar mass \citep[e.g.,][]{Hillebrandt2000}.
However, it is still controversial which mass (near-Chandrasekhar
vs. sub-{Chandrasekhar) of the WD is the actual progenitor. For the
presupernova evolution in close binaries, 
both the single degenerate (SD)
scenario and the double degenerate (DD) scenario have been discussed,
but the actual evolutionary path remains unclear
\citep[e.g.,][]{Nomoto1997Sci,Maoz2011}.

Further, recent observations have shown the diversified properties of
light curves and spectra of SNe Ia including very peculiar ones
\citep[e.g.,][]{Li2001,Taubenberger2017,Jha2017,Jiang2017}.
The diversity can be characterized by
a wide range of $^{56}$Ni and also differences
in the ejecta composition and abundance. To understand this
diversity, a wide range of theoretical models become
necessary in order to extract the effects of each 
model parameter to explosion properties and nucleosynthesis yields.

To understand the origin of such diversities, we are computing
SNe Ia models for wide ranges of model and environmental
parameters. In \cite{Nomoto2017,Leung2018} we have studied how the model parameters,
including the central density, metallicity, initial
flame structure, and C/O ratio affect the explosion properties of 
near-Chandrasekhar mass WD models.  
For example, we have demonstrated how some well-observed SNe Ia and
SNRs can be explained by tracing the variation of
isotopes in the yields with respect to the change of model parameters.

In this paper, we present our
parameter survey for the sub-Chandrasekhar mass WD model. The
sub-Chandrasekhar mass explosions could occur in both SD and DD
scenarios as follows.

\subsection{Sub-Chandrasekhar Mass Models in the Single Degenerate Scenario}

In the SD scenario, C+O WDs accrete matter from the non-degenerate
companion stars, which include slightly evolved main-sequence stars,
red-giant stars, He-main-sequence stars, evolved He stars.  As a result
of H-burning in the H-rich accreted material or a direct accretion of
He, the mass of a He layer increases above the C+O core, leading to
eventual He ignition \citep[e.g.,][]{Nomoto2017,Nomoto2018}.

If the accretion rate of He, $\dot M_{\rm He}$, is higher than $\sim
10^{-8} M_\odot$ yr$^{-1}$, He burning shell burning makes weak
flashes which recur many times to increase the WD mass toward the
Chandrasekhar mass \citep[e.g.,][]{Nomoto1982a,Woosley2011}.  For lower rates of
$10^{-10} M_{\odot}$ yr$^{-1} \lesssim \dot M_{\rm He} \lesssim 10^{-8}
M_\odot$ yr$^{-1}$, the compressional heating rate is lower and thus
the temperature of the He layer is lower, which causes a delay in the
He ignition until the mass of He layer becomes large enough and the
density at the bottom of He layer high enough for He burning to grow
into detonation.  It eventually leads to double detonation
\citep{Nomoto1982b,Woosley1986}.  The double detonation
model has been widely studied in 1D and multi-D simulations for
various model parameters
\citep[e.g.,][]{Livne1990a,Livne1990b,Livne1991,Livne1995,Arnett1996,
Fink2007,Fink2010,Sim2012,Moore2013,Moll2013,Shen2018,Polin2018}.

The important property of the sub-Chandrasekhar mass progenitors in
SD scenario is that the mass of the He layer exceeds $\sim 0.05
M_\odot$ to induce a He detonation
\citep[e.g.,][]{Nomoto1982b,Woosley1986}.  This is in contrast to the
sub-Chandrasekhar mass models in the
DD scenario as discussed below.

\subsection{Sub-Chandrasekhar Mass Models in the Double Degenerate Scenario} 
 
In the DD scenario, the detonation near the surface of the primary WD can be triggered during the violent
merging of two WDs for suitable binary parameters
\citep[e.g.,][]{Rasio1995,Segretain1997,Guerrero2004,Yoon2007,Fryer2010,
Dan2011,Raskin2012,Raskin2014,Moll2014,Sato2015}.  However, if there exists no He, the occurrence of
the surface C detonation may still depend on the numerical resolution \citep[e.g.,][]{Sato2015,Pakmor2017}.
Then \cite{Pakmor2013,Dan2015} presented a He ignited double detonation model where the
He detonation near the surface is triggered because a certain mass of He rich
envelope is assumed to exist on both WDs. 
In contrast to the double detonation in the SD scenario, the He-ignited
detonation could be triggered for a smaller mass He-rich envelope because of
shock compression.

In the above DD model, the collision point can
reach a sufficiently high temperature for triggering a He detonation.
The He detonation can produce a shock wave which propagates through the He
envelope and into the CO core.  The shock-heating in the C-rich
matter can induce a central or off-center C detonation.  The WD is then
disrupted by the C detonation. This model may produce the diversity
of the different brightness,
%delay times, and relative rates of normal and fast declining SNe Ia, 
depending on the masses of the CO core and
the He envelope \citep[e.g.,][]{Arnett1996,Sim2010,Woosley2011,Pakmor2013,Shen2018, Polin2018}.

\subsection{Motivation}

In \cite{Leung2017a,Nomoto2017} we have studied how the model parameters,
including the central density, metallicity, initial
flame structure and C/O mass fraction ratio, affect the 
chemical yield of SN Ia evolved from a 
near-Chandrasekhar mass WD. By tracing the variations
of isotopes with respect to the change of model
parameters, we have demonstrated 
how some well-observed SNe Ia can be explained by 
the near-Chandrasekhar mass model.

However, the occurrence rate of SNe Ia evolved from
sub-Chandrasekhar WD is suggested to be higher
than the near Chandrasekhar mass WD in population synthesis 
\citep[See, e.g.,][]{Yungelson2005,Maoz2014}.
It becomes necessary to ask whether the double detonation model 
can explain SNe Ia similarly to the Chandrasekhar mass model,
or can even replace the Chandrasekhar mass model in certain
parameter space.

Furthermore, through multi-dimensional hydrodynamics
simulations, one can draw constraints on how to trigger the
C detonation by the He detonation in the aspherical configuration systematically. 
This will set constraints on the criteria in triggering the
C detonation through the aspherical He detonation with
or without geometrical convergence.

To investigate possible origins of large diversities of SNe Ia, we
perform two-dimensional hydrodynamical studies of explosions and
associated nucleosynthesis in the sub-Chandrasekhar mass CO WDs for
wide ranges of parameters.  
%In all simulations we use the the
%two-dimensional hydrodynamics code \citep{Leung2015a}.
All simulations use the code based on the two-dimensional
hydrodynamics code developed for the explosion phase of supernovae
\citep{Leung2015a}.

This work is a continuation of our previous work 
in \cite{Leung2017a, Nomoto2017}, where we have
studied the dependence on model parameters of 
SNe Ia using the Chandrasekhar
mass models \citep{Leung2015b,Nomoto2017,Nomoto2018,Leung2018}. 
%and sub-luminous SNe Ia \cite{Leung2015b}. 
In \cite{Leung2018} we 
covered the density, metallicity, 
flame structure and detonation criteria. We have also 
shown that some of the chemical abundances features observed in
recently observed SNe Ia can be reproduced by our models. 

In the present paper, we want to extend our understanding
to the sub-Chandrasekhar mass models for even a wider parameter region.
We intend not only to explain the observed diversities of SNe Ia
but also to provide the predictions of nucleosynthesis properties
for coming observations \citep[e.g.,][]{Fryer2019}.

In Section \ref{sec:methods} we summarize the 
numerical methods used in this work and the input physics
specific to model the sub-Chandrasekhar mass model.
In Section \ref{sec:models}
we describe our two-dimensional simulations to study
the exploding WDs starting from the He detonation
at the envelope. 
In Section \ref{sec:benchmark} we describe the
benchmark model which is regarded as the representation
of a typical sub-Chandrasekhar mass model.
In Section \ref{sec:nucleo} 
we describe nucleosynthesis yields and their
dependence on the model parameters,
including the WD mass, He envelope mass and 
initial He detonation pattern. We also present our
cross-comparison with the classical double detonation model
with spherical symmetry, and 
its possible impacts on galactic chemical evolution.
In the Appendix we provide further numerical details
and tests we have done for this work. We also
discuss the implications of our models,
including a comparison from models in the literature.

%At last we give our conclusion.

%%%%%%%%%%%%%%%%%%%%%%%%%%%%%%%%%%%%%%%%%%%%%%%%%%%%%%%%%%%%%%%%%%%%%%%%%
%%%%%%%%%%%%%%%%%%%%%%%%%%%%%%%%%%%%%%%%%%%%%%%%%%%%%%%%%%%%%%%%%%%%%%%%%

\section{Methods}
\label{sec:methods}

\subsection{Input Physics}

Here we briefly review the structure of our hydrodynamics
code and then we describe the change done to describe the He 
detonation and the onset of C deflagration or C detonation. 
We use the same two-dimensional hydrodynamics code as reported
in \cite{Leung2015a} for our simulations. The code solves the 
Euler equations in cylindrical coordinates where the spatial 
discretization is done by the fifth-order weighted
essentially non-oscillatory (WENO) scheme and the time-discretization
is done by the five-step third-order non-strong-stability preserving
Runge-Kutta scheme. We use the helmholtz subroutine \citep{Timmes1999,Timmes1999c} 
as the matter equation of state (EOS). This EOS includes
the arbitrarily relativistic and degenerate electron gas, ions
as a classical ideal gas, Planckian photon gas and electron-positron
annihilation pairs. In the hydrodynamics section, we describe the
chemical composition by a 7-isotope network, which includes
$^{4}$He, $^{12}$C, $^{16}$O, $^{20}$Ne, $^{24}$Mg, $^{28}$Si
and $^{56}$Ni. 

The one-equation model \citep{Niemeyer1995b} is used to 
model the velocity fluctuations in the sub-grid scale due 
to turbulence. To describe the geometry of the two 
detonation fronts, we use the individual level-set functions 
\citep{Osher1988} as used in \cite{Reinecke1999a}. The geometry 
of the fronts are constructed by locating zero-value points
in the level-set function, and then the fractional volume in 
each mesh being burnt by flame or detonation is 
extracted. The energy from nuclear burning
is injected instantaneously to Eulerian grids which have an 
increase in the area (volume) fractions $\alpha$ in the 2- (3-)
dimensional models enclosed by the contours and their boundaries.
To prevent double-counting the energy released by burnt matter, $\alpha$
is set to be a monotonic increasing throughout the simulations.
In each step, the code calculates the area fraction
based on how the zero-contour intersects with the grid mesh.
A fraction of $\alpha^*$ is obtained where $\alpha = 1$ for a
completely burnt cell and 0 for pure fuel. The value of 
$\alpha^*$ is compared with that in the previous step $\alpha_{{\rm old}}$
and the larger one is taken, i.e. $\alpha = \max(\alpha^*, \alpha_{{\rm old}})$.
For He detonation, we assume the detonation is in form
of Chapman-Jouget detonation, where the detonation propagates
in sound speed given by $\sqrt{\gamma_2 (\partial p / \partial \rho)_s}$,
where $\gamma_2$ is the adiabatic index, $p$ and $\rho$ are
the pressure and density. For CO detonation, we
used the same prescription as \cite{Sharpe1999} by the
numerical speed of pathological detonation. 

All simulations are done by a resolution size of $400^2$
using the cylindrical coordinate. We choose reflecting and outgoing
boundaries for the inner and outer boundaries for both 
$r$ and $z$ axis. The resolution is fixed 
at either 15 or 23 km. The lower one is for higher mass 
white dwarf (mass $> 1 M_{\odot}$) while the higher one is
for lower mass white dwarfs. The grid size is chosen such that the simulation box
is about 2 - 3 times of the initial WD radius. We do that because
we want most exothermic reactions, which rely on the 
level-set method, can finish before the star reaches
the outer boundary of the simulation box. We follow \cite{Roepke2005} 
by implementing the moving-boundary technique so that, when the 
stellar outer radius reaches the simulation box boundary, 
the grid expands with a similar speed so that most 
matter can be contained in the simulation box. 
In our calculation, we choose the averaged radial velocity of
the low density matter (defined by 1 - 10 times of the atmospheric 
density) to be the expansion velocity. Following with the 
expansion, we also adjust the atmosphere density such that 
the total mass of matter in the "atmosphere" (also the 
minimum density allowed in the simulation) remains 
$\sim 10^{-4}$ of the star. 
We model only one quadrant
of the WD by the use of reflecting
boundary.

In our computation, it takes typically 2 - 3 days for 
a hydrodynamics simulation for a quadrant from the onset of 
He-detonation until homologous expansion develops by a single 
CPU run. The assumed symmetry allows only two He-detonation bubbles
to be ignited simultaneously. 
A more general detonation form
as a single He-detonation bubble requires hydrodynamics
simulations of a hemisphere. The computational time
can be lengthened by a factor of $\sim 4$ times with also a 
factor of 2 larger in memory. 
Thus only a small number of models 
are computed. In Appendix \ref{sec:sym_bound} we present some
exploratory models using the relaxed symmetry
and compare with our "quadrant" models. We also 
check the dependence of the general
detonation model on the chosen resolution and report 
in Appendix \ref{sec:test1}.

\subsection{Nuclear Reaction Scheme}

For the nuclear reactions of He-rich matter,
the region swept by the level-set contour is regarded as 
burning from $^4$He to $^{56}$Ni. For CO-rich matter, similar to 
previous works \citep{Leung2018}, we use the three-step burning scheme
so as to include more flexible nuclear reactions,
especially when there are contributions from 
shock wave collisions. In this work, we assume
this process is instantaneous regardless of the local density.
We follow the use of burning timescale as an approximation 
to burning where density is low (i.e. $\rho < 5 \times 10^7$ g cm$^{-3}$).
That includes the nuclear quasi-statistical
equilibrium (NQSE) timescale and the NSE timescale,
given by, respectively (see \cite{Calder2007,Townsley2007}), 
\begin{eqnarray}
\tau_{{\rm NQSE}} = {\rm exp}(182 / T_{f,9} - 46.1) ~{\rm s}, \\
\tau_{{\rm NSE}}  = {\rm exp}(196 / T_{f,9} - 41.6) ~{\rm s},
\end{eqnarray}
where $T_{f,9} = T_f / 10^9 {\rm ~K}$ is the final temperature of the ash. 
For a timestep shorter than these two timescales, we assume that a fraction of
matter given by linear interpolation with $\tau$ is burnt.
For a timestep longer than those, complete conversion of
fuel to ash is assumed. Similar treatment is done for 
He detonation. We describe more details in Appendix A. 

To determine whether a detonation wave can start, we follow the 
scheme in \cite{Fink2007}. For an Eulerian grid of 
CO matter, when the temperature exceeds the threshold temperature
as a function of density (see Table 1 and 2 in \cite{Fink2007}),
a bubble or ring of hot ashes (i.e., NSE matter from CO and $^{56}$Ni from He)
is put artificially around that grid of 1.5 times the grid size. 
In practice, we set the level-set scalar field $S$ in the way
that $S(r,z) = -\sqrt{(r - r_0)^2 + (z - z_0)^2} + 1.5 \Delta x$.
Here $r_0$ and $z_0$ are the center coordinates of the bubble and $\Delta x$
is the resolution size. However, when multiple detonation seeds 
are triggered, those within 10 $\Delta x$ from existing ones
are discarded.
At a density between $2 \times 10^7$ and $10^9$ g cm$^{-3}$, 
detonation propagates in the form of pathological detonation,
where behind shock front matter with a speed below
the frozen sound speed appears \citep{Sharpe1999}. The propagation 
velocity is obtained by solving the detonation structure 
explicitly. To prevent double-counting in the burnt material
due to numerical diffusion, once a grid reached NSE, it
is forbidden to carry out $^{16}$O and $^{24}$Mg burning
in the second burning step.
In the NSE state, the final composition is changed by solving iteratively
by requiring that the change in the internal energy equals to the change
in the binding energy up to the required precision. Matter in
the NSE state is also allowed to carry out electron capture with 
a rate obtained by interpolating the 
pre-computed rate table using the prescription described in \cite{Seitenzahl2010}. 

To apply NSE calculation in the modeling, after each hydrodynamics
step, we obtain a current density $\rho$, current electron fraction
$Y_{{\rm e},i}$, trial temperature $T_i$, specific internal energy
density $\epsilon_i$ and the nuclear binding energy per mass $q_i$.
We look for the electron capture rate $\dot{Y}_{{\rm e}}$
and its corresponding neutrino energy loss rate per mass $\dot{q}_{\nu}$.
To obtain the thermodynamics state in NSE, we solve the 
implicit equation
\begin{eqnarray}
\epsilon_i - q_i = \epsilon_f(\rho,T_f,X_f) - q_f(\rho,T_f) + \nonumber \\ 
(m_n - m_p - m_e) N_A c^2 \dot{Y}_e(\rho, T_i, Y_{{\rm e},i}) + \dot{q}_{\nu} (\rho, T_i, Y_{{\rm e},i}).
\end{eqnarray}
Here, $\epsilon_f(\rho,T_f,X_f)$ and $q_f(\rho,T_f)$
are those for the final state. $(m_n - m_p - m_e) N_A c^2$
is the energy loss due to mass difference between neutron and
electron-proton pair per mass. The above equation is solved by
implicitly finding the $T_f$ and its corresponding
$q_f$ such that the energy is balanced. 
The approximation $\dot{Y}_e(\rho, T_i, Y_{{\rm e},i})$
is true when the electron capture rate is much slower
than dynamical timescale, which is true for SN Ia.

\section{Initial Models}
\label{sec:models}

In this section we first describe the arrays of models
we have performed for the SNe Ia using the 
sub-Chandrasekhar mass WD. Then we describe the explosion
thermodynamics for each class of explosion. 

In Tables \ref{table:models} and \ref{table:modelsb} we tabulate the models studied for the 
double detonation model. 
The initial WD consists of a 
CO core and a He envelope. We regard the total WD mass $M$, the He envelope mass $M_{\rm He}$,
the initial metallicity $Z$, 
and the position of the He detonation seeds as
input parameters. The initial WD is assumed to be 
isothermal at a temperature of $10^8$ K \footnote{In general the 
WD can be away from isothermal profile due to 
the hydrostatic burning and convection. 
The exact profile depends on the competition between the compressional heating due to mass accretion
and radiative cooling. In view of uncertainties during accretion,
we neglect this factor and prepare identical initial models.}.
For $Z = 0.02$, we choose $49 \%$ $^{12}$C 
and $49 \%$ $^{16}$O and $2 \%$ $^{22}$Ne in mass fractions, 
and for smaller $Z$, the mass fraction of $^{22}$Ne
is smaller and C and O have larger mass fractions in equal.
For the He envelope, pure $^{4}$He is assumed.
Notice that the prescription of $^{22}$Ne is not
necessary the only element that represents metallicity. For example, 
in \cite{Shen2018}, the $^{22}$Ne mass fraction $X(^{22}$Ne)
scales as $Z = 1.1 X(^{22}$Ne). A more precise matching
between the abundances from the stellar evolutionary models
and the hydrodynamics simulations will require a more detailed
isotope network. 

It is shown that the 
actual C/O ratio can be sensitive to 
$M$ and $Z$ \citep{Umeda1999}. 
We remark that a direct extension 
for different C/O ratio is not straightforward 
since it requires first a quantitative study on
how C-detonation is triggered
as a function of density with a given composition. 

To start the He detonation, we place a spherical
detonation seed along the rotation symmetry axis.
Due to resolution limit, the initial detonation
seed is 1.5 times of the grid size in radius, i.e. 22 km. The position
of the seed is regarded as an input parameter of the
model, which ranges from 30 km to 300 km.
The detonation seed consists of hot ashes of $^{56}$Ni. 

We notice that starting the explosion near
the boundary may not be ideal in the 
two-dimensional models due to the possibility 
of enhancing nuclear burning along the 
symmetry boundary. But for our case, the detonation
propagates much faster than typical fluid velocity.
The hydrodynamical instabilities, especially Rayleigh-Taylor
instability, do not have adequate time to grow before
the fuel is swept by the detonation wave. 
So the boundary effect is less 
significant compared to the turbulent deflagration scenario. 
To construct the initial model, 
we choose models with a total mass $M =$ 0.9 - 1.2 $M_{\odot}$ and
$M_{\rm He} =$ 0.05 - 0.35 $M_{\odot}$.

%\begin{longtable}{|c|c|c|c|c|c|c|c|c|c|c|c|c|c|}
\begin{table*}
\caption{The model parameters and the global properties of
the energetics and nucleosynthesis of the SNe Ia Model
performed in this article. $E_{{\rm nuc}}$ and $E_{{\rm tot}}$
are the energy released by nuclear reaction and the total
energy, in units of $10^{50}$ erg. $M$, $M_{{\rm He}}$ and 
$M_{{\rm Ni}}$ are the masses of the initial WD model,
initial He envelope and the final synthesized $^{56}$Ni
in units of $M_{\odot}$. $R$, $R_{{\rm seed}}$ and $R_{{\rm det}}$ 
are the radii of the initial WD model, the distance of the initial
detonation seed from the He/CO interface and the radius where the second
detonation is started, in unit of km. $t_{{\rm det}}$ is the time when the 
second detonation is triggered. $\rho_c$ and $\rho_{{\rm det}}$ are
the initial central density and the density at which the second 
detonation is triggered, in units of $10^7$ g cm$^{-3}$. The category "Type" 
classifies the final results into five types. "N" stands for 
no second detonation induced. "Y" stands for the second
detonation which starts at location closer to the z-axis (the rotation
symmetry axis). "X" stands for the second detonation which starts at 
a location closer to the r-axis (the symmetry plane axis).
"D" stands for the second detonation which starts at somewhere between 
"Y" and "X" and "S" stands for the central detonation.}
\begin{center}
\label{table:models}
\begin{tabular}{|c|c|c|c|c|c|c|c|c|c|c|c|c|c|}
\hline
Group & Model & $\rho_c$ & $M_{{\rm He}}$ & $R_{{\rm seed}}$ & $M$ & $R$ & $E_{{\rm fin}}$ & $E_{{\rm nuc}}$ & 
$t_{{\rm det}}$ & $\rho_{{\rm det}}$ & $R_{{\rm det}}$ & $M_{{\rm Ni}}$ & Type \\ \hline

A & 090-050-2-B50 & $1.67$ & 0.05 & 50 & 0.90 & 7160 & 3.82 & 5.08 & 2.71 & 1.00 & 4170 & $< 10^{-2}$ & "Y" \\
A & 095-050-2-B50 & $2.26$ & 0.05 & 50 & 0.95 & 6710 & 4.70 & 6.17 & 2.29 & 1.00 & 3860 & 0.11 & "Y" \\
A & 100-050-2-B50 & $3.21$ & 0.05 & 50 & 1.00 & 6180 & 7.62 & 9.34 & 1.74 & 1.07 & 2870 & 0.31 & "Y" \\ 
A & 110-050-2-B50 & $6.17$ & 0.05 & 50 & 1.10 & 4930 & 10.8 & 13.1 & 1.18 & 1.24 & 3770 & 0.68 & "Y" \\ \hline

B & 090-050-2-S50 & $1.67$ & 0.05 & 50 & 0.90 & 7160 & 3.53 & 4.68 & 1.35 & 6.53 & 20 & 0.02 & "S"\\ 
B & 095-050-2-S50 & $2.26$ & 0.05 & 50 & 0.90 & 6710 & 7.28 & 8.56 & 1.18 & 6.15 & 40 & 0.45 & "S" \\ 
B & 100-050-2-S50 & $3.21$ & 0.05 & 50 & 1.00 & 6180 & 8.70 & 10.2 & 0.98 & 6.03 & 70 & 0.60 & "S" \\ 
B & 110-050-2-S50 & $6.71$ & 0.05 & 50 & 1.10 & 4930 & 11.7 & 13.8 & 0.83 & 11.7 & 40 & 0.82 & "S" \\ \hline

C & 090-100-2-50 & $1.67$ & 0.10 & 50 & 0.90 & 7160 & -0.50 & 0.77 & nil & nil & nil & $< 10^{-2}$ & "N"\\ 
C & 095-100-2-50 & $2.26$ & 0.10 & 50 & 0.90 & 6710 & -0.43 & 0.94 & nil & nil & nil & $< 10^{-2}$ & "N" \\ 
C & 100-100-2-50 & $3.21$ & 0.10 & 50 & 1.00 & 6180 & -0.36 & 1.38 & nil & nil & nil & $< 10^{-2}$ & "N" \\ 
C & 110-100-2-50 & $6.71$ & 0.10 & 50 & 1.10 & 4930 & 11.1 & 13.0 & 0.94 & $1.04$ & 3430 & 0.62 & "X" \\ \hline

D & 090-100-2-50  & $1.67$ & 0.100 & 50 & 0.90 & 7160 & -0.50 & 0.77 & nil  & nil & nil & $< 10^{-2}$ & "N" \\
D & 090-150-2-50  & $1.67$ & 0.150 & 50 & 0.90 & 7160 &  5.65 & 7.04 & 2.35 & $1.51$ & 2100 & 0.14 & "Y" \\
D & 090-200-2-50  & $1.67$ & 0.200 & 50 & 0.90 & 7160 &  7.92 & 9.02 & 1.21 & $1.03$ & 3940 & 0.28 & "X" \\
D & 090-300-2-50  & $1.67$ & 0.300 & 50 & 0.90 & 7160 &  11.6 & 12.9 & 0.83 & $1.00$ & 3370 & 0.54 & "D" \\ \hline

E & 100-050-2-50  & $3.21$ & 0.050 & 50 & 1.00 & 6180 & -1.45 & 0.27 & nil  & nil & nil & $< 10^{-2}$ & "N" \\ 
E & 100-075-2-50  & $3.21$ & 0.075 & 50 & 1.00 & 6180 & -1.08 & 0.63 & nil  & nil & nil & $< 10^{-2}$ & "N" \\
E & 100-100-2-50  & $3.21$ & 0.100 & 50 & 1.00 & 6180 & -0.36 & 1.39 & nil  & nil & nil & $< 10^{-2}$ & "N" \\
E & 100-150-2-50  & $3.21$ & 0.150 & 50 & 1.00 & 6180 &  8.64 & 10.3 & 0.99 & $1.06$ & 3370 & 0.47 & "X" \\
E & 100-200-2-50  & $3.21$ & 0.200 & 50 & 1.00 & 6180 &  15.0 & 13.3 & 0.75 & $1.00$ & 3360 & 0.61 & "D" \\ \hline

F & 110-050-2-50  & $6.17$ & 0.050 & 50 & 1.10 & 4930 & -1.89 & 0.39 & nil  & nil & nil & $1.1 \times 10^{-2}$ & "N" \\
F & 110-075-2-50  & $6.17$ & 0.075 & 50 & 1.10 & 4930 &  9.88 & 12.1 & 1.11 & $1.04$ & 3560 & 0.56 & "X" \\
F & 110-100-2-50  & $6.17$ & 0.100 & 50 & 1.10 & 4930 &  11.1 & 13.0 & 0.94 & $1.04$ & 3430 & 0.62 & "X" \\
F & 110-125-2-50  & $6.17$ & 0.125 & 50 & 1.10 & 4930 &  14.4 & 14.9 & 0.69 & $1.10$ & 3270 & 0.78 & "X" \\
F & 110-150-2-50  & $6.17$ & 0.150 & 50 & 1.10 & 4930 &  14.8 & 15.4 & 0.45 & $1.03$ & 3190 & 0.69 & "D" \\ 
F & 110-200-2-50  & $6.17$ & 0.200 & 50 & 1.10 & 4930 &  15.7 & 17.9 & 0.32 & $1.09$ & 3000 & 0.79 & "D" \\ \hline
%C & 110-400-2-50  & $6.17$ & 0.400 & 50 & 1.10 & 4930 &  17.2 & 19.4 & 0.01 & $1.18$ & 2400 & 0.90 & M \\ \hline

G & 120-050-2-50  & $14.8$ & 0.050 & 50 & 1.20 & 4250 &  14.4 & 17.5 & 0.90 & $1.10$ & 3010 & 0.83 & "X" \\
G & 120-100-2-50  & $14.8$ & 0.100 & 50 & 1.20 & 4250 &  16.7 & 19.8 & 0.39 & $1.00$ & 2790 & 0.92 & "D" \\
G & 120-150-2-50  & $14.8$ & 0.200 & 50 & 1.20 & 4250 &  18.9 & 22.1 & 0.26 & $1.56$ & 2570 & 0.96 & "D" \\ 
G & 120-200-2-50  & $14.8$ & 0.150 & 50 & 1.20 & 4250 &  20.2 & 23.2 & 0.27 & $1.61$ & 2440 & 1.00 & "D" \\ \hline

H & 090-150-0-50  & $1.67$ & 0.150 & 50 & 0.90 & 7160 &  5.68 & 6.95 & 2.35 & $1.63$ & 2080 & 0.15 & "Y" \\
H & 090-150-2-50  & $1.67$ & 0.150 & 50 & 0.90 & 7160 &  5.77 & 7.04 & 2.35 & $1.51$ & 2100 & 0.14 & "Y" \\
H & 090-150-6-50  & $1.67$ & 0.150 & 50 & 0.90 & 7160 &  5.79 & 7.06 & 2.35 & $1.47$ & 2100 & 0.14 & "Y" \\
H & 090-150-10-50 & $1.67$ & 0.150 & 50 & 0.90 & 7160 &  5.80 & 7.07 & 2.35 & $1.73$ & 2080 & 0.12 & "Y" \\ \hline

I & 110-100-0-50  & $6.17$ & 0.050 & 50 & 1.10 & 4930 &  11.3 & 13.8 & 0.94 & $1.02$ & 3430 & 0.67 & "X" \\
I & 110-100-2-50  & $6.17$ & 0.100 & 50 & 1.10 & 4930 &  11.1 & 13.0 & 0.94 & $1.04$ & 3430 & 0.62 & "X" \\
I & 110-100-6-50  & $6.17$ & 0.050 & 50 & 1.10 & 4930 &  11.2 & 13.4 & 0.93 & $1.01$ & 3410 & 0.51 & "X" \\
I & 110-100-10-50 & $6.17$ & 0.050 & 50 & 1.10 & 4930 &  11.8 & 13.9 & 0.93 & $1.01$ & 3410 & 0.52 & "X" \\ \hline

\end{tabular}
\end{center}
\end{table*}

\begin{table*}
\caption{$(cont'd)$ The initial models and their parameters.}
\begin{center}
\label{table:modelsb}
\begin{tabular}{|c|c|c|c|c|c|c|c|c|c|c|c|c|c|}
\hline
Group & Model & $\rho_c$ & $M_{{\rm He}}$ & $R_{{\rm seed}}$ & $M$ & $R$ & $E_{{\rm fin}}$ & $E_{{\rm nuc}}$ & 
$t_{{\rm det}}$ & $\rho_{{\rm det}}$ & $R_{{\rm det}}$ & $M_{{\rm Ni}}$ & Type \\ \hline

J & 105-125-0-50  & $4.33$ & 0.125 & 50 & 1.05 & 5300 &  9.85 & 12.2 & 0.96 & $1.06$ & 3580 & 0.56 & "X" \\
J & 105-125-2-50  & $4.33$ & 0.125 & 50 & 1.05 & 5300 &  10.3 & 15.2 & 0.94 & $1.04$ & 3560 & 0.57 & "X" \\
J & 105-125-6-50  & $4.33$ & 0.125 & 50 & 1.05 & 5300 &  10.3 & 12.4 & 0.96 & $1.03$ & 3560 & 0.50 & "X" \\
J & 105-125-10-50 & $4.33$ & 0.125 & 50 & 1.05 & 5300 &  10.0 & 11.9 & 0.96 & $1.04$ & 3560 & 0.43 & "X" \\ \hline

K & 110-100-2-50  & $6.17$ & 0.100 & 50  & 1.10 & 4930 & 11.1 & 13.0 & 0.94 & $1.04$ & 3430 & 0.62 & "X" \\
K & 110-100-2-100 & $6.17$ & 0.100 & 100 & 1.10 & 4930 & 11.1 & 13.1 & 0.93 & $1.04$ & 3430 & 0.62 & "X" \\
K & 110-100-2-150 & $6.17$ & 0.100 & 150 & 1.10 & 4930 & 11.8 & 13.6 & 0.93 & $1.06$ & 3410 & 0.65 & "X" \\ \hline

L & 105-050-2-S50  & $4.33$ & 0.050 & 50 & 1.05 & 5300 &  12.9 & 14.7 & 0.94 & $1.73$ &   40 & 0.50 & "S" \\
L & 105-050-2-50   & $4.33$ & 0.050 & 50 & 1.05 & 5300 & -1.73 & 0.31 & nil  & nil & nil & $8.6 \times 10^{-3}$ & "N" \\
L & 105-050-2-2R50 & $4.33$ & 0.050 & 50 & 1.05 & 5300 &  9.61 & 11.6 & 1.48 & $4.63$ & 3090 & 0.48 & "Y" \\ 
L & 105-050-2-3R50 & $4.33$ & 0.050 & 50 & 1.05 & 5300 & -1.67 & 0.33 & nil & nil & nil & $9.86 \times 10^{-2}$ & "N" \\ \hline

M & 090-150-2-50  & $1.67$ & 0.150 & 50 & 0.90 & 7160 &  5.65 & 7.04 & 2.35 & $1.51$ & 2100 & 0.14 & "Y" \\
M & 095-150-2-50  & $2.23$ & 0.150 & 50 & 0.95 & 6710 &  8.64 & 10.1 & 1.33 & $1.00$ & 4000 & 0.32 & "X"\\
M & 100-150-2-50  & $3.21$ & 0.150 & 50 & 1.00 & 6180 &  8.64 & 10.3 & 0.99 & $1.06$ & 3370 & 0.47 & "X" \\ 
M & 105-150-2-50  & $4.33$ & 0.150 & 50 & 1.05 & 5300 &  13.6 & 14.7 & 0.73 & $1.04$ & 3360 & 0.71 & "X" \\
M & 110-150-2-50  & $6.17$ & 0.150 & 50 & 1.10 & 4930 &  14.8 & 15.4 & 0.45 & $1.03$ & 3190 & 0.69 & "D" \\
M & 115-150-2-50  & $9.19$ & 0.150 & 50 & 1.15 & 4550 &  16.5 & 17.7 & 0.33 & $1.09$ & 2910 & 0.86 & "D" \\
M & 120-150-2-50  & $14.8$ & 0.150 & 50 & 1.20 & 4250 &  18.9 & 22.1 & 0.26 & $1.56$ & 2570 & 0.96 & "D" \\ \hline

\end{tabular}
\end{center}
\end{table*}
%\end{longtable}

\subsection{Model Names}

First we describe how these models are named and 
how they are chosen.
Each model is named by their parameters, including $M$, $M_{{\rm He}}$, $Z$,
and the initial position of the detonation bubble (sphere).  
For example, Model 105-050-2-50 stands for 
a WD with $M =$ 1.05 $M_{\odot}$, $M_{{\rm He}} =$ 
$0.05 M_{\odot}$, $Z = 0.02$ and
the initial He detonation triggered at 50 km above the core-envelope
interface. 

The endings "-S50" and "-B50" stand for 
different initial He detonations. The term "S50" stands
for a spherical detonation triggered at 50 km
above the He/CO interface and "B50" stands 
for a belt (ring) around the "equator" of the WD. 
"R50" stands for a bubble triggered at 50 km above
the He/CO interface. For "2R50" or "3R50" we put two or three "bubbles"
(a combination of torus and bubble) in the He-envelope. 
Note that
with the rotation and reflection symmetry, a bubble in the two-dimensional plane  
can be a "ring"  if the bubble is away from the rotation-axis,
in its three-dimensional projection.
The distance 50 km is chosen such that the surface of the initial bubble
is slightly separated by at least one grid from the interface. 
We find that this separation is necessary to avoid overlapping
the initial He-detonation bubble with the CO-rich matter.

In Groups A - M in Table 1, the following effects are studied:

\noindent (1) Initial mass $M$:}
Groups A, B, C and M 
study the effects of progenitor
mass on nucleosynthesis. For initial detonation with 
higher symmetry ("-S50" and "-B50" series), a
lower He envelope mass $0.05 ~M_{\odot}$ is used
while for that with lower symmetry, the He envelope
mass is fixed at $M_{{\rm He}} = $ 0.10 or 0.15 $M_{\odot}$. 
Metallicity is fixed at the solar metallicity. The progenitor varies from 
0.9 to 1.2 $M_{\odot}$. 

\noindent (2) He envelope mass $M_{{\rm He}}$: In Groups
D, E, F and G the effects of He envelope mass $M_{{\rm He}}$.
Each group includes models of the same mass
from 0.9 - 1.2 $M_{\odot}$ in a 0.1 $M_{\odot}$ interval,
but with a different $M_{{\rm He}}$ from 0.05
to 0.2 $M_{\odot}$. In all models, solar metallicity
is assumed. We remark that different 
$M_{{\rm He}}$ masses are used for different 
initial detonation geometries. It is because
for a He detonation with a lower symmetry,
the effects of shock convergence by geometry
is smaller. To make sure the second detonation 
can be triggered for comparison, a higher $M_{{\rm He}}$
are studied.

\noindent (3) Metallicity $Z$: In Groups H, I 
and J we study the effects of metallicity to 
the explosive nucleosynthesis.
Each group consists of models of the same
$M$, $M_{{\rm He}}$ and detonation configuration.  
Models vary by their metallicity from 0 to 
5 $Z_{\odot}$. We choose this large metallicity
because in \cite{Leung2018} we have already
shown that such high metallicity model can 
be a clue to explain the observe SN remnants.
For Groups H and I we pick these models because
they are the benchmark models of our 
sub-Chandrasekhar mass SN Ia models.

\noindent (4) Initial detonation geometry:
In Groups K and L we study the effects of 
initial He detonation geometry. All
models have the same $M$, $M_{{\rm He}}$
and at the solar metallicity. 
Group K consists of models with the detonation 
seed at different positions.
Group L consists of models
with different detonation geometry. It spans from
different number of detonation "bubbles"  
to those with a higher symmetry, such as spherical detonation.

\section{Detonation Trigger}

In the last column we classify the trigger mechanisms
into four types. In all simulation groups (from Group C to M except Group L), 
all He detonations are started by placing an 
detonation spot at the radius 50 km along the rotation-axis.
This mimics a single hot spot that induces
thermonuclear runaway in the
form of a bubble. In general, the detonation propagates
along the He envelope without penetrating into
the CO core. Depending on $M_{{\rm He}}$
and the interface density, different detonation types
are observed. 

\subsubsection{Type "N"}
Type "N" (no detonation) stands for no second detonation occurring. 
Type "N" can be found in models with a thin He
envelope, the shock wave sent by the He detonation is not 
strong enough to compress the matter at both the center and
the surface of the CO core. The CO core
has a temperature always below the threshold temperature. 

\subsubsection{Type "D"}
Type "D" (diagonal) stands for that the detonation first appears along elsewhere
other than the symmetry axis.
Type "D" occurs for models with high progenitor masses. In these cases,
the typical density of interface can be as high 
as $\sim 10^{7}$ g cm$^{-3}$. 

\begin{figure}
\centering
\includegraphics*[width=8cm,height=7cm]{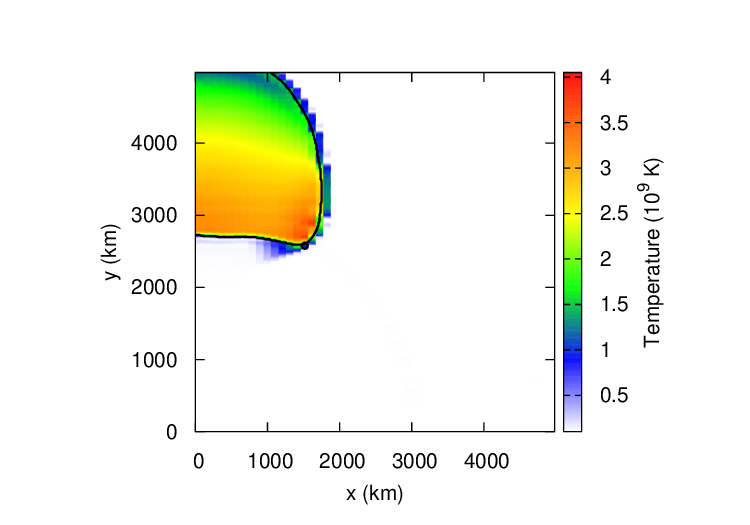}
\includegraphics*[width=8cm,height=7cm]{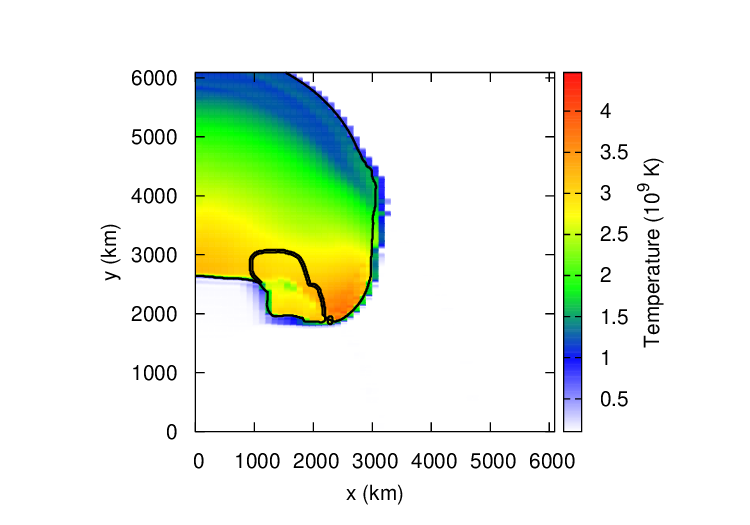}
\includegraphics*[width=8cm,height=7cm]{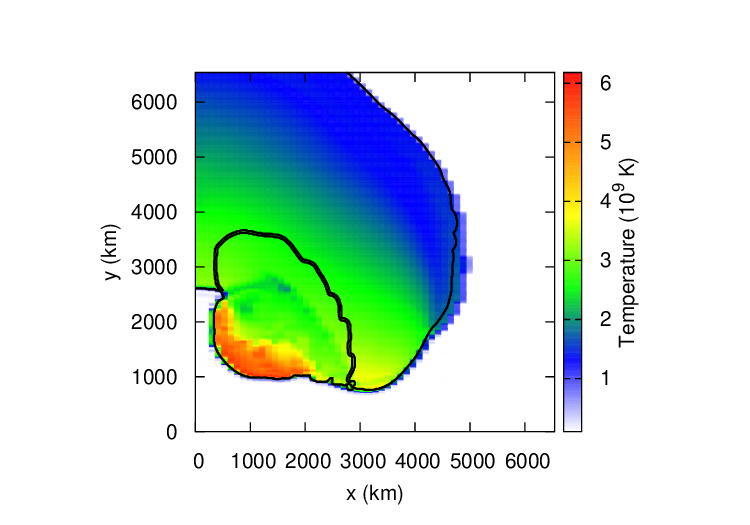}
\caption{The flame and detonation geometry and the temperature 
for Model 110-200-2-R50 (D). The detonation is captured at -0.34, 0.03, 
0.15 and 0.40 s from the detonation transition.}
\label{fig:det_M}
\end{figure}

In Figure \ref{fig:det_M}
we plot the temperature and explosion geometry for the
Model 110-150-2-50 (D). When the detonation reaches
the interface, the temperature of the CO matter
can easily reach the critical temperature to start the CO detonation. 

The temperature can reach $3 \times 10^9$ K
where the shock penetrates. Notice that even
the temperature in other detonated part can reach
$\sim 2 \times 10^9$ K. The propagation is along the iso-density 
contour, where there is almost no heating in the radial
direction. This makes no heating in the CO material.
Therefore, while the He detonation is still burning the 
matter in envelope, the second detonation is already triggered.

\subsubsection{Type "X"}

Types "Y" and "X" stand for the detonation which is first started 
along the rotation axis (in x-y plane the y-axis) and symmetry axis
(in x-y plane the x-axis).

Type "X" occurs when Type "D" cannot be started. This applies to
models with lower $M$. Notice that in our simulations, 
a quarter of the star is simulated. When the detonation propagates, its
burning rate increases due to the ring-shape structure, which 
has a local volume proportional to $r$. 
When the detonation 
approaches the symmetry-axis, the high velocity flow creates a 
strong compression of the remaining fuel. By symmetry, part of the fuel
is compressed towards the core. This heats up the near-interface
material and provides the required temperature for the first spot. 

\begin{figure}
\centering
\includegraphics*[width=8cm,height=7cm]{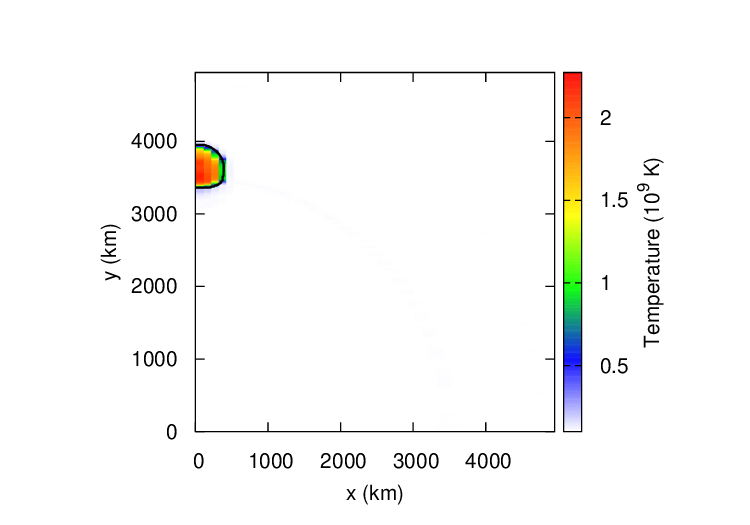}
\includegraphics*[width=8cm,height=7cm]{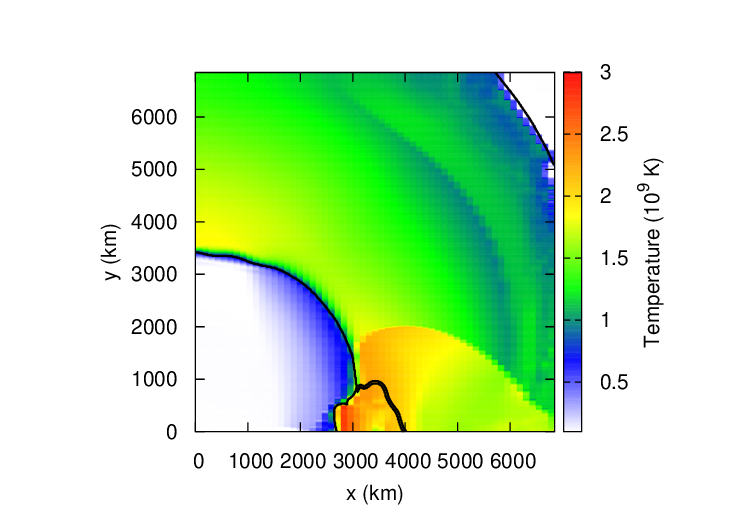}
\includegraphics*[width=8cm,height=7cm]{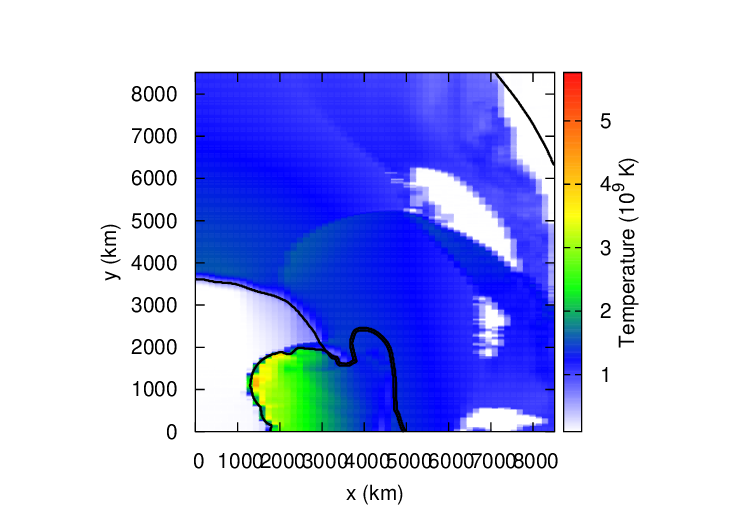}
\caption{Similar to Figure \ref{fig:det_M}, but for Model
110-100-2-R50 (Type "X"). The detonation is captured at -0.82, %0.02,
0.15 and 0.40 s from the detonation transition.}
\label{fig:det_R}
\end{figure}

Figure \ref{fig:det_R} shows a typical "X"-type detonation
for Model 110-100-2-50 (X). The 
second detonation is triggered at the $r$-axis,
where the detonation wave compresses materials. The temperature 
due to the compression at the r-axis can be higher than the 
temperature rise in other regions due to detonation heating. 
As an example, the actual temperature 
can reach $3 \times 10^9$ K near the $r$-axis compared to 
other region which is $\sim 2 \times 10^9$ K. 
We remark that this shock heating is not related
to the geometric convergence. Here, the detonation
waves approach the symmetry boundary, i.e. 
two laminar detonation waves approaching each other
(the collision site along the equator is locally flat).

\subsubsection{Type "Y"}

\begin{figure}
\centering
\includegraphics*[width=8cm,height=7cm]{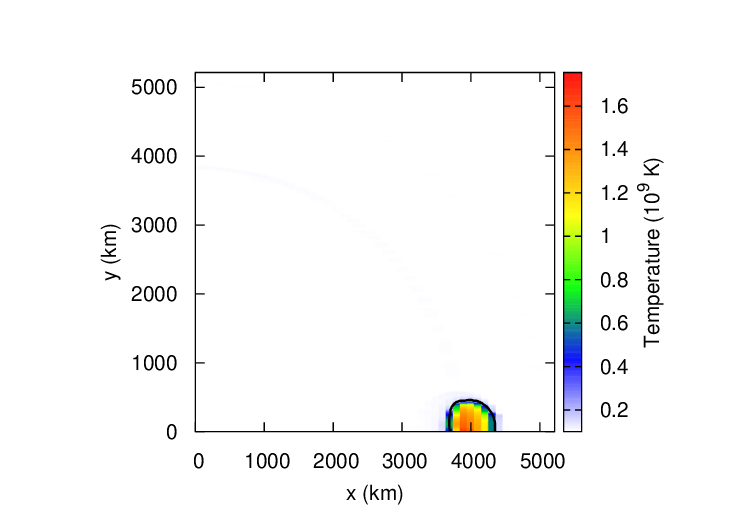}
\includegraphics*[width=8cm,height=7cm]{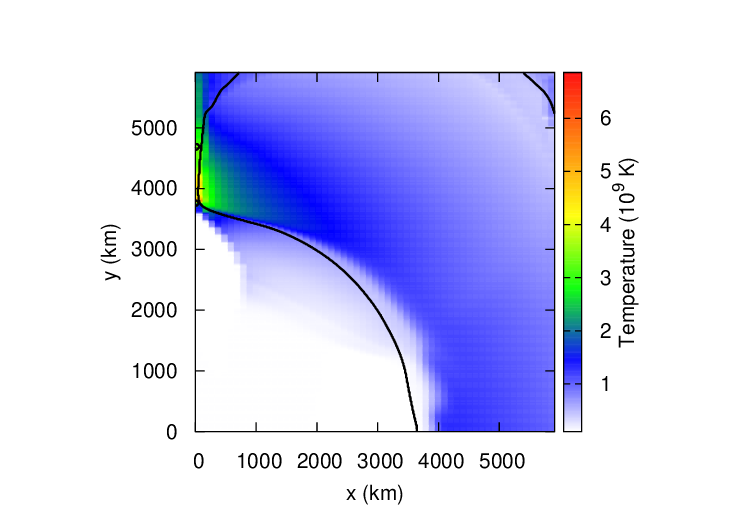}
\includegraphics*[width=8cm,height=7cm]{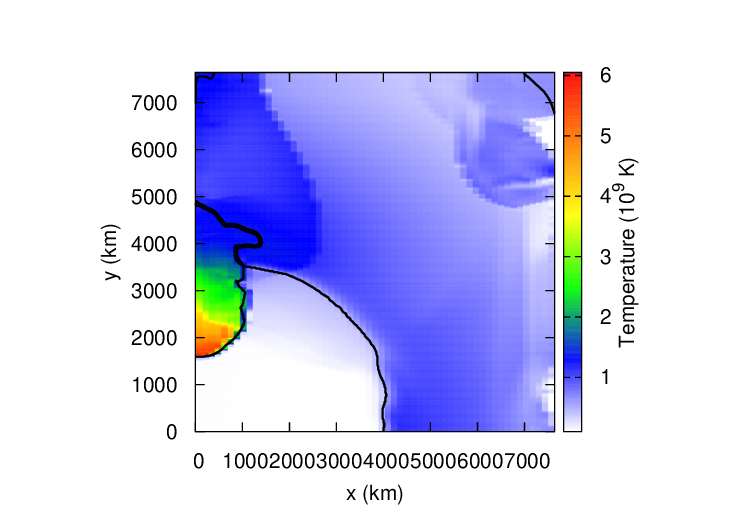}
\caption{Similar to Figure \ref{fig:det_R}, but for Model
110-050-2-B50 (Y). The detonation is captured at -0.98, 
0.05 and 0.22 s from the detonation transition.}
\label{fig:det_Z}
\end{figure}

In Figure \ref{fig:det_Z}
we demonstrate the "Y"-Type detonation by using 
Model 110-050-2-B50 (Y) as an example.
Type "Y" occurs when both Type "X" and Type "D" cannot be 
triggered beforehand. After the He shell is fully burnt, the 
first converging shock is not strong enough to detonate CO matter near
interface. Instead, the mild shock continues to travel along 
the density-contour in the envelope. 
The flow creates another converging shock when the shock front
returns to the rotation axis, which again creates the first
hot spot for the C detonation. 
 
\subsubsection{Type "S"} 
 
Type "S" can be found in models with detonation seeds which have
spherical symmetry, while the He envelope is not massive enough to 
ignite the near-interface C. The converging shock creates the
hot spot at the center, where the geometric enhancement is
the strongest. 

\begin{figure}
\centering
\includegraphics*[width=8cm,height=7cm]{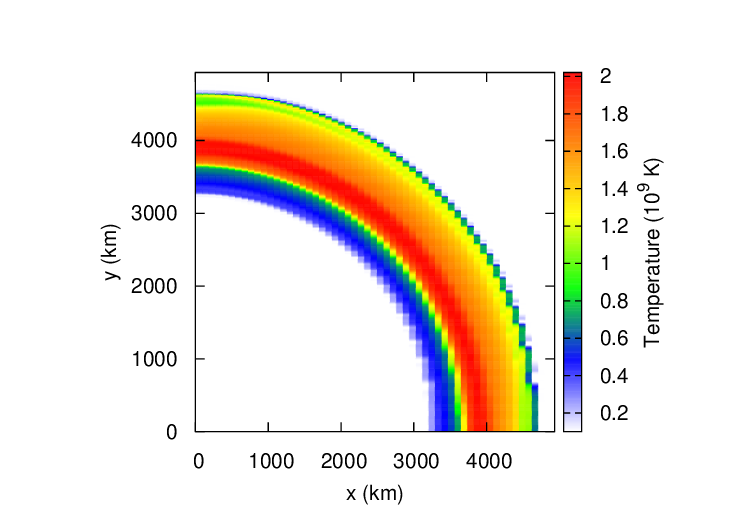}
\includegraphics*[width=8cm,height=7cm]{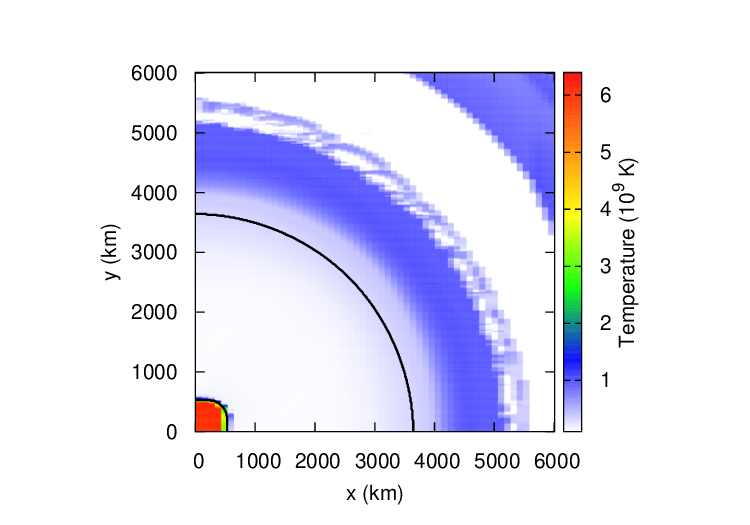}
\includegraphics*[width=8cm,height=7cm]{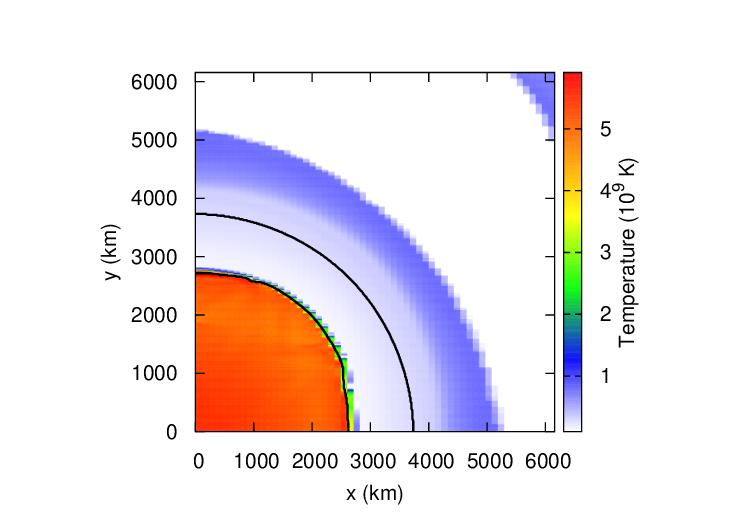}
\caption{Similar to Figure \ref{fig:det_R}, but for Model
110-050-2-S50 (S). The detonation is captured at -0.72, 
0.08 and 0.32 s from the detonation transition.}
\label{fig:det_S}
\end{figure}

In Figure \ref{fig:det_S} we plot Model
110-050-2-S50 (S). In contrast to the other 3 cases,
the spherical detonation allows the envelope to be burnt
much faster. In the plot, the He shell has expanded 
and cools down mostly, leaving almost a mild trace in the 
temperature distribution. On the contrary, the center, where
the C detonation starts, can reach as high as 
$6 \times 10^9$ K, sufficiently high for the burnt matter to reach NSE. 

\subsection{Thermodynamics}

\begin{figure}
\centering
\includegraphics*[width=7cm,height=5cm]{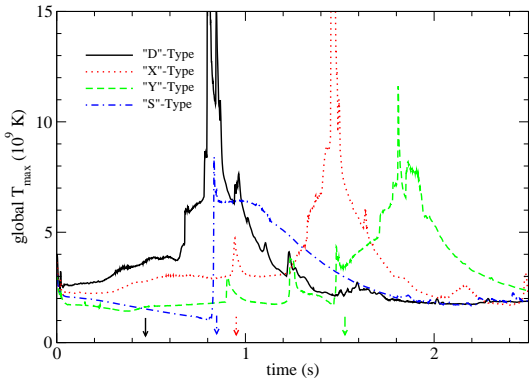}
\caption{Maximum temperature in the simulations against
time for Models 110-150-2-50 (D), 110-100-2-50 (X), 110-100-2-2R50 (Y)and
110-050-2-S50 (S) respectively. The arrows stand for the 
time where DDT is triggered for each model.}
\label{fig:temp_max}
\end{figure}

In Figure \ref{fig:temp_max} we plot 
the maximum temperature against time for the four models
presented. The letters in the figure correspond to the 
threshold temperature where the C detonation is triggered.
The temperature needed to trigger 
the C detonation is the global maximum temperature in the simulations
for Type "S", "X", "Y" detonations but not for Type "D" detonation. 
The global maximum temperature reaches its 
peak during the trigger of second detonation for Types
"S", "X" and "Y". No such peak is observed for Type "D" detonation.
Furthermore for Type "S" detonation, the maximum temperature
when the C-detonation is triggered is the highest temperature
reached in the simulation.
This means that 
for a non-spherical trigger, even the hot ash can be
higher than the threshold temperature, unless certain shock convergence occurs,
the CO matter near interface can remain a temperature
below the critical temperature.

Another feature
is that in most cases, when the C detonation
approaches the center, nuclear burning, despite at
its low density, can be enhanced when the convergence effect
is strong. This effect is robust under different 
resolution, and is even stronger when a finer resolution is used.
It is because the shock strength can increase in the 
way $\sim 1/r$ for a cylindrical detonation 
and $\sim 1/r^2$ for a spherical. Locally, the 
density growth in the core will be higher for a 
finer resolution run, which allows more rapid 
reactions. However, globally the energy production
will not diverge because the finer the resolution
is, the smaller contribution such temperature peak 
gives. On the other hand, by using different 
geometry (e.g. Cartesian Coordinate) or higher 
dimensions (i.e. three-dimensional model), the level
of shock convergence will be changed because it 
depends on how the geometry describes the structure
with a high symmetry such as a ring or a sphere.

The peak temperature,
albeit contributing to an extremely small 
amount of mass $(\sim 10^{-8-11} M_{\odot})$, 
can reach above $10^{10}$ K. One feature in "Y"-Type detonation
does not appear in other types of detonation, namely the 
multi-peaks prior to detonation. This reflects the shock
interaction from multiple detonations. For example, they 
correspond to the first collision of He detonations,
the arrival of the reflected shock on the $r$-axis
and the $z$-axis respectively. 

This shows that the exact peak temperature
can vary a lot depending on the geometric convergence.
But how the convergence of shock and its subsequent divergence
in temperature take place are related to the spatial resolution.
In Appendix \ref{sec:test2} we preform a numerical study to see
how the spatial resolution affects the thermodynamics properties
in local and global properties in some of the explosion models.

Below, we discuss the hydrodynamics behaviour of these models.

\subsubsection{Effects of He Envelope Mass} 
In Groups D, E, F and M we cover the effects of $M_{{\rm He}}$ 
for different progenitor masses from 0.9 - 1.2 $M_{\odot}$. 
Some common trends can be seen in these series. 
At low $M_{{\rm He}}$, no second detonation can be triggered.
By increasing $M_{{\rm He}}$, the second detonation can be
triggered by "Y"-Type, "X"-Type and then "D"-Type
in ascending $M_{{\rm He}}$. The created $^{56}$Ni increases
with $M_{{\rm He}}$. The explosion energy follows the same trend. 
Due to the change of detonation channel, the detonation 
trigger time becomes earlier for a higher $M_{{\rm He}}$.
We remark that "S"-Type is independent from other
three detonation types because it requires always a
spherical He-detonation independent of $M_{{\rm He}}$. 
Models with a high $M_{{\rm He}}$
favours the "D"-Type detonation. 
For $M \geqslant 1.0 ~M_{\odot}$, transition from 
"X"-Type to "D"-Type detonation occurs when $M_{{\rm He}} 
> 0.1 - 0.15 ~M_{\odot}$. For $M \leqslant 1.0 ~M_{\odot}$, 
transition from "Y"-Type to "X"-Type and then "D'-Type 
takes place for the transition $M_{{\rm He}}$ at 
0.15 and 0.2 $M_{\odot}$ respectively. 
The critical $M_{{\rm He}}$ where no second detonation
takes place depends on $M$, which decreases when $M$ increases,
and the detonation geometry. He-detonation with rotation or 
spherical geometry can trigger the second detonation
with $M_{{\rm He}}$ as low as 0.05 $M_{\odot}$.
For other types of He-detonation, 
the minimum value of $M_{{\rm He}}$ is $\sim 0.15 M_{\odot}$ for $M = 0.9 - 1.0 ~M_{\odot}$,
0.1 $M_{\odot}$ for $M = 1.1 ~M_{\odot}$ and 0.05 $M_{\odot}$ for 
$M = 1.2 ~M_{\odot}$.

\subsubsection{Effects of Metallicity}
In Groups H, I and J we cover the effects of $Z$ for
three different models. The latter two are the benchmark
models decided by its $M_{{\rm Ni}}$ at solar metallicity.
We can see that metallicity has a very mild influence
on the explosion energy and final energy. The detonation
position, its channel and its trigger time are 
insensitive to $Z$.
The major difference can be seen from the the $^{56}$Ni
mass, which drops when $Z$ increases.

\subsubsection{Effects of detonation pattern}
In Groupd K and L we explore the effects of the detonation pattern,
determined by its initial He detonation spot and 
its geometry. Again, all models share the same $M$, $M_{{\rm He}}$ and $Z$.
The initial He detonation spot has 
almost no impact on the explosion energetics and 
explosion properties. The Ni-production is 
also insensitive to the change of detonation 
position. On the other hand, the choices of
detonation geometry is very influencing
to the explosion properties. Some models (Models
105-050-2-50 (N) and 105-050-2-3R50 (N)) cannot trigger
C detonation spontaneously while some (Models 105-050-2-S50 (S)
and 105-050-2-2R50 (Y)) can. This reflects that the 
symmetry of the initial shock and how the detonation
waves collide with each other determine the 
final fate of the WD. The resultant $M_{{\rm Ni}}$ 
can vary from $\sim 10^{-2}$ $M_{\odot}$ in 
a failed detonation to $\sim$ 0.5 $M_{\odot}$ 
in a successful detonation. 
We note that Models 105-050-2-3R50 (N)
and Model 105-050-2-2R50 (Y) behave differently. 
To show that the result is robust in our study, in 
Appendix \ref{sec:test3} we do a resolution study
to demonstrate how the trigger of C-detonation
depends on the spatial resolution.

\subsubsection{Effects of Initial Mass}
In Group J we explore the effects of $M$ on the
explosion energetics. Compared to the near-Chandrasekhar
mass WD studied in \cite{Leung2018}, the mass range for 
sub-Chandrasekhar mass is much wider (from
0.9 to 1.2 $M_{\odot}$). We do not explore mass
below 0.9 $M_{\odot}$ since the central density 
of these models is below $10^7$ g cm$^{-3}$,
where the incomplete burning dominates. We also do not
extend the upper bound to 1.3 $M_{\odot}$ since 
it is unclear, if there is nuclear runaway, 
whether the explosion is carried out as
deflagration or detonation \citep{Nomoto1976,Nomoto1982b,Nomoto1984b}.
We can see that when $M$ increases, some effects are
similar as increasing $M_{{\rm He}}$. The explosion 
energy increases. Also, the explosion time becomes
earlier with its position being closer to the core. 
The detonation channel also changes from Type "Y"
to Type "X" and then Type "D".

%%%%%%%%%%%%%%%%%%%%%%%%%%%%%%%%%%%%%%%%%%%%%%%%%%%%%%%%%%%%%%%%%%%%%%%%

\section{Benchmark Models}
\label{sec:benchmark}

In this section, we study in details some models
which behave most similar to a standard Type Ia supernova, 
determined by their $^{56}$Ni production, which should 
be $\sim$ 0.6 $M_{\odot}$ as observed in the majority of 
normal SNe Ia. Since there is a degeneracy in the models 
to produce this feature, we pick the one with the 
lowest amount of $M_{{\rm He}}$.
We selected Models 110-100-2-50 (X), 100-050-2-S50 (S) and 
110-050-2-B50 (Y). All of them
have a healthy explosion of $^{56}$Ni mass $\sim 0.6$ $M_{\odot}$.

In contrast, for sub-Chandrasekhar
mass WD, we do not impose the constraints of
Mn and Ni as what we have done in \cite{Leung2018}
because all models we built always underproduce
Mn and Ni. Chosen by the $^{56}$Ni production, 
there exists a degeneracy of models which satisfy
this constraint. As a result, from each 
detonation trigger, we choose one model with 
$M_{{\rm Ni}} = 0.6 ~M_{\odot}$. They include
Models 110-100-2-50 (X), 105-050-2-B50 (Y) and 100-050-2-S50 (S).
No model with Type "D" produces an explosion  
with $M(^{56}$Ni) $\sim 0.6 M_{\odot}$.

\subsection{Energy Evolution}

\begin{figure}
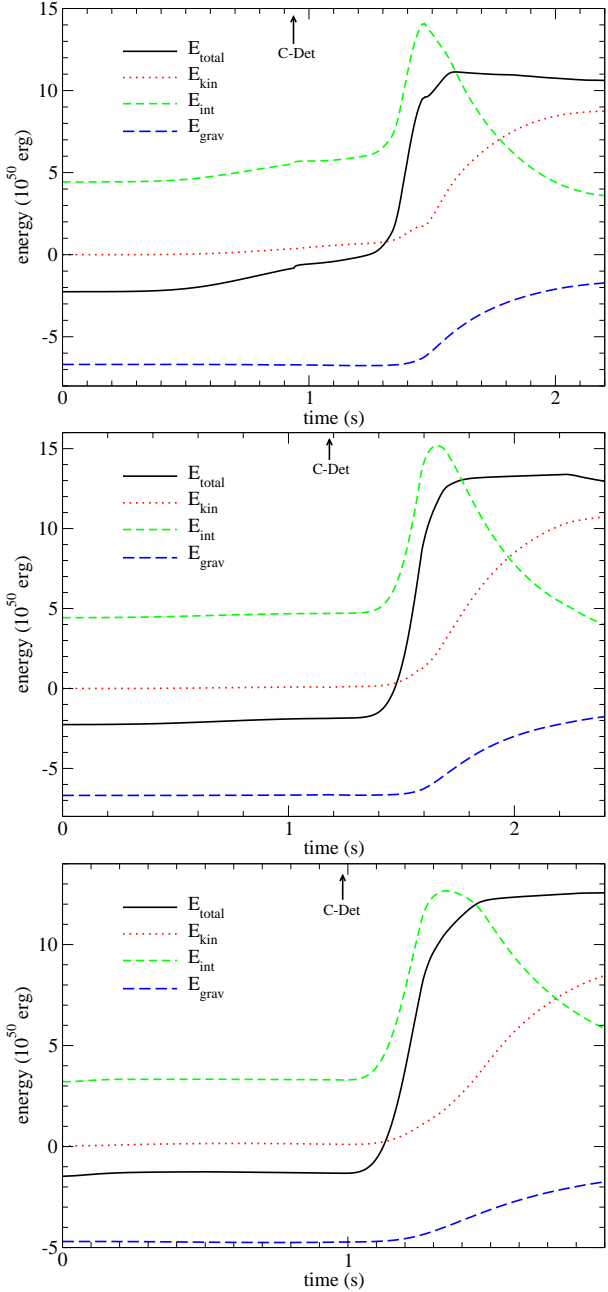

\centering
\includegraphics*[width=8cm,height=5.7cm]{fig6a.eps}
\includegraphics*[width=8cm,height=5.7cm]{fig6b.eps}
\includegraphics*[width=8cm,height=5.7cm]{fig6c.eps}
\caption{The total energy, kinetic energy, internal energy
and gravitational energy against time for the Models 110-100-2-50 (X)
(top panel), 110-050-2-B50 (Y) (middle panel) and 100-050-2-S50 (S)
(bottom panel).}
\label{fig:energy_benchmark1}
\end{figure}

In Figure \ref{fig:energy_benchmark1} we plot the time evolution
of the total energy, kinetic energy, internal energy 
and gravitational energy for the benchmark 
models. 
Here we give an analysis on the energy evolution
of only Model 110-100-2-50 (X). The other two benchmark
models have similar evolution as this one, except at
different detonation triggers and different 
He detonation convergence effects.

Before 0.9 s, there is
only He detonation. The energy release  
$\sim 1 \times 10^{50}$ erg is insufficient
to unbound the star due to the small amount of He
and its low density. There is almost no change in the gravitational
energy and kinetic energy. Almost all the energy change is 
reflected in the change of internal energy, showing that 
the He detonation does not influence the global dynamics. 
From 0.9 to 1.6 s C detonation takes place. The total energy 
sharply increases by $10^{51}$ erg at $\sim$ 1.3 s, showing that the 
C detonation is rapidly turning the CO fuel into 
ash. At the same time, the internal energy, gravitational 
energy and kinetic energy increase. The C detonation
is strong enough to heat up the WD, and causes the subsequent
expansion. Beyond 1.6 s, the total energy
remains a constant, signifying the end of both He and C detonations. 
Simultaneously, the internal energy drops while internal energy
and gravitational energy increase and reach their equilibrium
values at $\sim$ 2 s. This corresponds to the phase that the 
thermalized ash is quickly expanding to accelerate the matter
outwards until homologous expansion is developed.

\subsection{Luminosity evolution}

\begin{figure}
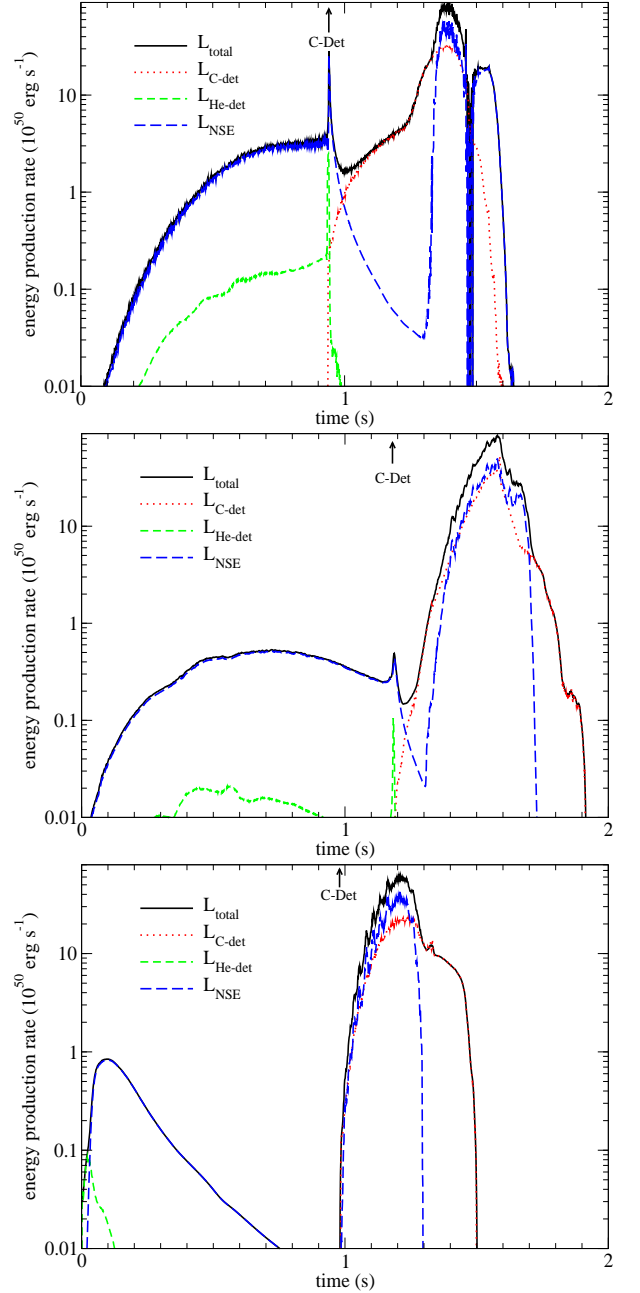

\centering
\includegraphics*[width=8cm,height=5.7cm]{fig7a.eps}
\includegraphics*[width=8cm,height=5.7cm]{fig7b.eps}
\includegraphics*[width=8cm,height=5.7cm]{fig7c.eps}
\caption{The total luminosity, C detonation luminosity, 
He detonation luminosity and NSE burning luminosity
against time for Models 110-100-2-50 (X) (top panel), 110-050-2-B50 (Y)
(middle panel) and 100-050-2-S50 (S) (bottom panel).}
\label{fig:lumin_benchmark}
\end{figure}

In Figure \ref{fig:lumin_benchmark} we plot
the luminosity of the three benchmark models
110-100-2-50 (X), 110-050-2-B50 (Y), 100-050-2-S50 (S) in 
the top, middle and bottom panels respectively.
First, we analyze the evolution of Model 110-100-2-50 (X).

Before 0.9 s, the 
total luminosity and the He detonation overlap with 
each other. This means most energy is produced directly from
detonation where NQSE and NSE do not actively contribute to
the energy evolution. There is a peak at 0.9 s, which is the 
moment where the He detonation reaches the symmetry
axis. The compression causes a sudden jump in the density
and temperature, which allows He burning to proceed
much more efficiently. After that the He detonation
ceases as there is not any pure He left but 
only partially burnt He in the ash. At $t = 0.9$ s, the 
C burning takes place to the major nuclear reactions. 
But there is no advanced burning, showing that the 
detonation is still incinerating material in the low
density region. At $t = 1.5$ s, the advanced burning 
exceeds the C detonation to become the major energy production
channel. This shows that the detonation has finally reaches the
center, which is dense and hot enough to carry out 
silicon burning up to NSE. Around 1.5 s, the C detonation
begins to cease. Also, beyond 1.6 s, all matter becomes too 
cold or of too low density for further exothermic nuclear reactions to occur. 

Model 110-050-2-B50 (Y) has a similar evolution to Model 110-100-2-50 (X)
but has the "Y"-Type detonation. The shape of the energy production
rates are similar but with two major differences. However
the delay between the C-detonation and NSE luminosity rise is shorter
than the Type-"X" detonation model. This feature is similar to 
the "S"-Type detonation (See below), despite its off center
ignition. One reason is that during the geometric convergence, 
not only it heats up the CO-rich matter below the interface,
but also generating a strong inward flow, which helps to guide 
the detonation reaching high density region. Such channeling
is weaker in the "X"-Type model due to the absence of geometric
convergence.

Model 110-050-2-S50 (S) is the "S"-Type detonation.
It has a different structure from the other two by the absence of 
He-burning peak at the onset of second detonation and the 
similarity between the total luminosity and that by the NSE burning. 
Due to the detonation symmetry, there is no geometric convergence 
for the He-detonation. The He-detonation creates an inward
moving shock while propagating outwards to burn the remaining He. 
Hence no luminosity peak during the transition is observed. 
Then, after the C-detonation is triggered, the 
total energy release, NSE burning and C-burning closely 
follow each other. It is because the detonation starts from
the center. The higher density compared to the envelope allows
the burning reaching NSE much shorter than dynamical timescale. 
This feature is not observed in Type "X"- or "Y"-Type detonation.
At 1.3 s the energy production by C-burning drops rapidly,
showing that the detonation wave has finished sweeping 
all C-fuel in the star. Accompanying with the expansion of the 
star, the recombination of $^{4}$He into $^{56}$Ni becomes the 
only energy production, which also ceases at 1.5 s.

\subsection{Chemical abundance}

\begin{figure}
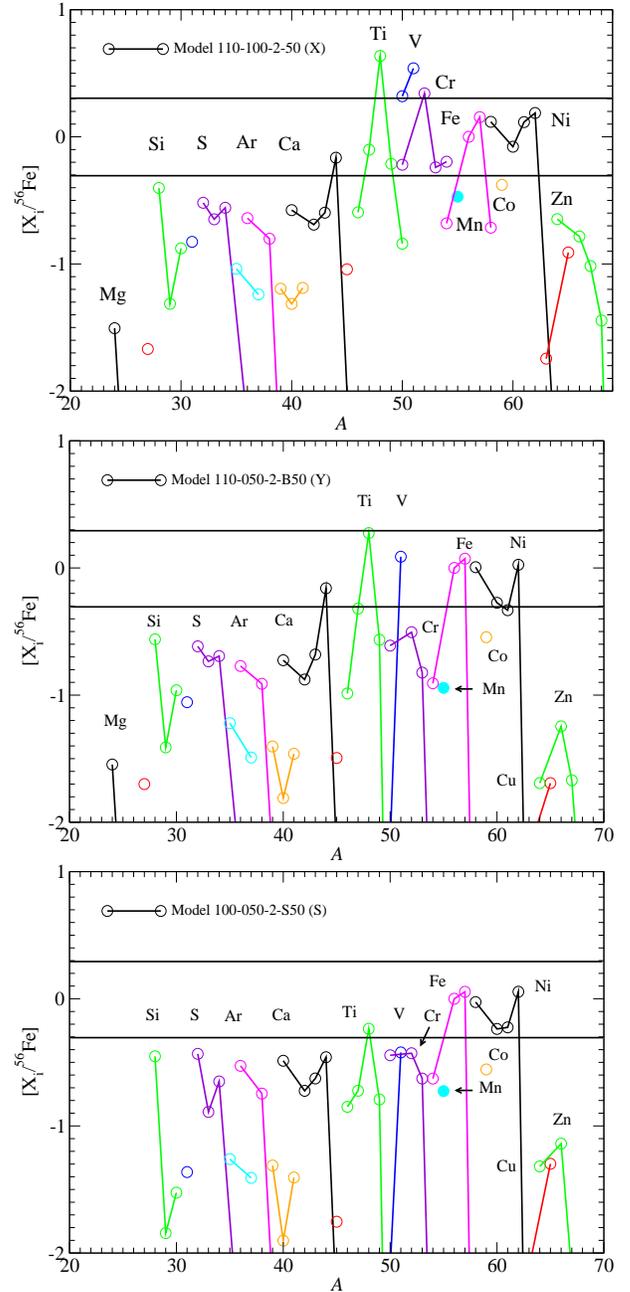

\centering
\includegraphics*[width=8cm,height=5.7cm]{fig8a.eps}
\includegraphics*[width=8cm,height=5.7cm]{fig8b.eps}
\includegraphics*[width=8cm,height=5.7cm]{fig8c.eps}
\caption{$[X_{i}/^{56}$Fe]
after all short-lived isotopes have decayed for Models
110-100-2-50 (X) (top panel) and 110-050-2-B50 (Y) (middle panel)
and 100-050-2-S50 (S) (bottom panel). $[X_{i}/^{56}$Fe]
is defined as
$\log_{10} (X_i/X({{\rm ^{56}Fe}})) - \log_{10} (X_i/X({{\rm ^{56}Fe}}))_{\odot}$.
The upper and lower horizontal lines stand for 
two times and half of the solar value.}
\label{fig:final_benchmark}
\end{figure}

We use the tracer particle scheme to reconstruct the detailed
nucleosynthesis. The massless tracers are advected by 
the fluid motion, but have no effect on the fluid. 
They record the local density and temperature accordingly. 
Here we examine the typical chemical abundances of the 
three benchmark models presented in previous parts.

In Figure \ref{fig:final_benchmark} we plot the final chemical 
abundance of the three benchmark models mentioned above. 
$[X_{i}/^{56}$Fe]
is defined as
$\log_{10} (X_i/X({{\rm ^{56}Fe}})) - \log_{10} (X_i/X({{\rm ^{56}Fe}}))_{\odot}$.

For Model 110-100-2-50 (X), the intermediate mass elements (IMEs)up to $^{40}$Ca
are underproduced. Starting from Ti, the production becomes 
similar to the solar abundance, where some of which are
even overproduced, including $^{48}$Ti, $^{51}$V and 
$^{52}$Cr. They are from 3 to 6 times higher than the 
observed solar values. Most Fe and Ni isotopes are
very close to the solar values. Isotopes beyond Ni
are underproduced. The pattern for Ni where 
$^{60}$Ni and $^{62}$Ni are more abundant can be observed.
Also, as expected, $^{55}$Mn, which comes mostly from 
the low electron fraction matter, 
is underproduced. In order to produce Mn,
two channels are possible. First, Mn can be 
directly formed from NSE when the electron fraction
of the matter is $Y_{{\rm e}} = 0.45$. Second, it is 
formed during alpha-chain burning of $^{52}$Fe, where
$^{52}$Fe$(\alpha,p)^{55}$Co. The $^{55}$Co will later 
decay by $^{55}$Co$(e^-,\nu_e)^{55}$Fe$(e^-,\nu_e)^{55}$Mn. 
The formation of $^{55}$Co is favourable at 
$Y_{{\rm e}} = 0.49$. For pure C+O matter, the $Y_{{\rm e}} = 0.5$,
therefore directly NSE burning without electron capture
or alpha-chain burning cannot form seeds of $^{55}$Mn,
which is the case of sub-Chandrasekhar mass model.

For Model 110-050-2-B50 (Y), the nucleosynthesis pattern
is very similar to the previous model. There are minor 
variations such as the much lower $^{50}$V, and no trace
of $^{54}$Cr. This is because there is no shock convergence
by the C detonation in the core due to the propagation
direction. The effects of hot spot become
less significant in this benchmark model. 

For Model 100-050-2-S50 (S), the nucleosynthesis pattern is 
very different from the previous two models. Due to the
imposed He detonation symmetry, much lower amount of 
He envelope mass is needed to trigger the C detonation.
As a result, the resultant chemical pattern, related
to He burning, is highly suppressed.
A major drop of the abundances in $^{47-48}$Ti, $^{51}$V,
and $^{52}$Cr becomes solar or even sub-solar. Other 
abundances, which are basically the C detonation products, 
remain the same as the two other models.

\subsection{Ejecta Composition}

\begin{figure}
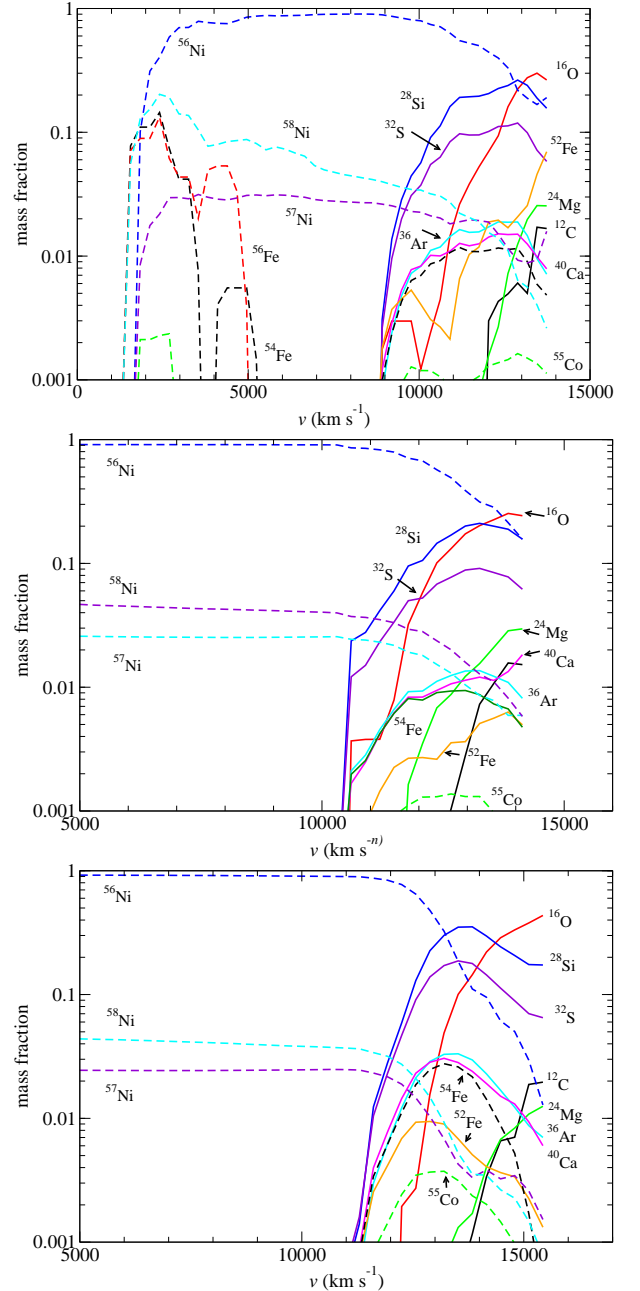

\centering
\includegraphics*[width=8cm,height=5.7cm]{fig9a.eps}
\includegraphics*[width=8cm,height=5.7cm]{fig9b.eps}
\includegraphics*[width=8cm,height=5.7cm]{fig9c.eps}
\caption{The mass fraction against velocity of the final 
abundance before the decay of short-live isotopes
for the benchmark models 110-100-2-50 (X) (upper panel),
105-100-2-B50 (Y) (middle panel) and 100-050-2-S50 (S) (lower panel).}
%The data is obtained from the tracer particles within a 
%slice from $0^{\deg}$ to $9^{\deg}$.}
\label{fig:slice1_benchmark}
\end{figure}

In Figures \ref{fig:slice1_benchmark}
we plot the 
velocity distribution of some representative isotopes
for the benchmark Models 110-100-2-50 (X), 110-050-2-B50 (Y)
and 100-050-2-S50 (S) in the left, middle and 
right panels respectively. We extract the 
chemical abundances and velocities of the tracer particles.

In Model 110-100-2-50 (X), this benchmark model possesses
both the typical sub-Chandrasekhar mass SN Ia 
ejecta profile with asymmetric effects. 
Here, we refer to e.g. \cite{Shigeyama1992} 
for a typical sub-Chandrasekhar mass SN Ia.
For the typical one, we can see in the core, 
up to 10000 km s$^{-1}$, the ejecta is made of
mainly $^{56-58}$Ni. Beyond that, IMEs,
including $^{28}$Si, $^{32}$S, $^{36}$Ar
and $^{40}$Ca become more abundant. 
However, in this model $^{56}$Ni remains
the most abundant almost throughout the 
star within $v < 13000$ km s$^{-1}$. 
Traces of $^{12}$C and $^{16}$O can be found
when $v > 12000$ km s$^{-1}$. They correspond 
to the products from the incomplete C burning. 
These features are common to all three
benchmark models presented here. 

In Model 110-100-2-50 (X) and 110-050-2-B50 (Y), 
we can see a mild rise of $^{52}$Fe 
near the surface. Also, $^{54}$Fe remains
to maintain a few percent mass fraction
even to the outermost ejecta. They come from
He burning, especially when there is
shock convergence or detonation wave collision.
The further compressional heating with this
hydrodynamical origin can enhance the formation
of these isotopes. On the other hand, we see 
a clear cut of $^{54}$Fe inside the outermost
ejecta of Model 100-050-2-S50 and $^{52}$Fe
has a clear falling trend when $v > 13000$ km s$^{-1}$.

Major differences appear in the innermost part of 
the ejecta because of the C detonation convergence. 
As discussed in previous sections, the further 
C detonation induced geometric convergence can
create hot spot which allows the matter to be heated
up to a temperature above it can normally reach 
through simple detonation. In that sense, this allows
a small part of matter to undergo complete
burning and even electron capture. This property can be 
found in Model 110-100-2-50 (X). We can see that at the innermost
part of the ejecta, neutron-rich isotope including
$^{54}$Fe and $^{56}$Fe are produced. Some $^{55}$Co 
can be even produced. Notice that these features are 
usually found in the Chandrasekhar mass model (See
e.g. \cite{Nomoto1984, Iwamoto1999} for the detailed
ejecta profile of some classical models).
This demonstrates that the asphericity of the 
He detonation and hence the C detonation can be 
reflected by the low-velocity ejecta.  

\subsection{Thermodynamics}

\begin{figure}
\centering
\includegraphics*[width=8cm,height=5.7cm]{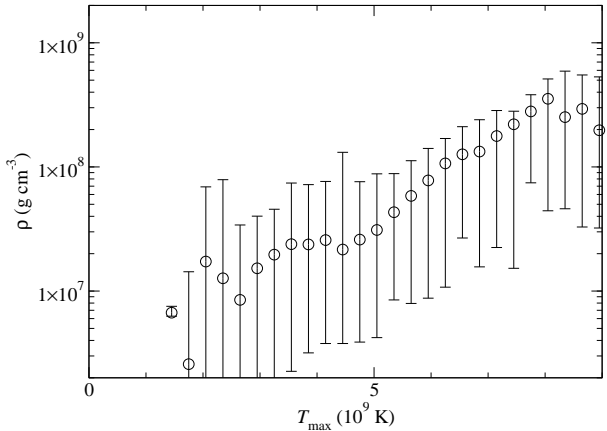}
\caption{The $\rho_{{\rm max}}$ against $T_{{\rm max}}$ for
the benchmark model obtained from the tracer particle 
thermodynamics histories for the Model 110-100-2-50 (X).
The error bars stand for the temperature ranges of
the tracer particles at a specific density bin
with the circle being the average.}
\label{fig:traj_summary_benchmark}
\end{figure}

In Figure \ref{fig:traj_summary_benchmark} we plot the
$\rho_{{\rm max}}$ against $T_{{\rm max}}$ for the
benchmark model obtained from the tracer particles. 
The sampling is done by grouping the tracer particles
into bins according to their $\rho_{{\rm max}}$, which is 
defined by their individual thermodynamics history.
Then the average, upper and lower limits of $T_{{\rm max}}$ in 
each density bin is taken. $T_{{\rm max}}$ is also the maximum
value in the thermodynamics history. In most cases, the particle
achieves its $\rho_{{\rm max}}$ and $T_{{\rm max}}$ is 
at the same time, when the detonation wave swept across the 
particle. However, in the case where multiple detonation shocks
appear, the two moments can be non-simultaneous.
Notice that the initial central
density of this model is $\sim 6 \times 10^7$ g cm$^{-3}$.
Due to the shock wave compression, which is further enhanced 
by the geometric convergence as well as shock wave
collision, the matter can reach a maximum density as high 
as $3 \times 10^8$ g cm$^{-3}$. Together with the rise of the 
density, the temperature can rise as high as $7 \times 10^9$ K.
Certain particles which are directly under shock interaction,
can reach a maximum density $5 \times 10^8$ g cm$^{-3}$
with a maximum temperature of $9 \times 10^9$ K. 
This can compared with Figure 12 in \cite{Leung2018}. 
In that figure, the tracer particles show a uniform 
$\rho_{{\rm max}}$ against $T_{{\rm max}}$ for the particles inside
deflagration zones and a spread of $T_{{\rm max}}$ in the 
detonation zone. Our model here shows a similar behaviour 
for the detonation, except that the effects are more pronounced
because of the inward motion during the shock propagation.

\begin{figure}
\centering
\includegraphics*[width=8cm,height=5.7cm]{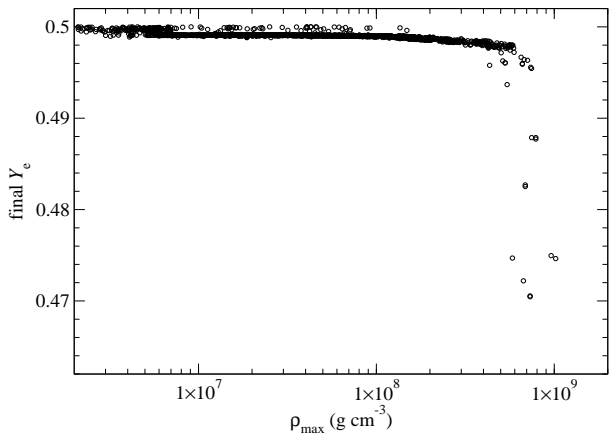}
\caption{Same as Figure \ref{fig:traj_summary_benchmark} 
but for the final $Y_{{\rm e}}$ against $\rho_{{\rm max}}$ for
the benchmark model 110-100-2-50 (X) obtained from the tracer particle 
thermodynamics histories.}
\label{fig:traj_ye_benchmark}
\end{figure}

At last in Figure \ref{fig:traj_ye_benchmark} we plot also
the final $Y_e$ of the tracer particles against $T_{{\rm max}}$.
We can see three groups of particles. The first group is 
the particle from the He envelope. It has a uniform
final $Y_{{\rm e}} = 0.5$ which has a density from $10^{6-8}$ g cm$^{-3}$.
This shows that the He envelope has in general low
density where electron capture processes are inefficient. The second 
group is the $10^6 - 5 \times 10^8$ g cm$^{-3}$. This 
corresponds to the tracer particles experiencing single
pass of detonation wave. The final $Y_e$ shows a mildly decreasing
function as $\rho_{{\rm max}}$, which suggests that 
electron capture becomes important at near $10^8$ g cm$^{-3}$. The third
group of particles are those with $Y_e$ from 0.47 - 0.495
with a $\rho_{{\rm max}}$ from $5 - 10 \times 10^8$ g cm$^{-3}$.
This corresponds to tracer particles which are excited by
shock compression. There are much fewer particles of this types
since it occurs to the particles very close to the 
symmetry boundary or lying inside the collision site of
C detonation shock. Again, this figure can be 
compared with Figure 12 in \cite{Leung2018}. In that figure,
the distribution of particles is more uniform and there exists a
one-one correspondence for a given $\rho_{{\rm max}}$ to 
final $Y_{{\rm e}}$. In this work, this correspondence is broken down
because of the He envelope. Also, the pronounced shock
interactions provide a wider diversity to the thermodynamics
history in the tracer particles.

%%%%%%%%%%%%%%%%%%%%%%%%%%%%%%%%%%%%%%%%%%%%%%%%%%%%%%%%%%%%%%%%%%%%%%%%

\section{Nucleosynthesis}
\label{sec:nucleo}

To calibrate the nucleosynthesis yield, we use the post-process
scheme as described in \cite{Travaglio2004, Seitenzahl2010}. In the 
hydrodynamics simulations we place massless particles which record
the thermodynamics history of the local density and temperature 
of the Eulerian grid. The density and temperature evolution, together
with the initial chemical composition depending on its initial 
position, are sent to the nuclear reaction network to calculate
the chemical abundance of the corresponding model. Similar to 
previous works, we use the nuclear reaction network as developed 
in \cite{Timmes1999b}. It includes a network of 495 isotopes ranging
from $^1$H to $^{91}$Tc. The nuclear reaction rates are updated by the 
values provided in \cite{Rauscher2000}. We include the electron screening 
by \cite{Kitamura2000} and \cite{Benvenuto2015}. The formula aims for 
strong electron screening, and it reduces to the weak electron 
screening given in \cite{Abe1959}.
We include the corresponding
free energy for the calculation of NSE as described in \cite{Seitenzahl2010}.
The chemical potential assumes classical ideal gas form, which is  
suitable for the density ($\sim 10^9$ g cm$^{-3}$) and temperature ($10^9$ K)
used here.
We have also updated electron capture rate table by including 
the rate table from \cite{Nabi1999,Nabi2004}. We use this rate table when 
there is no rate given in the original version of the nuclear
reaction network, although we remark that the electron capture and 
its related weak interaction processes are of less importance
due to the typically lower density than the near-Chandrasekhar
mass explosion model. 

\subsection{Dependence on WD Mass}

\subsubsection{One-Bubble Configuration}

\begin{figure*}
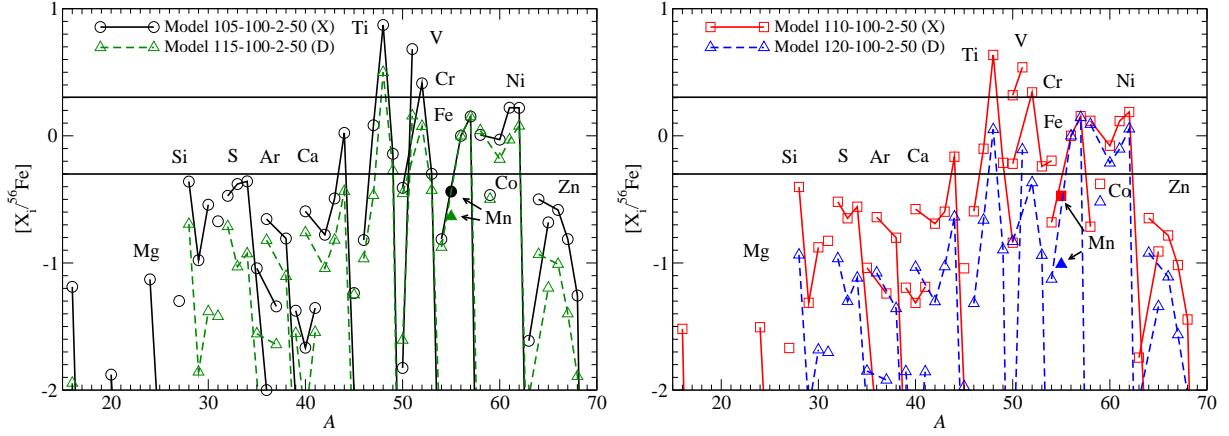

\centering
\includegraphics*[width=8cm,height=5.7cm]{fig12a.eps}
\includegraphics*[width=8cm,height=5.7cm]{fig12b.eps}
\caption{$[X_i/^{56}$Fe] for Models 
105-100-2-50 (X) $(M = 1.05 ~M_{\odot})$ and 115-100-2-50 (D) $(M = 1.15~M_{\odot})$ in the left panel and 
Models 110-100-2-50 (X) $(M = 1.10~M_{\odot})$ and 120-100-2-50 (D) $(M = 1.20~M_{\odot})$ in the right panel.
All models assume $M_{{\rm He}} = 0.1 ~M_{\odot}$, $Z = 0.02$ and a bubble-shape initial He-detonation
at 50 km above the CO-envelope interface.}
\label{fig:final_M}
\end{figure*}

In Figure \ref{fig:final_M} we plot $[X_i/^{56}$Fe]
for the isotopes from Models 105-100-2-50 (X), 110-100-2-50 (X),
115-100-2-50 (D) and 120-100-2-50 (D). The isotopes are obtained from the nucleosynthesis
by the post-processing as described above, but after all 
short-live isotopes decayed. In practice 
after the nucleosythesis yield is computed by post-processing, 
we allow further radioactive decays by computing the network 
while suppressing thermonuclear reactions. We fixed the period to 
be $10^6$ years. The period is chosen to be long enough to allow
certain long lived isotopes, such as $^{59}$Ni, to completely
decay to compute the asymptotic chemical yield. But we also 
note that there are still some isotopes with even longer 
half-lives, such as $^{27}$Al and $^{60}$Fe.

These models in this figure have the same configurations by
setting the same $M_{{\rm He}}$, initial He detonation pattern and 
metallicity. As a result, the mass of the 
CO fuel increases when the total mass increases. By increasing the
mass, there is a systematic decrease in $[X_i/^{56}$Fe].
This is because when the mass increases, the produced $^{56}$Ni
increases as shown in Table \ref{table:models}. The final
$^{56}$Fe yield thus increases. For IME, there is a drop 
from Mg to Ca by a factor of $\sim O(1)$. Similar effects 
are observed in Fe-peak isotopes. This shows that when the 
mass increases, the extra ash contributes
to the production of $^{56}$Fe. Therefore, the qualitative
features of the mass fraction remain. 

Nevertheless, even for the 
massive progenitor like Model 115-100-2-50 (D), the over-production
of $^{48}$Ti cannot be resolved as they are mostly produced 
in the He envelope. Some of the isotopes, such as 
$^{51}$V, $^{52}$Cr and $^{70}$Zn, become comparable to 
the solar abundance when $M = 1.15$ $M_{\odot}$. However, 
compared to the Chandrasekhar mass SN Ia, none of
the isotopes shows a drastic boost when $M$ increases. 
This can be compared to the Chandrasekhar mass 
WD scenario, by increasing the mass from 1.30 to 1.37 $M_{\odot}$,
some of the isotopes, such as $^{54}$Cr and $^{60}$Fe, 
can be drastically enhanced.
One reason is that the density related to 
the sub-Chandrasekhar mass model is low that the electron capture
does not play an important role in most parts of the star. 
The major changes come from the increment of 
$^{56}$Ni, which systematically lowers all mass
fractions of all isotopes. 

\subsubsection{One-Ring Configuration}

\begin{figure*}
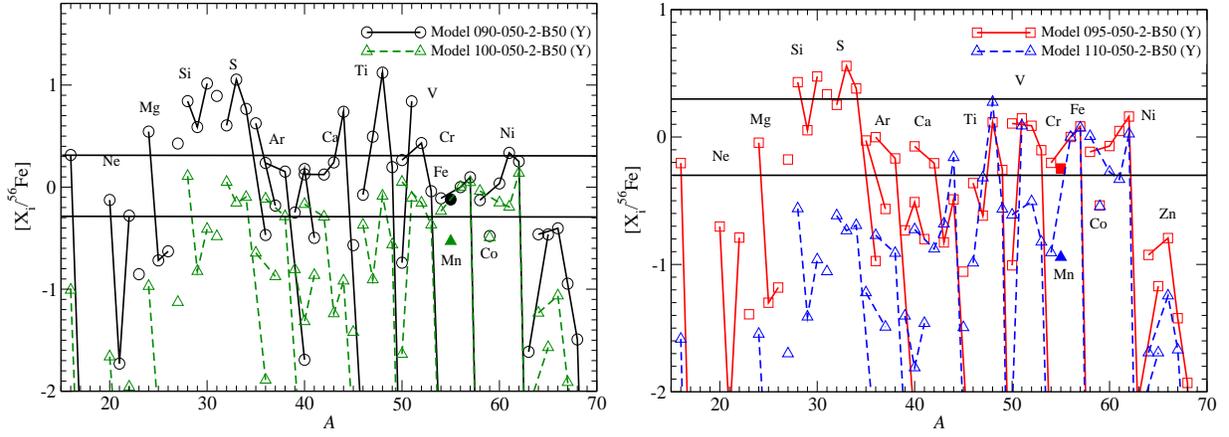

\centering
\includegraphics*[width=8cm,height=5.7cm]{fig13a.eps}
\includegraphics*[width=8cm,height=5.7cm]{fig13b.eps}
\caption{$[X_i/^{56}$Fe] fractions for Models 
090-050-2-B50 (Y) ($M = 0.90 ~M_{\odot}$) and 100-050-2-B50 (Y) ($M = 1.00 ~M_{\odot}$) in the left panel and 
Models 095-050-2-B50 (Y) ($M = 0.95 ~M_{\odot}$) and 110-050-2-B50 (Y) ($M = 1.10 ~M_{\odot}$) in the right panel.
All models assume $M_{{\rm He}} = 0.05 ~M_{\odot}$, $Z = 0.02$ and a belt-shape initial He-detonation
at 50 km above the CO-envelope interface.}
\label{fig:final_M_B50}
\end{figure*}

In Figure \ref{fig:final_M_B50} we plot similar to 
Figure \ref{fig:final_M} but for Models 
090-050-2-B50 (Y), 095-050-2-B50 (Y), 100-050-2-B50 (Y) and 110-050-2-B50 (Y)respectively.
These models correspond to the series of models of the 
same $M_{{\rm He}}$ but at different masses, each with 
the same initial He detonation by a He ring.
Due to the detonation symmetry which may trigger the 
second detonation with a lower He mass, the effects of
the He incomplete burning products, such as Ti, V and Cr
become better fit to the solar abundance. 
The qualitative trends for an increasing mass 
can be observed.

At lower mass, the lower production of $^{56}$Ni causes
a strong overproduction of elements like Si, S.
Ti and V are overproduced but this feature is suppressed
at Model 090-050-2-B50 (Y). As mass increases, the relative
productions of IMEs drop. This includes
Si, S, Ar and Ca. Relative productions of Ti, V and Cr also 
decrease when the mass increases, but they remain 
saturated around the solar values. Fe and Ni are overall
insensitive to the mass change. 

\subsubsection{Spherical Configuration}

\begin{figure*}
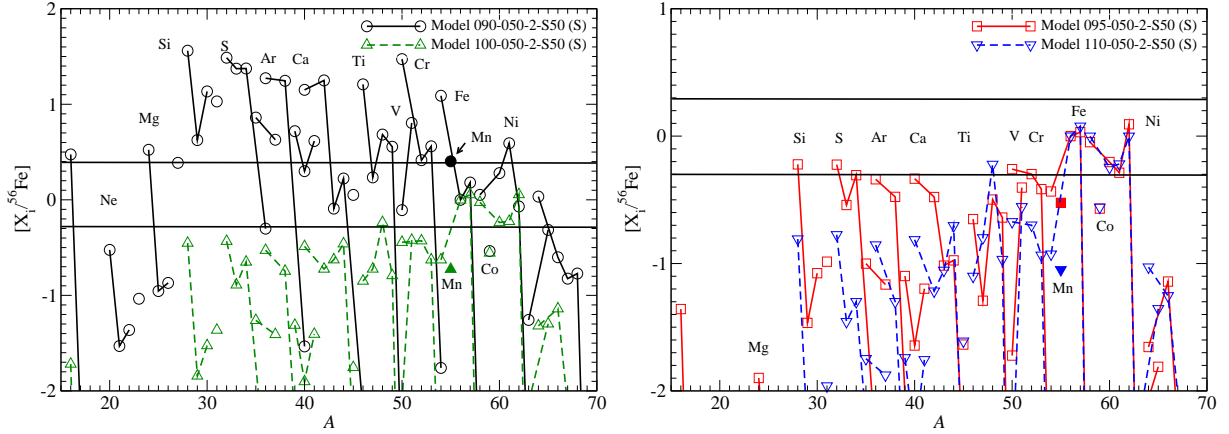

\centering
\includegraphics*[width=8cm,height=5.7cm]{fig14a.eps}
\includegraphics*[width=8cm,height=5.7cm]{fig14b.eps}
\caption{$[X_i/^{56}$Fe] for Models 
090-050-2-S50 (S) ($M = 0.90 ~M_{\odot}$) and 100-050-2-S50 (S) ($M = 1.00 ~M_{\odot}$) in the left panel and 
Models 095-050-2-S50 (S) ($M = 0.95 ~M_{\odot}$) and 110-050-2-S50 (S) ($M = 1.10 ~M_{\odot}$) in the right panel.
All models assume $M_{{\rm He}} = 0.05 ~M_{\odot}$, $Z = 0.02$ and a spherical initial He-detonation
at 50 km above the CO-envelope interface.}
\label{fig:final_M_S50}
\end{figure*}

In Figure \ref{fig:final_M_S50} we plot similar to 
Figure \ref{fig:final_M} but for Models 
090-050-2-S50 (S), 095-050-2-S50 (S), 100-050-2-S50 (S) and 110-050-2-S50 (S) respectively,
using the spherical He detonation as the initial trigger. 
Again, the higher He detonation symmetry allows 
triggering the second detonation at a lower He envelope.
The overproduction of intermediate $\alpha$-chain burning
production is less severe for the models with a 
normal amount of $^{56}$Ni ($\sim 0.6 ~M_{\odot}$). 
Due to the spherical symmetry, the second detonation
all starts at the core for all models, so that 
the variations of elements become more regular. 

The IMEs Si, S, Ar and Ca show
a flat distribution and decrease with an increasing
mass. The over-productions of $^{50}$Ti, $^{51}$V and $^{52}$Cr
as the major products in He detonation before reaching
$^{56}$Ni, are largely suppressed once the mass reaches
above 0.95 $M_{\odot}$. The isotopes of Fe remain
non-sensitive to the variation of mass except for $^{54}$Fe.
A systematic drop of $^{55}$Mn can also be seen, showing that
the amount of $^{55}$Mn is not increased significantly
when the mass increases.

\subsection{Dependence on He Envelope Mass}

\subsubsection{One-Bubble Configuration}

\begin{figure*}
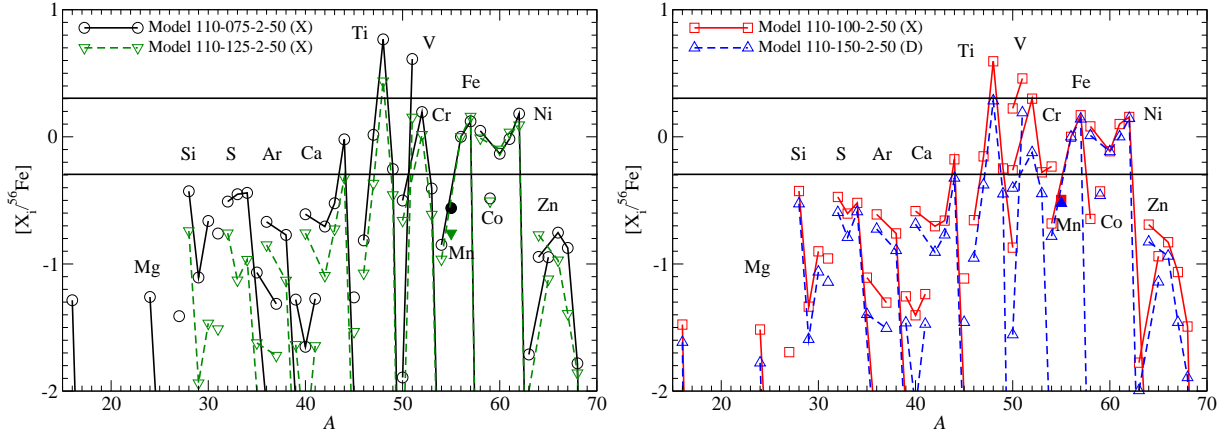

\centering
\includegraphics*[width=8cm,height=5.7cm]{fig15a.eps}
\includegraphics*[width=8cm,height=5.7cm]{fig15b.eps}
\caption{$[X_i/^{56}$Fe] for models comparing the
effects of He envelope mass including Models 110-075-2-50 (X) ($M_{{\rm He}} = 0.075 ~M_{\odot}$), 
and 110-125-2-50 (X) ($M_{{\rm He}} = 0.125 ~M_{\odot}$) in the left panel and 
Models 110-100-2-50 (X) ($M_{{\rm He}} = 0.100 ~M_{\odot}$) and 110-150-2-50 (D) ($M_{{\rm He}} = 0.150 ~M_{\odot}$) in the right panel.
All models assume $M = 1.10~M_{\odot}$, $Z = 0.02$ and a bubble-shape initial He-detonation
at 50 km above the CO-envelope interface.}
\label{fig:final_MHe}
\end{figure*}

In Figure \ref{fig:final_MHe} we plot similar to Figure 
\ref{fig:final_M} but for Models 110-075-2-50 (X), 
110-100-2-50 (X), 110-125-2-50 (X) and 110-150-2-50 (D). This series of 
model have also the same configurations except for the 
He envelope mass. Notice that among these models,
Model 110-150-2-50 (D) has a different detonation mechanism 
as it has "D"-Type detonation instead "X"-Type detonation.
By increasing $M_{{\rm He}}$, the mass fractions of IMEs
reduce. However, by comparing Models 110-125-2-50 (X)and 
110-150-2-50 (D), the IME mass fractions increase. 
This is because the "Y"-Type detonation allows an earlier
detonation, which ensures that the low density matter
is well detonated before it expands and the density becomes too low
for nuclear reaction. For Fe-peak elements, 
clear trends can be seen in elements
like Ti, Cr and V. Again a decreasing trend is observed when
$M_{{\rm He}}$ increases but there is not much difference
in Fe and Ni. 

\subsubsection{One-Ring Configuration}

\begin{figure*}
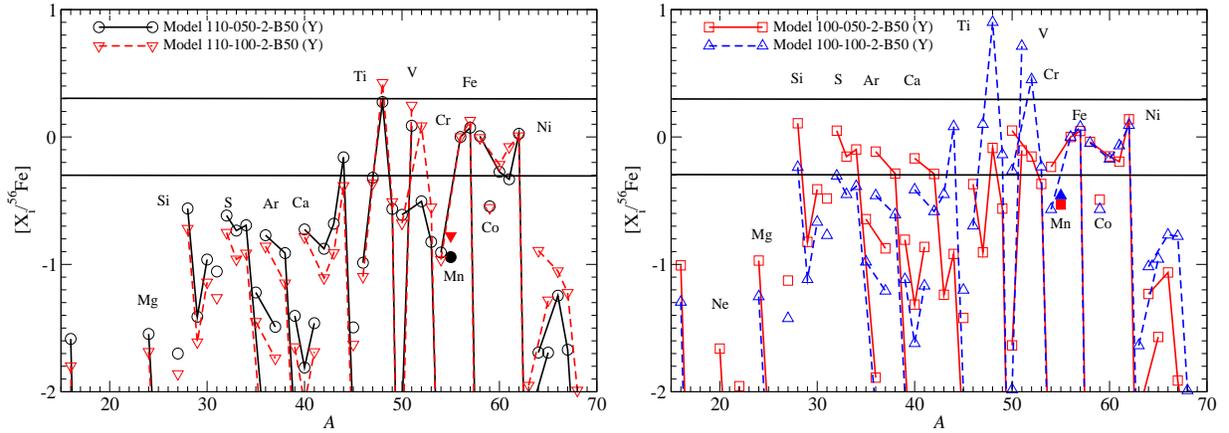

\centering
\includegraphics*[width=8cm,height=5.7cm]{fig16a.eps}
\includegraphics*[width=8cm,height=5.7cm]{fig16b.eps}
\caption{Similar to Fig. \ref{fig:final_MHe}, but for 
Models 100-050-2-B50 (Y) ($M = 1.00~M_{\odot}$) and 110-050-2-B50 (Y) ($M = 1.10~M_{\odot}$) in the left panel and 
Models 100-100-2-B50 (Y) ($M = 1.00~M_{\odot}$) and 110-100-2-B50 (Y) ($M = 1.10~M_{\odot}$) in the right panel.
All models assume $M_{{\rm He}} = 0.05~(0.10)~M_{\odot}$ in the left (right) panel, 
$Z = 0.02$ and a belt-shape initial He-detonation at 50 km above the CO-envelope interface.}
\label{fig:final_MHe_B50}
\end{figure*}

In Figure \ref{fig:final_MHe_B50} we plot similar to Figure \ref{fig:final_MHe}
but for Models 100-050-2-B50 (Y), 100-100-2-B50 (Y), 110-050-2-B50 (Y) and 110-100-2-B50 (Y),
where all models share the same initial masses $M = 1.00$ and 1.10 $M_{\odot}$ 
and He detonation configuration. Different He envelope masses are used.
We remind that Model 110-050-2-B50 (Y) is the benchmark model and 
we choose a progenitor mass for comparison to extract the effects
of $M_{{\rm He}}$ at different mass. 

For $M = 1.10$ $M_{\odot}$, the chemical abundances do not change
strongly with $M_{\rm He}$. It is because the overall production is 
dominated by $^{56}$Ni. A small suppression of IMEs
for $^{28}$Si, $^{32}$S and $^{36}$Ar can be observed. 
Almost no change can be found for Fe-peak elements from Ti to Ni.
On the other hand for $M = 1.00$ $M_{\odot}$, the chemical abundances scale
strongly with $M_{{\rm He}}$. Besides the more obvious drop in the 
IMEs, there is a huge jump in $^{48}$Ti. $^{51}$V and 
$^{52}$Cr when $M_{{\rm He}}$ increases. Again, Fe-peak elements like 
Fe, Mn and Ni are less changed by $M_{{\rm He}}$. 

\subsubsection{Spherical Configuration}

\begin{figure*}
\centering
\includegraphics*[width=8cm,height=5.7cm]{fig17a.eps}
\includegraphics*[width=8cm,height=5.7cm]{fig17b.eps}
\caption{Similar to Fig. \ref{fig:final_MHe}, but for 
Models 100-050-2-S50 (S) ($M = 1.00~M_{\odot}$) and 110-050-2-S50 (S) ($M = 1.10~M_{\odot}$) in the left panel and 
Models 100-100-2-S50 (S) ($M = 1.00~M_{\odot}$) and 110-100-2-S50 (S) ($M = 1.10~M_{\odot}$) in the right panel.
All models assume $M_{{\rm He}} = 0.05~(0.10)~M_{\odot}$ in the left (right) panel, 
$Z = 0.02$ and a spherical initial He-detonation at 50 km above the CO-envelope interface.}
\label{fig:final_MHe_S50}
\end{figure*}

In Figure \ref{fig:final_MHe_S50} we plot similar to Figure 
\ref{fig:final_MHe} but for the Models 
090-050-2-S50 (S), 090-100-2-S50 (S), 100-050-2-S50 (S) and 100-100-2-S50 (S).
The models consist of initial masses of 1.00 and 1.10 $M_{\odot}$.
All models assume a spherical He detonation as the initial trigger. 
Again we remind that Model 100-050-2-S50 is the benchmark model
of this work. 

For the spherical model, due to its stronger explosion, a lower
mass model is used for the benchmark model. So compared to the 
one-ring structure, the effects of He envelope are larger. 
Besides a more prominent decrease in IMEs, 
the $\alpha$-chain products including $^{48}$Ti, $^{52}$Cr
and $^{51}$V are vastly increased
for a more massive He envelope. No significant change
is observed for Fe, Mn and Ni. 
The effects are more significant for the lower mass cases
due to a smaller $^{56}$Ni mass. A flat distribution in $^{28}$Si, $^{32}$S, 
$^{36}$Ar, $^{40}$Ca, $^{42}$Ca, $^{44}$Ti and $^{48}$Cr 
can be seen. The $^{55}$Mn is even over-produced because
of the suppressed $^{56}$Ni and hence $^{56}$Fe. 
A higher $M_{{\rm He}}$ results in a global suppression of this
relative production rate.

\subsection{Dependence on Metallicity}

\subsubsection{One-Bubble Configuration}
\begin{figure*}
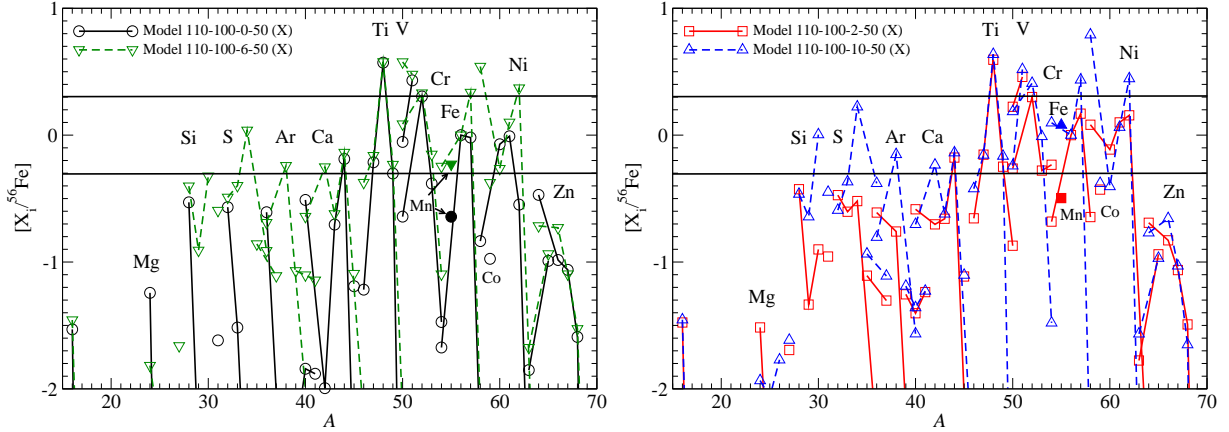

\centering
\includegraphics*[width=8cm,height=5.7cm]{fig18a.eps}
\includegraphics*[width=8cm,height=5.7cm]{fig18b.eps}
\caption{Similar to Fig. \ref{fig:final_M}, but for 
Models 110-100-0-50 (X) ($Z = 0$) and 110-100-6-50 (X) ($Z = 0.06$) in the left panel 
and Models 110-100-2-50 (X) ($Z = 0.02$) and 110-100-10-50 (X) ($Z = 0.10$) in the right panel.
All models assume $M = 1.10~M_{\odot}$, $M_{{\rm He}} = 0.10~M_{\odot}$, 
and a bubble shape initial He-detonation at 50 km above the CO-envelope interface.}
\label{fig:final_Z}
\end{figure*}

In Figure \ref{fig:final_Z} we plot $[X_i/^{56}$Fe]
of stable isotopes after all short-lived radioactive 
isotopes have decayed for Models 110-100-0-50 (X), 
110-100-2-50 (X), 110-100-6-50 (X) and 110-100-10-50 (X).
Similar to Chandrasekhar mass SNe Ia \citep{Leung2018},
metallicity is important to the production of isotopes
with a neutron-proton ratio close to the unity. Also, the
presence of $^{22}$Ne slightly lowers the energy
release of C detonation. We can observe a 
boost of IMEs including $^{30}$Si, $^{34}$S, 
$^{38}$Ar, $^{42}$Ca. The boost factors can be as large
as beyond two orders of magnitude when the metallicity 
increases from 0 to 5 $Z_{\odot}$. 
For Fe-peak elements, we also
observe a boost in the production $^{46}$Ti, $^{50}$Cr
$^{54}$Fe, $^{55}$Mn, $^{58}$Ni. The boost can range from
ten to hundred when contrasting the Models 110-100-0-50 (X)
and 110-100-10-50 (X). In Tables \ref{table:stable_yield}, \ref{table:stable_yieldb},
\ref{table:stable_yield2}, \ref{table:stable_yield2b},
\ref{table:stable_yield3} and \ref{table:stable_yield3b} we tabulate the masses
of the stable isotopes in different metallicity for 
the benchmark models based on Models 110-100-2-50 (X),
100-050-2-B50 (Y) and 110-050-2-S50 (S). By comparing models with the 
same configuration but different metallicity, it can be
seen that when metallicity increases, IMEs with a high neutron
ratio are boosted sharply. For example, we observe a
clear increasing trend for $^{29-30}$Si, $^{34,36}$S, 
$^{37}$Cl, $^{38}$Ar and $^{40}$K and so on. The jump 
can be as high as four orders of magnitude from 
zero metallicity to 5 $Z_{\odot}$. For Fe-peak elements,
we have $^{46}$Ti, $^{50}$V, $^{50}$Cr, $^{54}$Fe and 
$^{58}$Ni to be the representative isotopes. The results
here are very similar to those in Chandrasekhar mass WD. 

\subsubsection{One-Ring Configuration}

\begin{figure*}
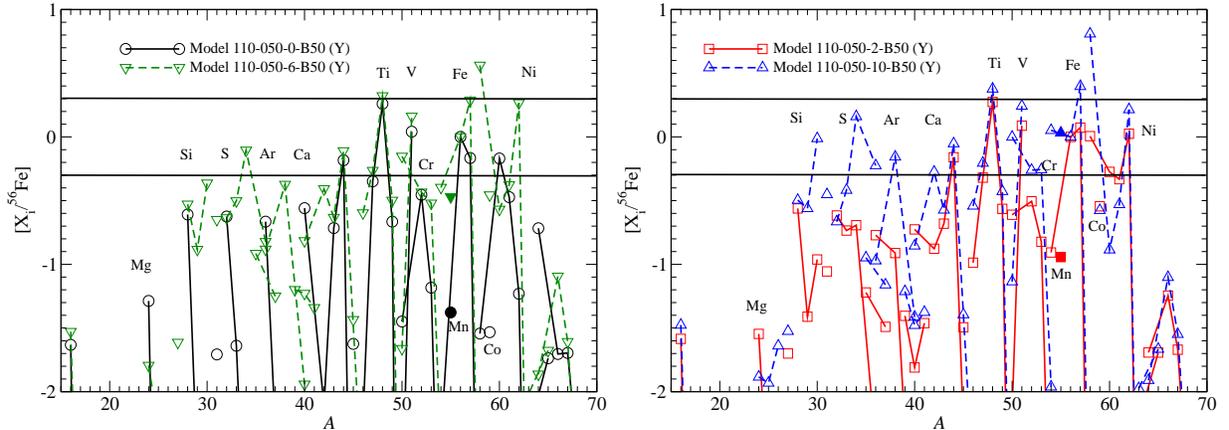

\centering
\includegraphics*[width=8cm,height=5.7cm]{fig19a.eps}
\includegraphics*[width=8cm,height=5.7cm]{fig19b.eps}
\caption{Similar to Fig. \ref{fig:final_Z}, but for 
Models 110-050-0-B50 (Y) ($Z = 0$) and 110-050-6-B50 (Y) ($Z = 0.06$) in the left panel and 
Models 110-050-2-B50 (Y) ($Z = 0.02$) and 110-050-10-B50 (Y) ($Z = 0.10$) in the right panel.
All models assume $M = 1.10~M_{\odot}$, $M_{{\rm He}} = 0.05~M_{\odot}$, 
and a belt shape initial He-detonation at 50 km above the CO-envelope interface.}
\label{fig:final_Z_B50}
\end{figure*}

In Figure \ref{fig:final_Z_B50} we plot similar to Figure \ref{fig:final_Z}
but for the Models 110-050-0-B50 (Y), 110-050-2-B50 (Y), 110-050-6-B50 (Y) and 110-050-10-B50 (Y).
This series of models focus on the effects of metallicity for the 
same progenitor mass $M = 1.1~M_{\odot}$, He mass at $0.05~M_{\odot}$
and with the same He detonation trigger. 

The general trends of isotopes on metallicity are similar to the 
one-bubble case. There is no significant change for 
the $\alpha$-chain isotopes such as $^{28}$Si, $^{32}$S, $^{36}$Ar
and $^{44}$Ca. But the slightly low-$Y_{{\rm e}}$ isotopes, 
such as $^{30}$Si and $^{34}$S, are strongly enhanced at high
metallicity. There are smaller changes for the Fe-peak elements
except for $^{52}$Cr, $^{54}$Fe, $^{55}$Mn and $^{58}$Ni. 
Minor increases can be observed for isotopes like $^{48}$Ti and 
$^{51}$V. 

\subsubsection{Spherical Configuration}

\begin{figure*}
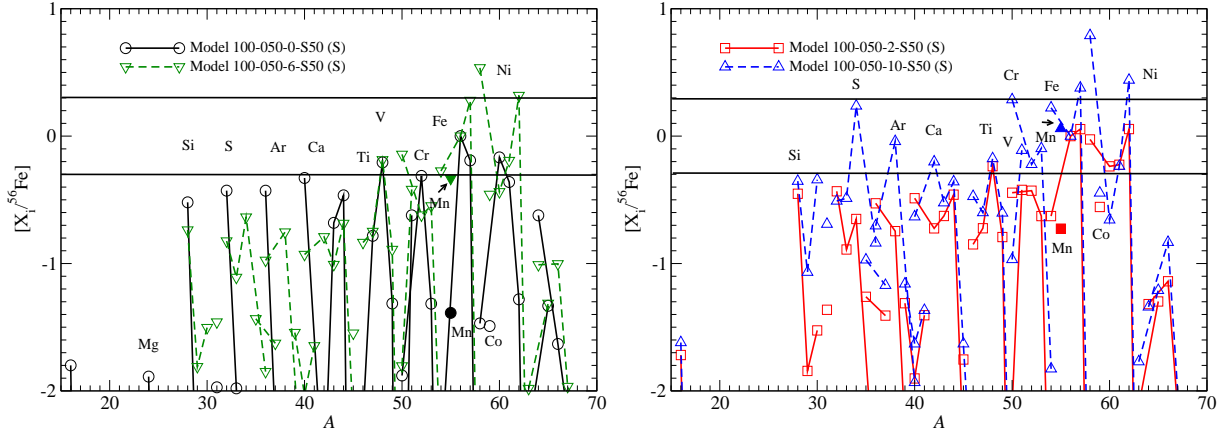

\centering
\includegraphics*[width=8cm,height=5.7cm]{fig20a.eps}
\includegraphics*[width=8cm,height=5.7cm]{fig20b.eps}
\caption{Similar to Fig. \ref{fig:final_Z}, but for 
Models 110-050-0-S50 (S) ($Z = 0$) and 110-050-6-S50 (S) ($Z = 0.06$) in the left panel and 
Models 110-050-2-S50 (S) ($Z = 0.02$) and 110-050-10-S50 (S) ($Z = 0.10$) in the right panel.
All models assume $M = 1.10~M_{\odot}$, $M_{{\rm He}} = 0.05~M_{\odot}$, 
and a spherical initial He-detonation at 50 km above the CO-envelope interface.}
\label{fig:final_Z_S50}
\end{figure*}

In Figure \ref{fig:final_Z_S50} we plot similar to Figure \ref{fig:final_Z}
but for Models 110-050-0-S50 (S), 110-050-2-S50 (S), 110-050-6-S50 (S) and 110-050-10-S50 (S).
Again, the models here share the same initial progenitor mass at 1.10 $M_{\odot}$, 
He mass at 0.05 $M_{\odot}$ and an initial spherical He detonation. 
The overall pattern remains compatible with the single one-bubble case. 

The metallicity plays an important role to the slightly low-$Y_{{\rm e}}$ 
isotopes (defined by the neutron number $N$ comparable but
not larger than atomic number $Z$) including $^{30}$Si, $^{34}$S, $^{38}$S, $^{42}$Ca for the
IMEs, and $^{51}$V, $^{52}$Cr, $^{55}$Mn, $^{54}$Fe
and $^{58}$Ni for Fe-peak isotopes. The variations of isotopes against metallicity
are similar to the previous two cases. This shows that the metallicity dependence
is not sensitive to the explosion energetics.

\subsection{Dependence on He Detonation Pattern}

\begin{figure*}
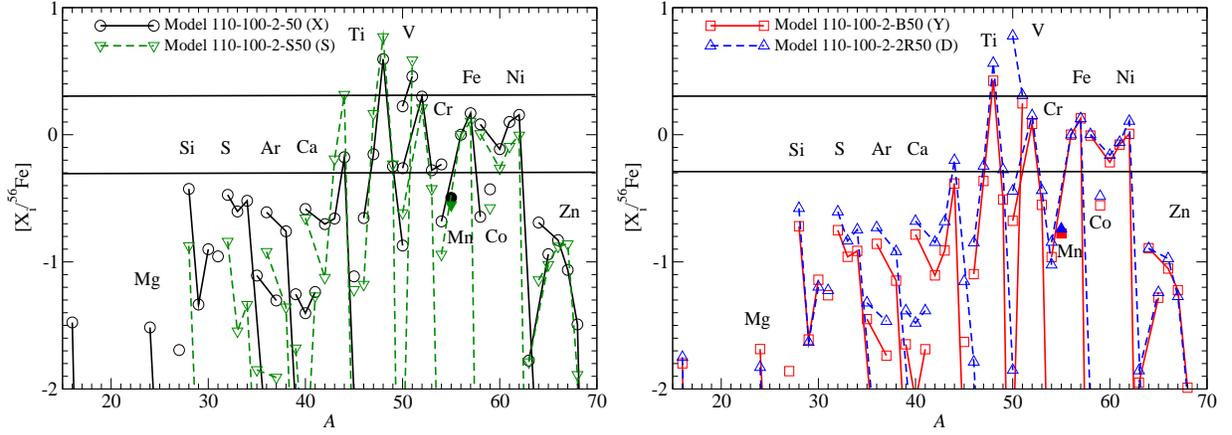

\centering
\includegraphics*[width=8cm,height=5.7cm]{fig21a.eps}
\includegraphics*[width=8cm,height=5.7cm]{fig21b.eps}
\caption{$[X_i/^{56}$Fe] for the series of models
studying the effects of initial He detonation structure. 
Similar to Fig. \ref{fig:final_M}, but for 
Models 110-100-2-50 (X) (bubble shape) and 110-100-2-B50 (Y) (belt shape) in the left panel and 
Models 110-100-2-2R50 (D) (bubble$+$belt shapes) and 110-100-2-S50 (S) (spherical) in the right panel.
All models assume $M = 1.10~M_{\odot}$, $M_{{\rm He}} = 0.10~M_{\odot}$, 
and $Z = 0.02$.
}
\label{fig:final_det}
\end{figure*}

Here, we analyze the final chemical abundance for different
types of detonations. In Figure \ref{fig:final_det} we plot
$[X_i/^{56}$Fe] for Models 110-100-2-50 (X) and 110-100-2-B50 (Y), 
110-100-2-2R50 (D) and 110-100-2-S50 (S). 
They represent the typical detonation of Type "D", "X", "Y"
and "S" respectively. All of the four models have 
$^{56}$Ni closest to 0.6 $M_{\odot}$ among all models 
we have. 
We observe that in general "S"-Type is the strongest
that it has more isotopes with abundances closer to solar
values. "X"- and "Y"-Type are the intermediate ones and 
"D"-Type is the weakest among the four models. The difference
for IME can be as large as a factor of $\sim O(1)$. 
For Fe-peak elements, differences can be found to Ti to Cr. 
The "Y"-Type model tends to produce less $^{47-50}$Ti, $^{50-51}$V,
and $^{64-70}$Zn. The major difference between "Y"-Type and
other detonation types is that there is no shock-convergence induced 
heating along the $r$-axis. This shows that $^{48}$Ti 
is a sensitive indicator on how the He detonation
propagates along the surface of the WD. 
Similar to previous cases, Fe and Ni are less sensitive 
to the detonation mechanism owing to the difference in production 
site.

\subsection{Differences from spherical detonation}

\begin{figure}
\centering
\includegraphics*[width=8cm,height=5.7cm]{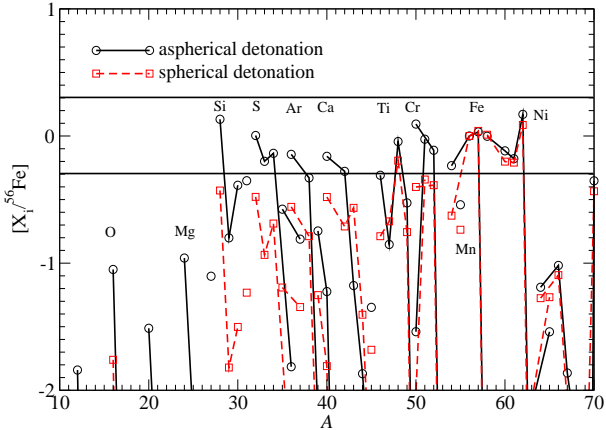}
\caption{$[X_i/^{56}$Fe] for
the two models 105-050-2-S50 (S) and 110-100-2-50 (X).}
\label{fig:final_assymetry}
\end{figure}

One theoretical uncertainty in the He detonation
is that it is unclear whether the pre-explosion fluid 
motion is strong enough to alter the first detonation site. 
In the case with a strong fluid motion background, heat 
generated can be distributed by the eddy motion or 
be further enhanced by the local turbulent motion.
This breaks the initial symmetry and creates some detonation
bubbles. On the other hand, in the quiescent star, 
the whole He layer can simultaneously burn and 
reach the explosive temperature together. Thus, 
the initial detonation can preserve the symmetry. 
To derive constraints on the initial detonation profile, 
we examine the scaled
mass fraction again in Figure \ref{fig:final_assymetry}
for both spherical and aspherical detonation model. 
Both models produce a very similar distribution for Fe and Ni
since they are chosen to produce $\sim$ 0.6 $M_{\odot}$.
For lighter Fe-peak elements, differences appear. 
The aspherical model produces more Ti, V and Cr
than the spherical one for at least one order of magnitude. 
In particular, the $^{48}$Ti, $^{50-51}$V and $^{52}$Cr
are $\sim$ 2 - 5 times higher than solar values. This suggests
that observations of non-aspherical detonation model can 
be characterized by the excess of these light Fe-peak 
elements. 

\begin{figure}
\centering
\includegraphics*[width=8cm,height=5.7cm]{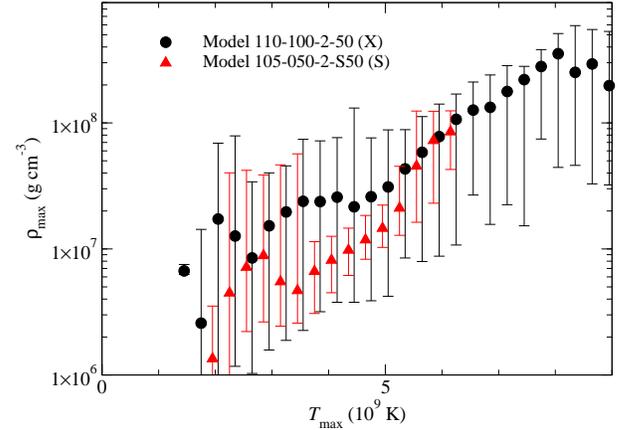}
\caption{Similar to Figure \ref{fig:final_assymetry}, but 
for $\rho_{{\rm max}}$ against $T_{{\rm max}}$.}
\label{fig:traj_summary_assymetry}
\end{figure}

In Figure \ref{fig:traj_summary_assymetry} we plot 
the $\rho_{{\rm max}}$ against $T_{{\rm max}}$ for the
tracer particles of the two models. It can be seen that 
even for the same $^{4}$He mass and total mass, 
the spherical model, whose evolution contains 
no oblique shock and the detonation wave propagates 
radially outward only, provides a uniform element distribution.
This can be contrasted with the aspherical model, 
where the scattering in density and temperature 
is much pronounced. 

\subsection{Constraints on progenitor model}

The double detonation model is one of the well accepted
physical models due to the robustness of initiating the 
detonation and its variability in producing the 
dispersion in the observed SNe Ia brightness. However, 
one major concern, in contrast to the near-Chandrasekhar
mass white dwarf, is that the detonation nature can produce 
a considerable amount of $^{56}$Ni if the detonation
is triggered too early, which produces over-luminous SNe Ia.
This is incompatible to the majority of SNe Ia,
where $\sim$ 0.5 - 0.7 $M_{\odot}$ of $^{56}$Ni 
is observed as induced by their light curves. In view of
that it becomes important to understand, at which condition
we could obtain realizations which can resemble with 
the typical SNe. This may provide constraints on the 
progenitor model, including the typical mass, the 
He envelope mass, and the initial detonation seed.
In particular the position of the initial detonation 
seed is not yet well constrained. 

\begin{figure}
\centering
\includegraphics*[width=8cm,height=5.7cm]{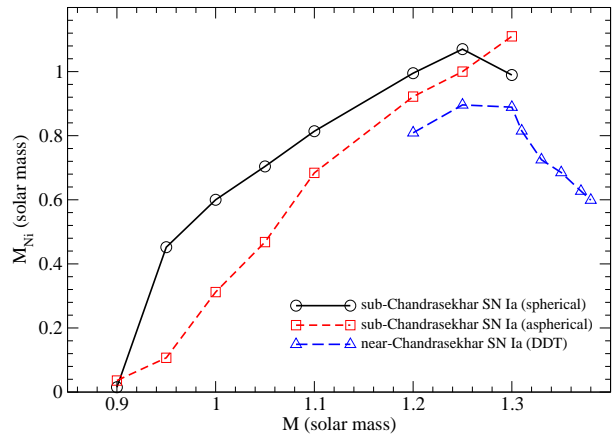}
\caption{The synthesized $^{56}$Ni mass at the end of
the simulations against the initial WD
mass. Both Chandrasekhar and sub-Chandrasekhar
mass models are included. For sub-Chandrasekhar mass
models, we use Models $C$-050-2-50 for aspherical model
and Models $C$-050-2-B50 for spherical model,
where $C$ is the initial mass shown in the figure.
The results for sub-Chandrasekhar
mass and Chandrasekhar mass WD are selected from
this work and in \cite{Leung2017a}.}
\label{fig:Mni}
\end{figure}

To do so, we plot the $^{56}$Ni mass against progenitor
WD mass for different progenitor masses and different
explosion mechanisms. In Figure \ref{fig:Mni} we plot
that for the double detonation models for both the
spherical detonation and the aspherical one, which 
we choose the one bubble pattern along the z-axis. 
For the near-Chandrasekhar mass WD, we use the 
standard DDT model with turbulent deflagration
as reported in \cite{Leung2017a}. 

The sub-Chandrasekhar mass WD model corresponds
to both single and double degenerate scenarios. 
We remind that for the violent merger model, 
due to the compactness of CO core, the 
secondary WD is disintegrated when the 
He detonation starts. Thus effectively
it has a structure similar to the double detonation 
model in the single degenerate scenario.
The Chandrasekhar mass model corresponds
to the near-Chandrasekhar mass models presented
in \cite{Leung2017a,Nomoto2017}. In particular, we choose
the configuration identical to the benchmark model
but for different central density from 
$5 \times 10^8$ to $5 \times 10^9$ g cm$^{-3}$.

In the sub-Chandrasekhar mass models (0.9 -- 1.2 
$M_{\odot}$), $M_{{\rm Ni}}$
increases with $M$ for both spherical (Model 100-050-2-S50)
and aspherical (Model 110-050-2-B50)
models. This is because in principle the whole
star is burnt. How complete the nuclear burning
depends only on the density. For a lower mass WD,
there is less matter with sufficient density to
reach complete burning (typically $5 \times 10^7$ g cm$^{-3}$).
Therefore, the $^{56}$Ni scales almost linearly with
$M$. On the other hand, in the Chandrasekhar mass
branch, $M_{{\rm Ni}}$ decreases with $M$. 
This is related to the more efficient electron capture
in the matter burnt by deflagration, which lowers
the matter electron fraction. As $^{56}$Ni is produced
in NSE while $^{56}$Ni has an electron fraction 0.5, 
any electron capture in the matter will only suppress
the production of $^{56}$Ni. 
 
We note that we compute both Chandrasekhar and sub-Chandrasekhar mass models  
from both scenarios for a mass 1.2 -- 1.3
$M_{\odot}$. This is because in this intermediate
regime, it is unclear 
whether the thermonuclear runaway occurs in the form of 
deflagration or detonation, because
the pressure jump becomes close to the initial
pressure \citep{Nomoto1976,Nomoto1982b,Nomoto1984b}. 
Therefore, both scenarios cannot be ruled out.
By examining the overlapping mass range for all three curves, 
it can be seen that the sub-Chandrasekhar mass branch
has a $M_{{\rm Ni}}$ lower than the Chandrasekhar mass
branch. Future statistics of observed SNe Ia 
for this pair of quantities will resolve the 
uncertainty here. 

At last we explain the difference of $M_{{\rm ^{56}Ni}}$
between the spherical and aspherical models.
The spherical model in general produces more 
$^{56}$Ni than the aspherical model for the same
$M$. This is because the C-detonation starts in the 
center for the spherical model and off-center for the aspherical 
one. However, most the $^{56}$Ni is produced near the center, where 
the density is the highest. This means, for aspherical detonation
to produce $^{56}$Ni, it needs to overcome the 
density gradient and the outward motion of the white dwarf during
expansion. This requires more time for the detonation to reach 
the center to burn the matter for synthesizing $^{56}$Ni,
while the white dwarf has started its expansion. 
As a result, the matter density burnt by aspherical detonation in general
is lower, which suppresses the production of $^{56}$Ni. 
Future observations of SN Ia mass and $^{56}$Ni mass 
can provide further constraints on this degeneracy,
and hence the asphericity of the initial He-detonation.  

\subsection{Contribution to Galactic Chemical Evolution}

The single-degenerate (Chandrasekhar mass model)
versus double degenerate (sub-Chandrassekhar mass model) scenario
has been a long lasting theoretical tension remaining
unsolved. The Chandrasekhar mass model has been
favored because of its correspondence to an invariant 
model which can explain the similarity among observed SNe Ia.
However, the shock-companion star interaction is
shown to provide strong X-ray signal before the bolometric
maximum of the light curves \citep{Kasen2010}.
The absence or non-discovery of such feature 
leads to the consideration of using the sub-Chandrasekhar mass model
as an alternative to explain the origin of SNe Ia.

\begin{figure}
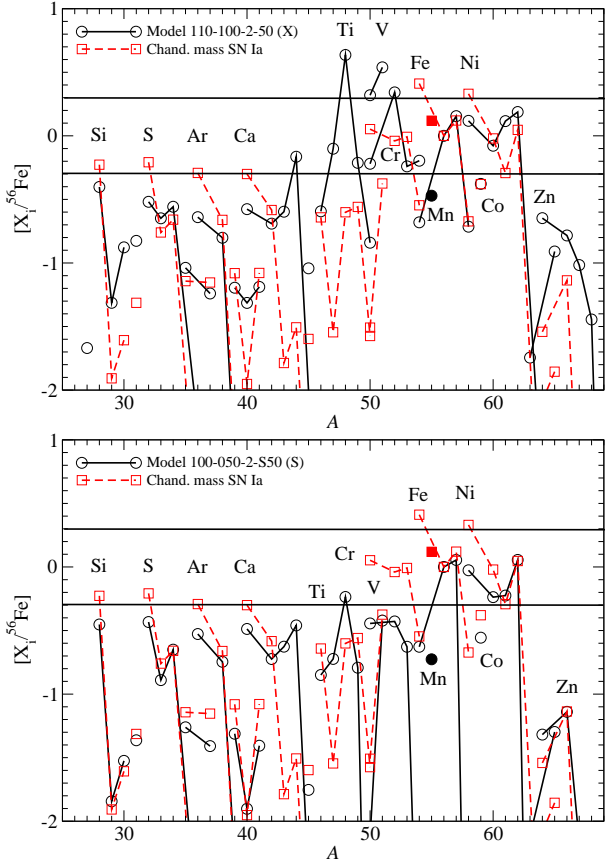

\centering
\includegraphics*[width=8cm,height=5.7cm]{fig26a.eps}
\includegraphics*[width=8cm,height=5.7cm]{fig26b.eps}
\caption{$[X_i/^{56}$Fe] for the benchmark Model
110-100-2-50 (X) (upper panel) and 100-050-2-S50 (S) (lower panel).}
\label{fig:DDvsSD}
\end{figure}

To compare how the sub-Chandrasekhar mass model influences
the metal enrichment process we first compare the chemical
yield directly. In Figure \ref{fig:DDvsSD} we compare
the chemical yield between the benchmark Models of this
work, namely Model 110-100-2-50 (upper panel) and
Model 100-050-2-S50 (lower panel) with the 
benchmark Chandrasekhar mass model. We can see that 
the Chandrasekhar mass model has its IME more closer
to the solar value. The Ti and V productions are
suppressed compared to the sub-Chandrasekhar mass model. 
The Fe and Ni patterns are similar for the two classes
of model, except the $^{54}$Fe and $^{58}$Ni are more
enhanced in the Chandrasekhar mass model. As remarked,
the amount of Mn in the sub-Chandrasekhar mass model is 
very small to explain the solar value owing to the 
differences in electron capture rates. In Model 100-050-2-S50,
a similar difference can be observed, except that the 
over-productions in Ti, V and Cr become regulated
due to its less massive He envelope.

In Figure \ref{fig:Mni_evol} we plot the evolution of 
$X(^{55}$Mn)/$X(^{56}$Fe), scaled with the solar value,
as a function of metallicity $Z$. To contrast with the results
of the sub-Chandrasekhar mass model, we also include 
the benchmark Chandrasekhar mass model from \cite{Leung2017a,Nomoto2017}. 
The stellar abundance
from galactic disk F and G dwarfs \citep{Reddy2003}, 
cluster and field stars \citep{Sobeck2006} 
and stars from thin discs \citep{Feltzing2007}. 
are included. As expected, at $Z < -1$ 
both models do not alter the results since the time-delay
of SNe Ia mutes the contribution of SNe Ia. After that, 
the two models deviate. The Chandrasekhar mass model,
which shows a healthy electron capture, provides sufficient
$^{55}$Mn to raise the ratio close to the solar value. 
On the other hand, the sub-Chandrasekhar mass model, which produces
only 30 \% of the solar ratio. The prolonged underproduction
of $^{55}$Mn makes the ratio even decreases in [Fe/H]$ = -0.2 - 0$
to $\approx 30 \%$ of the solar value. 

This suggests that even when sub-Chandrasekhar model can provide a
variety of model, with ranges of $^{56}$Ni to match observational
results of different peak luminosity and with ranges of progenitor
mass for different light curve widths. The Chandrasekhar mass model
contribution to the stellar evolution remains important. 
The nucleosynthesis suggests that $^{55}$Mn can be partially
produced owing to the strong compression heating of the matter
inside the star. The related mass is far from enough to explain 
the grow of $^{55}$Mn especially from $\log_{10} Z = -0.2 - 0$.

\begin{figure}
\centering
\includegraphics*[width=8cm,height=5.7cm]{fig27.eps}
\caption{The $[^{55}$Mn/$X(^{56}$Fe)]
against [Fe/H] for the benchmark model 110-100-2-50 (X) and the
typical Chandarsekhar mass SNe Ia \citep{Leung2018}.}
\label{fig:Mni_evol}
\end{figure}

For further application of our sub-Chandrasekhar SNe Ia 
yield in the context of GCE, we also present in Tables
\ref{table:stable_yield4}, \ref{table:stable_yield4b} and \ref{table:radioactive_yield4},
Tables \ref{table:stable_yield5}, \ref{table:stable_yield5b} and \ref{table:radioactive_yield5},
and Tables \ref{table:stable_yield6}, \ref{table:stable_yield6b} and \ref{table:radioactive_yield6},
the mass abundance of our representative SNe Ia models
with the minimum $M_{{\rm He}}$ necessary for triggering the 
second detonation based on the Models 110-100-2-50 (X), 110-050-2-50 (Y) and 
100-050-2-S50 (S) respectively.
Metallicity is obviously an important factor which 
contributes to $^{56}$Ni production and also the 
production of high neutron-ratio isotopes. 
$M_{{\rm He}}$ determines the minimum mass above which
the C detonation can be triggered in our aspherical
detonation models. $M_{{\rm Ni}}$ is the primary 
indicator of the explosion strength as derived from
the light curves.

It can be seen that, from the observational point of 
view, the sub-Chandrasekhar mass SNe Ia produce 
radioactive isotopes qualitatively different from
the conventional Chandrasekhar mass SNe Ia. Due 
to the He envelope burning, $\alpha$-chain elements 
are more pronounced. Among that, $^{44}$Ti is produced,
which has a half life of $\approx 60$ years
by electron capture to form $^{44}$Sc. A typical
amount of $\sim 10^{-3} M_{\odot}$ is found. Their 
abundance decreases when the $^{56}$Ni production
increases. 
The detection of the decay of $^{44}$Ti as a
long term energy of SNe Ia remnant may give very
stringent constraints on the progenitor type of SNe Ia.

%%%%%%%%%%%%%%%%%%%%%%%%%%%%%%%%%%%%%%%%%%%%%%%%%%%%%%%%%%%%%%%%%%%%%%%%%

\section{Comparisons with Observed Supernovae and Supernova Remnants}  

We have shown in \cite{Leung2017a,Nomoto2017,Leung2018} that the Chandrasekhar mass
turbulent deflagration model with delayed detonation 
transition can be constrained through the observational data
including the late-time light curves and the spectra. 
The late-time light curves can give indications to the amount
of minor isotopes which has a longer lifetime compared
to $^{56}$Ni with a half life 7.8 days. 
They include for example $^{56}$Co and $^{57}$Co, which 
have a decay lifetimes of 77.2 and 272 days respectively. The energy deposition
during the decay supports the light curve being observed. 
Another way to study SN chemical abundance 
is by the spectra of SN remnant. Through a comparison
of the X-ray line strengths of the radioactive elements, 
such as Cr, Mn, Fe and Ni, one can obtain the ratio 
among these elements and thus cast constraints on the 
explosion mechanism (See. e.g. \cite{Yamaguchi2014}). 

\subsection{Supernova Remnant 3C 397}

\begin{figure}
\centering
\includegraphics*[width=8cm,height=5.7cm]{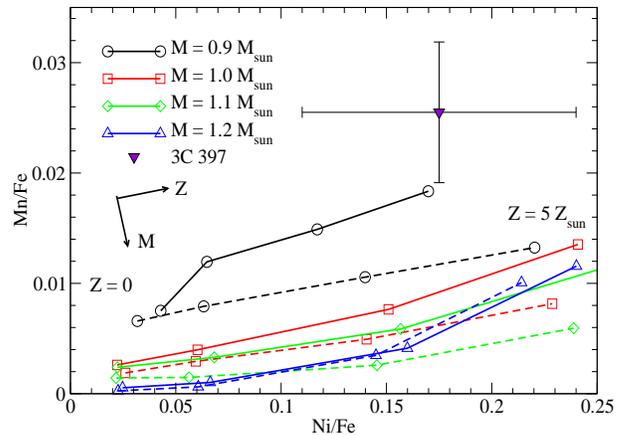}
\caption{The mass ratio Mn/Fe against Ni/Fe for the sub-Chandrasekhar 
mass models from $M = $ 0.9, 1.0, 1.1 and 1.2 $M_{\odot}$. 
The He envelope is fixed to be 0.1 (solid line) and 0.2
(dashed line) $M_{\odot}$. The metallicity ranges
from 0, 1, 3 and 5 $Z_{\odot}$ for models on the same
line from the left to right. The observational data
point of the SN remnant 3C 397 is included. The typical mass and metallicity
dependences of the models are shown by the arrows.}
\label{fig:Mn_Ni_plot}
\end{figure}

The first example we study is the SN remnant 3C 397 \citep{Yamaguchi2015}. 
This SN remnant has a remarkable X-ray spectrum in 
terms of its rich neutronized material compared to 
other SN remnants such as Tycho and Kepler. This remnant
is also shown that the Chandrasekhar mass model
is one of the feasible realizations of SN Ia 
explosions constrained by direct observational data. 
In the measurement, this remnant is found to have 
$0.027 \pm ^{+0.007}_{-0.006} M_{\odot}$ Cr, 
$0.025 \pm ^{+0.008}_{-0.007} M_{\odot}$ Mn and
$0.17 \pm ^{+0.07}_{-0.05}$ Ni. This corresponds
to the Mn/Fe and Ni/Fe ratios being 0.018 - 0.033
and 0.11 - 0.24 respectively. In \cite{Yamaguchi2015}
it is shown that by using one-dimensional models,
Chandrasekhar mass model ($M$ $\approx$ 1.37 $M_{\odot}$)
with a metallicity five times to the solar metallicity
is shown to produce the closet abundance ratio. 
In \cite{Nomoto2017,Leung2018} we reported similar discovery
based on a series of two-dimensional models of
turbulent deflagration model with delayed detonation transition. 
Here we shall examine our models to see in the 
sub-Chandrasekhar mass domain what kind of model 
is needed to explain this SN remnant. 

In Figure \ref{fig:Mn_Ni_plot} we plot  
Mn/Fe against Ni/Fe for our sub-Chandrasekhar mass
models with the observational data from the SN remnant. 
The SN Ia models of $M = $ 0.9 - 1.2 $M_{\odot}$ 
are included with a He envelope of $M_{{\rm He}} = 0.1 - 0.2$ $M_{\odot}$.
We pick $Z = $ 0 to 5 $Z_{\odot}$ as 
done in \cite{Leung2018}. It can be seen that in general 
when metallicity increases, Mn/Fe increases with 
Ni/Fe. However, when the total mass $M$ increases, 
the whole shifted downward, showing that the Mn/Fe
ratio drops but no significant change in Ni/Fe
observed. This is because when the mass increases,
the central density of the initial model increases, 
therefore, the C detonation becomes more energetic 
which can unbind the star more quickly. As a result, 
more $^{56}$Ni is produced which suppresses the ratio.

The model with a more massive He envelope has a 
lower [Mn/Fe] in general because of the higher $^{56}$Ni
as part of it can be produced in the envelope.
This relation is uniform for almost all models except 
for $Z = 5$ $Z_{\odot}$ at $M = 1.2$ $M_{\odot}$. 
The two models show a rapid jump in the [Mn/Fe] ratio.
This is because at high density, electron density 
becomes important. The C detonation, which can release 
adequate energy to burn the core matter into NSE,
is followed by electron capture before the matter
cooling down by adiabatic expansion. Certainly, the typical 
electron capture rate in the sub-Chandrasekhar mass 
model is considerably lower than those in typical
Chandrasekhar mass models. The lowered electron fraction
in the core matter, when in NSE, will be more favourable 
to produce $^{55}$Mn, which has a proton ratio of 0.454.
The data point of 3C 397 is included. It can be seen
that the data point lies very far from the other lines. 
This is consistent with the conclusion in \cite{Yamaguchi2015}
that the single degenerate Chandrasekhar mass SN Ia
channel is more likely to explain this peculiar SN Ia. 

Our models show that the $^{55}$Mn production is 
in general too low that even with a rather small 
$^{56}$Ni production at the end of simulation, the resultant 
[Mn/Fe] ratio remains insufficient to explain. 
The closest model is the $M = 0.9$ $M_{\odot}$ at 
$Z = 5$ $Z_{\odot}$. Our result is comparable with theirs. 

One may note that this object has raised interest in the 
literature owing to its predicted high metallicity and 
different proposals are raised in order to recover the 
high [Mn/Fe] ratio without invoking the high metallicity.
For example, in \cite{Shen2018} the sub-Chandrasekhar
mass SNe Ia in the spherical approximation is revisited.
The high [Mn/Fe] is shown to be viable if one consider
a subset of ejecta, namely by taking the effects of 
reverse shock heating into account. Another attempt is 
done in \cite{Dave2017}. The gravitational confined
detonation model is explored with extension to pure
turbulent deflagration with or without delayed
detonation transition for the Chandrasekhar mass model. 
It is shown that a combination of high central density, 
low [C/O] ratio and a high offset of initial deflagration 
can provide an alternative to this observation. 
These trends are consistent with our previous 
finding as reported in \cite{Leung2018}.

\subsection{SN 2012cg}

\begin{figure}
\centering
\includegraphics*[width=8cm,height=5.7cm]{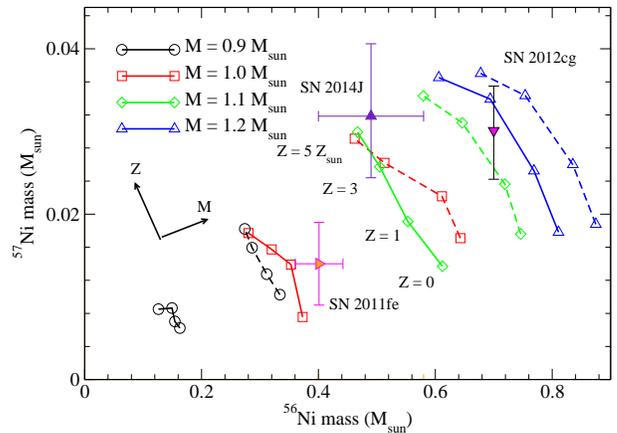}
\caption{$^{57}$Ni against $^{56}$Ni for our sub-Chandrasekhar 
mass models from $M = $ 0.9, 1.0, 1.1 and 1.2 $M_{\odot}$. 
The He envelope is fixed to be 0.1 (solid line) and 0.2
(dashed line) $M_{\odot}$. The metallicity ranges
from 0, 1, 3 and 5 $Z_{\odot}$ for model from bottom
to top. The observational data
point of the SN 2011fe, SN 2012cg and SN 2014J are included.
The typical mass and metallicity
dependences of the models are shown by the arrows.}
\label{fig:Ni56_Ni57_plot}
\end{figure}

The next application is on SN 2012cg. This SN
exploded at 2012 May 17 (UT) in the nearby spiral
galaxy NGC 4424, which is measured in the 
Lick Observatory Supernova Search \cite{Kandrashoff2012}.
The SN Ia nature is revealed in the spectral study found in 
\cite{Cenko2012, Marion2012}. This SN Ia is close
enough that the late-time light curve after $\sim 1000$
days can still be measured. The low-density ejecta 
becomes transparent to most $\gamma$-ray so that the 
$\gamma$-ray emitted during decay can escape freely
from the ejecta without significant heating. On the
other hand, the emitted $e^-$ is completely absorbed 
by the surrounding matter. This means that one can trace
its amount through its decay as a heat source in 
the light curve. In particular, the channels
$^{56}$Co $\rightarrow ^{56}$Ni (half life $\approx 113$ days) and
$^{57}$Ni $\rightarrow ^{57}$ Co (half life $\approx 272$ days)
can be readily measured. In \cite{Graur2016}, the 
B-band light curve of SN 2012cg is revisited at 900 
days after the B-band maximum. It is shown that 
this SN Ia has a high $^{57}$Ni/$^{56}$Ni 
ratio at 0.043$\pm ^{+0.012}_{-0.011}$, which is
twice to the corresponding solar ratio. 

In Figure \ref{fig:Ni56_Ni57_plot} we plot similar to 
Figure \ref{fig:Mn_Ni_plot} but for $^{57}$Ni 
against $^{56}$Ni for the same series of models
and with this SN Ia. For models with an 
increasing metallcity, $^{57}$Ni production increases
while $^{56}$Ni mildly decreases. This is because 
the initial electron fraction, as metallicity 
increases, deviates from the value 0.5, which most
favours the production of $^{56}$Ni in NSE. On the 
other hand, the lowered electron fraction enhances
production of $^{57}$Ni. Models with a thicker 
He envelope has higher $^{56}$Ni and $^{57}$Ni 
compared to models with the same mass but lower
$M_{{\rm He}}$. Similarly, for models with an
increasing $M$, the $^{56}$Ni and $^{57}$Ni productions are 
enhanced as a result of higher central density, which 
allow more matter to be burnt until NSE for producing
Fe-peak elements. 

Then, we compare our results with this SN Ia. 
The data point of SN 2012cg
is included. It can be seen that this SN has a rather high
$^{56-57}$Ni as a healthy explosion. In our models,
the high mass models $M = 1.2~M_{\odot}$ 
with high metallicity from $3 - 5~Z_{\odot}$ are
more likely to explain this data point. This is consistent
with our previous work \citep{Leung2017a,Nomoto2017,Leung2018}
that a high metallcity model from $1 - 5~Z_{\odot}$ 
with a central density from $5 \times 10^8$ - $1 \times 10^9$ 
g cm$^{-3}$ may fit this observational data the best. 
However, compared to our Chandrasekhar mass model, 
the sub-Chandrasekhar mass models can fit the 
upper range of this data point by models with $M = 1.2 ~M_{\odot}, M_{{\rm He}} = 0.1 M_{\odot}$
and fit the lower range of that by models with $M = 1.1 ~M_{\odot}, M_{{\rm He}} = 0.2 M_{\odot})$. 
The trend derived
here agress with the estimation from the analytic formula 
as done in \cite{Graur2016} that the Chandrasekhar 
mass model is more preferred for this
high $^{57}$Ni abundance. However, we also emphasized
that the sub-Chandrasekhar mass model is not excluded 
by this SN Ia as a physical picture. To further clarify
its origin, future spectral study in the remnant,
similar to the SNR 3C 397 will be necessary.

\subsection{SN 2011fe}

\begin{figure}
\centering
\includegraphics*[width=8cm,height=5.7cm]{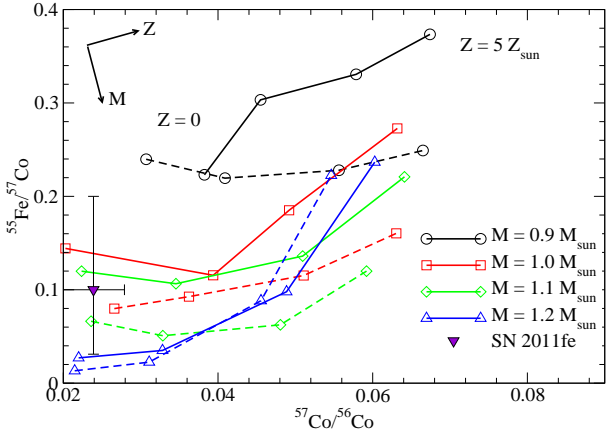}
\caption{$^{55}$Ni/$^{57}$Co against $^{57}$Ni/$^{56}$Co
for our sub-Chandrasekhar 
mass models from $M = $ 0.9, 1.0, 1.1 and 1.2 $M_{\odot}$. 
The He envelope is fixed to be 0.1 (solid line) or 0.2
(dashed line) $M_{\odot}$. The metallicity includes
0, 1, 3 and 5 $Z_{\odot}$. The observational data
point of the SN 2011fe is included. The typical mass and metallicity
dependences of the models are shown by the arrows.}
\label{fig:Mn55_Fe57_plot}
\end{figure}

The third example comes from the well observed SN 2011fe.
This recent SN exploded at 2011 August 24 in a rather
proximate Pinwheel galaxy \citep{Nugent2011}, 
which situated at 6.4 Mpc away \citep{Shappee2011}.
The close distance of this SN Ia has attracted
intensive study in different bandwidths (See \cite{Shappee2016}
for the references therein) and spectral studies. 
This also allows detection of 
light curves beyond $\sim 1000$ days. This provides
more abundance constraints compared to the previous SN 2012cg. 
This SN Ia is first probed with the decay of $^{55}$Fe 
($^{55}$Fe $\rightarrow ^{55}$Mn with a half life 
of 999.67 days) directly. 
Through taking ratios with other decaying channels,
they observe $\log_{10} (^{57}{\rm Co}/^{56}{\rm Co}) = -1.62 \pm ^{+0.08}_{-0.09}$.
In their best fit model they also showed
$\log_{10} (^{55}{\rm Fe}/^{57}{\rm Co}) = -1.0 \pm ^{+0.3}_{-0.5}$.

In Figure \ref{fig:Mn55_Fe57_plot} we plot similar 
to Figure \ref{fig:Ni56_Ni57_plot} but for 
the ratio $^{55}$Fe/$^{57}$Co against 
$^{57}$Co/$^{56}$Ni for our sub-Chandrasekhar mass models
and the observational data SN 2011fe. Our models 
show a less uniform variation with increasing
metallicity in the sub-Chandrasekhar mass range. 
The variation is non-uniform at low metallicity 
(0 - 1 $Z$). On the other hand, $^{55}$Fe/$^{57}$Co
increases with metallicity, showing that the 
abundance of $^{55}$Fe is more sensitive to metallicity.
This is expected, as shown in Figure \ref{fig:Mn_Ni_plot} that
the metallicity still plays an important role 
in the formation of stable Mn, which comes from the 
decay of $^{55}$Fe. The models tend to have a lower 
ratio when the He envelope becomes thicker. Also,
when the total mass increases, the ratio is 
also suppressed. This is because the growth of $^{57}$Ni,
which is very sensitive to the size of zone being burnt
into NSE, is faster than $^{55}$Fe. The 
much faster growth of $^{55}$Fe/$^{57}$Co for the 
model $M = 1.2~M_{\odot}$ and $Z = 5~Z_{\odot}$
is again related to the enhancement of electron capture
in the NSE region. The observational data point fits
our model much better than the previous two models. It can
be seen that a wide range of parameters can be used to explain
this SN Ia. SN Ia models from $M = 1.0 - 1.2$ $M_{\odot}$
and a low metallicity $Z = 0 - 1$ $Z_{\odot}$ are 
adequate to fit in this observational data. This is 
also consistent with our previous work \citep{Leung2017a,Nomoto2017,Leung2018}
that a low central density from $5 \times 10^8 - 7.5 \times 10^8$ 
g cm$^{-3}$ with a metallicity $0 - 1$ $Z_{\odot}$ can explain 
this data point using the turbulent deflagration model with 
DDT. 

Recent late time study of the light curve in the 
optical band has also revealed the $^{57}$Ni/$^{56}$Ni
ratio of this SN Ia. By measuring the shift of 
late time light curve after most $^{56}$Ni has decayed,
the decay of $^{57}$Ni $\rightarrow ^{57}$Co can 
be another important radioactive source. In \cite{Dimitriadis2017},
the pseudo-bolometric light curve is produced by combining
data of the optical and near-IR bandwidth in the literature 
from 200 to 1600 days after explosion. It is shown 
that this supernova, albeit with significant 
systematic uncertainties, $^{57}$Co/$^{56}$Co = $0.031 \pm 0.011$.

In Figure \ref{fig:Ni56_Ni57_plot} we also plot
this data point with our model sequences. 
The prediction of WD progenitor using the
explosion product has been discussed for 
SN 2012cg in the previous section. Here 
we further apply this technique for SN 2011fe. 
The WD sequence with a mass $\approx 1.0$ $M_{\odot}$
with a He envelope $\approx 0.1$ $M_{\odot}$
can explain the observed $^{57}$Ni/$^{56}$Ni
ratio. The data point can be the best explained
by the model with $\sim Z_{\odot}$. 

We remark that, from the first sight the SN 2011fe 
is very well explained by the sub-Chandrasekhar mass
model, in fact, in \cite{Leung2017a,Nomoto2017} we demonstrate
that this SN Ia can also be explained by the 
Chandrasekhar mass model in the high central density
(high mass) limit with a metallicity close to $Z_{\odot}$
in the centrally ignited model. This suggests that
to further constrain the progenitor, future follow-up 
observations will be essential to measure the abundances of 
other isotopes or elements, similar to the 
analysis done for the SNR 3C 397.

\subsection{SN 2014J}

The fourth application of our models to the SN Ia
observation is the candidate SN 2014J. This is 
an extremely well observed SN Ia owing to its vicinity. 
This SN Ia is observed from very early time since its
rising light curve \citep{Goobar2014}.
The multi-frequency light curve and spectra 
are observed ranging from the 
infrared spectra \citep{Telesco2015}, optical photometry 
and spectrography \citep{Ashall2014},
ultraviolet \citep{Foley2014, Brown2015}, to gamma-ray
light curve and spectra \citep{Diehl2014, Churazov2015, Diehl2015, Isern2016}.
This supernova is interesting for 
its peculiar gamma ray signals. 
It has an early gamma ray signal coming from the decay of 
$^{56}$Ni at about 20 days after explosion, which is 
10 days in advance of typical SNe Ia \citep{Diehl2014}.
The follow-up observation in its time-domain variations
shows that it has a non-monotonic variation
in the $^{56}$Co-decay gamma ray line. The 
Doppler shift analysis further shows the 
highly fluctuating Co-decay line frequency \citep{Diehl2015}. 
Such features are argued to be originated from the
He detonation and asymmetry in the detonation. 
The current work on the asymmetry double detonation
model appears to well match with this SN. 

Here we try to constrain its progenitor from 
some of its observable by its 
$^{57}$Ni/$^{56}$Ni mass fraction ratio. 
This ratio has been applied to other 
SNe Ia including the previous SN 2011fe and SN 2012cg.
The late time flattening of the late curve 
in the optical band is analysed, from 277 days to 1181
days after explosion. 
From the analysis of the late time light curve \citep{Yang2018},
the mass ratio of $^{57}$Co$/^{56}$Co $= 0.066 \pm ^{0.009}_{0.008}.$
The ratio is even higher than SN 2012cg. 
By using the $^{56}$Ni derived from gamma ray \citep{Isern2016},
where $^{56}$Ni $= 0.49 \pm 0.09$, we plot in 
Figure \ref{fig:Ni56_Ni57_plot} the data point
of SN 2014J.

From the figure we observe that the sub-Chandrasekhar
mass model is capable of reaching the high Ni-isotope
ratio at the high metallicity end. Two of the model
sequences can approach this observed data point, 
namely when $M = 1.1$ $M_{\odot}$ with $M_{{\rm He}} = 0.10$ $M_{\odot}$
and $M = 1.0$ $M_{\odot}$ with $M_{{\rm He}} = 0.20$ $M_{\odot}$. 
Both sequences require $Z \approx 5 ~Z_{\odot}$
to reach the high mass fraction ratio. 
Again, the more massive He envelope is capable
of producing the required $^{56}$Ni, however
such early surface $^{56}$Ni can be very different
from that produced through standard Chandrasekhar mass
WD. In the latter case, the $^{56}$Ni is mostly
produced by C detonation after deflagration-detonation
transition. But it is always covered by another layer
of IME when the detonation reaches the surface. As a result, 
the $^{56}$Ni-decay is not directly visible, but can
be seen as a heat source in the light curve. 
On the other hand, with the He envelope, there is 
almost no shielding for the synthesized $^{56}$Ni,
therefore it is expected that the early gamma-ray 
signal can be very different. 
We also note that such a massive He envelope with decaying 56Ni
should show rather strong He lines.

We also compare the 57Ni/56Ni ratio of SN 2014J with the Chnadrasekhar
mass models in Figure 20 of \cite{Leung2018}.  We note that the
Chandrasekhar mass models with $Z \approx 3 - 5 ~Z_{\odot}$ produces
the observed high 57Ni/56Ni ratio.

\section{Summary}

In this paper, we study the hydrodynamics and associated nucleosynthesis 
of sub-Chandrasekhar mass models for SNe Ia, where the 
C detonation is triggered by the surface He detonation of 
 a symmetric or an asymmetric structure. 
Such a double detonation can in both the single degenerate 
and the double degenerate scenarios.   
Our findings are summarized as follows.

(1) We find that whether C detonation triggered is strongly sensitive
to the He detonation pattern.  
We consider four possible structures: namely, one-bubble, 
one-ring, bubble-and-ring, and spherical, in view of the unresolved, 
inner fluid motion of the He shell before nuclear runaway.
The He detonation with higher
symmetry (one-ring and spherical structures) can result 
in geometric convergence, which can very robustly 
heat up the C fuel to the ignition temperature for 
the subsequent temperature. He detonation with lower
symmetry (one-bubble) requires a more massive He envelope $(> 0.1 ~M_{\odot})$
to trigger the second explosions. The case with multi-bubbles 
depends on how the shock wave propagates inside the WD.

(2) We carry out a parameter survey on the nucleosynthesis
for the sub-Chandrasekhar mass WD models with different
model parameters. We perform two-dimensional hydrodynamical
simulations using our own supernova simulation code from
the onset of the He detonation until all detonations quench
by the expansion. The following parameters are studied:
the metallicity, He envelope mass, total mass, the initial 
He detonation, and the initial C/O mass fraction ratio.
We pay attention to some representative elements, including 
intermediate mass elements (e.g. Si, S, Ar and Ca),
light iron-peak elements (Ti, V and Cr) and 
other iron-peak elements (Mn, Fe, Co, Ni). 
These elements are strongly sensitive to the total mass, metallicity
and He-envelope mass, 
but less sensitive to the initial He detonation and
C/O mass fraction ratio. Metallicity affects mostly
on the low-$Y_{\rm e}$ isotopes e.g., $^{55}$Mn
and $^{58}$Ni. He envelope mass affects light iron-peak
elements, especially $^{48}$Ti, $^{50,51}$V
and  $^{52}$Cr. Total mass affects $^{56}$Ni
and hence the mass fraction
$[X_i/^{56}$Fe] with respect to the Sun.

(3) We also compare our two-dimensional models
with the classical spherical double detonation model 
and show that the chemical signature due to asphericity
is very significant. The aspherical detonation can create hot spots
which produce distinctive abundance pattern in intermediate
mass elements and light iron-peak elements (Ti, V and Cr).
Explosion of progenitor with a mass $\sim 1.1 - 1.2 ~M_{\odot}$ may
help distinguish in the future the degeneracy of single
and double degenerate scenario. However, an exact
matching with the observed $^{56}$Ni distribution
will also require the stellar initial mass function.
We further show that 
the sub-Chandrasekhar mass WD models cannot substitute 
the Chandrasekhar mass one because of the persistent 
insufficiencies of Mn production. The final [Mn/Fe] can
be 0.4 dex lower than the model using Chandrasekhar mass
WD model.
We provide corresponding yield tables for the applications
to the galactic chemical evolution.

(4) We apply our models to provide constraints on some
well-observed SNe Ia, including 
SN 2012cg, SN 2011fe, SN 2014J, and SN Ia remnant 3C 397. The probable progenitor 
configurations are implied based on the derived 
chemical abundance of some Fe-peak isotopes.
We used the late time light curve to indicate the $^{57}$Ni/$^{56}$Ni
ratio. We find that SN 2014J can resemble with the sub-Chandrasekhar
mass model at $1.0 - 1.1 ~M_{\odot}$ with metallicity $Z = 3 - 5 ~Z_{\odot}$.
(Note that the Chandrasekhar mass models also resemble 
to SN 2014J if $Z = 3 - 5 ~Z_{\odot}$.)
SN 2011fe can be explained by models with $M \sim 1.0 M_{\odot}$
with near $Z_{odot}$. SN 2012cg can be approached by models
with $M = 1.1 - 1.2 M_{\odot}$ at $Z = 1 - 3 ~Z_{\odot}$.
For supernova remnant 3C 397, the high Mn/Ni ratio cannot
be resembled with any of our current sub-Chandrasekhar mass models.
The Mn/Fe ratios in our models are much lower than the observed value. 
Only models at the lower mass end $(0.9 M_{\odot})$ with $Z = 5 Z_{\odot}$
can approach the observe data point.

(5) The hydrodynamical structures and nucleosynthesis profiles provide useful
predictions for future observations of elemental abundances and line $\gamma$-rays.

\section{Acknowledgment}

This work has been supported by World Premier 
International Research Center Initiative 
(WPI), and JSPS KAKENHI Grant Number JP17K05382. 
S.C.L. acknowledges support from Grant HST-AR-15021.001-A.

We thank the anonymous referee for the very detailed 
and helpful suggestions to improve the manuscript. 

We thank Frank Timmes for
his open-source micro-physics subroutines
including the Helmholtz equation of state 
and the $torch$ subroutine for 
the post-process nucleosynthesis.  
We also thank Ken Shen for the ideas in 
the numerical modeling of the detonation physics.
We thank Amanda Karakas and Chiaki Kobayashi for the 
information about the galactic stellar abundances. 
At last we thank Roland Diehl and Jordi Isern for the
inspiring discussion about SN 2014J. 

\appendix

\section{Determination of the He Detonation Timescale}

In Section \ref{sec:methods} we mentioned that simplified 
schemes for C and He detonation are used. In this section, 
we describe in more details how they are implemented.
Unlike C detonation,
He detonation in the sub-Chandrasekhar mass WD scenario, 
occurs at a much lower density ($\sim 10^4 - 10^7$ g cm$^{-3}$
in the He envelope). The low density, as well as the 
non-degenerate property of the electron gas, lead to a 
lower final temperature, after all He is burnt. 
As a result, it becomes important to estimate more
precisely how much He is burnt in the reaction zone 
and in the post-reaction zone. In particular, we are
interested to know how He is burnt as a function 
of time, which is used to calibrate the amount of energy
released by the detonation. 

We calculate the detonation structure following the 
numerical scheme described in \cite{Sharpe1999}. Here we
give a brief summary about this method. In general,
detonation consists of three sections, 
the pre-shock region, the reaction zone
and the post-reaction region. We assume at
every point inside the detonation wave, 
thermodynamics equilibrium is maintained, such that 
the specific internal energy $\epsilon$, pressure $p$
are related by the thermodynamics input including 
the density $\rho$, temperature $T$ and the 
number fraction of each isotope $Y_i$ ($i = 1, N)$
in a network with $N$ isotopes. Therefore, 
\begin{equation}
\Delta \epsilon = \frac{\partial \epsilon}{\partial \rho}|_{T,X_i} + \frac{\partial \epsilon}{\partial T}|_{\rho,Y_i} + 
\sum_i \frac{\partial \epsilon}{\partial Y_i}|_{\rho,T}.
\end{equation}
The steady state Euler equation can be written as
\begin{eqnarray}
\frac{d \rho}{dx} = -\frac{\rho a^2_f}{v} 
\frac{{\bf \sigma} \cdot {\bf R}}{\iota}, \\
\frac{dT}{dx} = \left( \frac{\partial p}{\partial T} \right)^{-1}_{\rho, X} 
\{ \left[ u^2 - 
\left( \frac{\partial p}{\partial \rho} \right)_{T,X} \right]  
\frac{d \rho}{dx} - \nonumber \\ 
\sum^N_{i=1} \left( \frac{\partial p}{\partial X_i} 
\right)_{\rho,T,X_{j \neq i}} \frac{dX_i}{dx} \} , \\
\frac{dX_i}{dx} = \frac{R_i}{A_i v},
\label{eq:deton_struct}
\end{eqnarray}
where
\begin{equation}
\eta = a^2_f - v^2
\end{equation}
is the sonic parameter, 
$A_i$ is the atomic mass for the $i$-th isotope, 
\begin{equation}
a^2_f = \left( \frac{\partial p}{\partial \rho} \right)_{T,X} + 
\left[ \frac{p}{\rho^2} - \left( \frac{\partial \epsilon}{\partial \rho} \right)_{T,X} \right]
\left( \frac{\partial p}{\partial T} \right)_{\rho,T} \left( \frac{\partial \epsilon}{\partial T} \right)^{-1}_{\rho,X}
\end{equation}
is the sound speed of constant composition (also known
as frozen sound speed in the literature of detonation), 
\begin{eqnarray}
\sigma_i = \frac{1}{\rho a^2_f} \{ \left( \frac{\partial p}{\partial X_i} \right)_{\rho,T,X_{j \neq i}} - 
\left( \frac{\partial p}{\partial T} \right)_{\rho, X} 
\left( \frac{\partial \epsilon}{\partial T} \right)^{-1}_{\rho,X} \nonumber \\ 
\left[ \left( \frac{\partial \epsilon}{\partial X_i} 
\right)_{\rho,T,X_{j \neq, i}} - \left( 
\frac{\partial q}{\partial X_i} \right)_{X_{j \neq i}} \right] \}
\end{eqnarray} 
is the thermicity constant, such that 
${\bf \sigma} \cdot {\bf R}$ is the thermicity. 
In integrating this set of differential equations, we
use the boundary conditions at $x = 0$, $\rho = \rho_i$, 
$T = T_i$; at $x \rightarrow \infty$, $\rho = \rho_f$,
$T = T_f$ and $X = X_f$ with thermicity $= 0$. 
Notice that $\rho_i$, $T_i$ and $Y_i$ 
are the quantities after shock. They are related to 
the pre-shock quantities $(\rho_0, T_0, X_0)$ by
\begin{eqnarray}
\rho_0 D = \rho_i c_s, \\
\rho_0 D^2 = \rho_i c_s^2, \\
\rho_0 D^2 + \rho_0 \epsilon_0 + P_0 = \rho_i c_s^2 + \rho_i \epsilon_i + P_i.
\end{eqnarray}
$D$, $c_s$ and $P_0$ are the pre-shock matter density, 
speed of sound and pressure of the pre-shock matter. 

\begin{figure*}
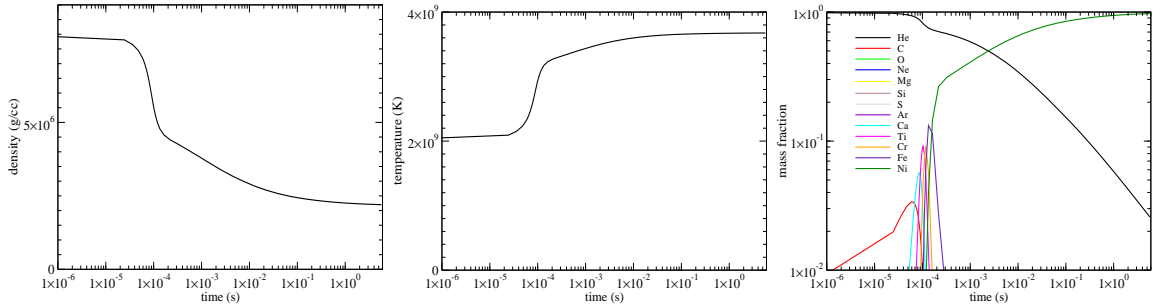

\centering
\includegraphics*[width=5cm,height=4cm]{fig31a.eps}
\includegraphics*[width=5cm,height=4cm]{fig31b.eps}
\includegraphics*[width=5cm,height=4cm]{fig31c.eps}
\caption{(left) The density evolution of He during detonation
for pure He fuel at a
density $10^6$ g cm$^{-3}$. The matter is assumed to 
be shock-heated to above $2 \times 10^9$ K. (middle)
The temperature evolution of the detonation
wave at an initial density $10^6$ g cm$^{-3}$.
(right) The isotope evolution of the detonation
wave at an initial density $10^6$ g cm$^{-3}$}
\label{fig:det_1e6_rho}
\end{figure*}

In Figure \ref{fig:det_1e6_rho}, we plot the density, 
temperature and chemical isotope profiles for 
a detonation wave at a density $10^6$ g cm$^{-3}$. 
To trigger the first incineration, the matter is
assumed to be shock heated to a temperature 
$\sim 2 \times 10^9$ K. Before $10^{-4}$ s. the temperature
does not rise considerately. Also, there is only very subtle 
drop in the density. There is also a slow change in 
the chemical composition from $^{4}$He to $^{12}$C.
At $\sim 10^{-4}$ s, the temperature rises drastically
from $2 \times 10^9$ K to $3 \times 10^9$ K. 
The density also drops by $\sim$ 30 \%. 
We can see a isotopes from 
$^{12}$C, $^{40}$Ca, $^{48}$Ti and $^{52}$Fe 
burst out one by one around $10^{-4}$ s. This means
that even at low temperature, the $\alpha$-chain
reaction can proceed efficiently, once the triple
$\alpha$ reactions have provided the first fuel 
for the subsequent reactions. Beyond $4 \times 10^{-4}$ s,
the productions of other isotopes are suppressed again,
except $^{56}$Ni. At that time, $^{4}$He is stably 
burnt into $^{56}$Ni, causing the temperature (density)
to grow (drop) to its equilibrium value. At $\sim$ 1 s,
the temperature and density reaches its equilibrium 
at $3.6 \times 10^9$ K and $4.6 \times 10^6$ g cm$^{-3}$. 

\begin{figure*}
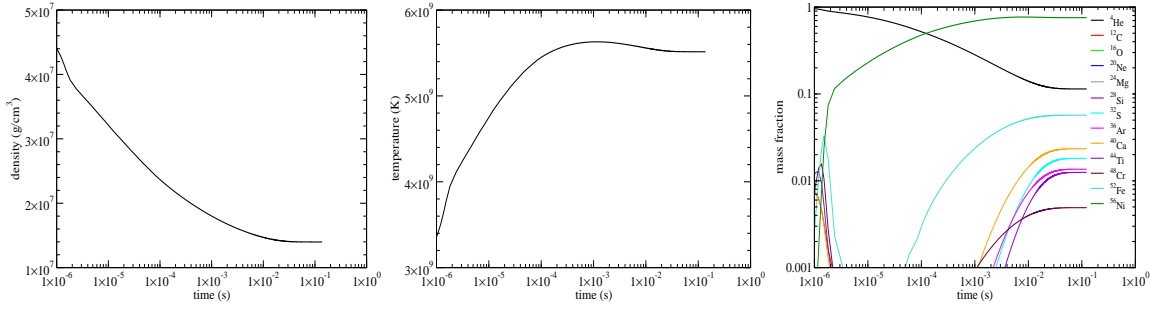

\centering
\includegraphics*[width=5cm,height=4cm]{fig32a.eps}
\includegraphics*[width=5cm,height=4cm]{fig32b.eps}
\includegraphics*[width=5cm,height=4cm]{fig32c.eps}
\caption{(left) The density evolution of He during detonation
for pure He fuel at a
density $10^7$ g cm$^{-3}$. The matter is assumed to 
be shock-heated to above $2 \times 10^9$ K. (middle)
The temperature evolution of the detonation
wave at an initial density $10^7$ g cm$^{-3}$.
(right) The isotope evolution of the detonation
wave at an initial density $10^7$ g cm$^{-3}$}
\label{fig:det_1e7_rho}
\end{figure*}

In Figure \ref{fig:det_1e7_rho}, we plot the temperature,
density and isotope abundance profiles for pure He fuel
at an initial density of $10^7$ g cm$^{-3}$. With
a high density, nuclear reactions can take place 
spontaneously. In the first $10^{-4}$ s, temperature
increases quickly from $3 \times 10^9$ K to 
$6 \times 10^9$ K. while the density drops 
from $4 \times 10^7$ g cm$^{-3}$ to 
$\sim 2 \times 10^7$ g cm$^{-3}$. The initial peaks 
for various isotopes except $^{56}$Ni can be found 
at the first $10^{-6}$ s, while the conversion of
$^{4}$He to $^{56}$Ni can be found at the first 
$10^{-4}$ s. After that, the temperature and 
density start to converge to their asymptotic values
at $\sim 5.5 \times 10^9$ K and $1.4 \times 10^7$ g cm$^{-3}$. 
At the same time, the temperature is sufficiently 
high that NSE emerges. $^{52}$Fe, $^{40}$Ca, 
$^{48}$Cr, $^{36}$Ar, $^{32}$S form one by one 
and reach their equilibrium value at $\sim$ 0.1 s. 

By comparing the two sets of results, we can see that
in the density range related to the sub-Chandrasekhar
mass double detonation models, the time necessary for 
He to completely release its energy 
into the system increases by two orders of magnitude
when the density drops from $10^7$ to $10^6$ g cm$^{-3}$.
(In the simulations we 
find the typical time steps has a size 
$\sim 10^{-4} - 10^{-3}$ s, depending on the global
velocity distribution). Therefore, especially for 
the He near the surface, once they are burnt 
they expand drastically, making their local density
much lower than those underneath. As a result, 
their energy release process is incomplete. To
mimic this effect, we use a density dependent time 
scale $\tau_{{\rm He}} (\rho)$ which is calibrated by the
detonation waves as demonstrated above. The time scale
corresponds to the time when 90 \% of energy is released
with respect to its equilibrium value. 
To establish the relation $\tau_{{\rm He}}$, we
repeat the above process for He-detonation wave at
different initial densities. Then we collect the 
necessary time scale by the above detonation. A 
simple power-law fitting provides us the formula:
\begin{equation}
\tau_{{\rm He}} = 1.72 \times 10^{-6}  \left( \frac{\rho}{10^8 ~ {\rm g~cm^{-3}} } \right)^{-2} {\rm s}.
\end{equation}
In the simulations,
when the current time step $\Delta t > \tau_{{\rm He}}$,
complete burning is assumed. Otherwise, only the fraction
of matter $\Delta t / \tau_{{\rm He}}$ is assumed to 
release its energy. We have only considered the 
effect of density because the reaction rate is very sensitive
to the input temperature. Below the ignition temperature
($\sim 10^9$ K), the reactions are so slow that the burning 
time scale is much longer than the dynamical timescale, which 
means no detonation can be formed. On the other hand, above the 
ignition temperature, the fuel burns instantaneously.
Also the energy generated by the nuclear reaction is 
much larger than the different choices of input temperature.
Thus, the product of the detonation wave is less insensitive to the
input temperature compared to the density.

Certainly a self consistent way, which is to calculate the 
network directly, can provide us the most accurate 
results regarding to the process of  partial burning. However, 
such inclusion is beyond the current capability of
our computing resource. Furthermore, in the 
hydrodynamics, acoustic waves are found everywhere
inside the star. These waves cause fluctuations 
in the local temperature. These fluctuations increase
the computation time significantly when a complete
network is used, 
since the nuclear composition always adjusts itself 
to the local temperature, where at high temperature
the typical time step is small.  

\section{Effects of symmetry boundary}
\label{sec:sym_bound}

In this work we have carried out simulations of sub-Chandrasekhar
SNe Ia in a quadrant of sphere. This uses a reflective boundary
along the symmetry plane $z = 0$. As a result, the initial He
detonation configuration, namely a one-bubble structure, 
corresponds to two synchronous ignitions of He detonation, 
one at the "north"-pole and one at the "south"-pole. 
It is unclear, prior to the runaway of He, how the 
velocity field, especially the turbulent velocity motion, 
may perturb the initial ignition of He. Certainly
it is more likely for two He detonation bubbles to 
have asynchronous ignition time, or even there is 
only one ignition before C detonation is triggered.
Therefore, it is unclear whether the C detonation 
can still be robustly triggered when there is a 
time-lapse between the two He detonation bubbles. 

To do the comparison, we develop a temporary extension of 
our hydrodynamics code to model the hemisphere of the 
WD by relaxing the reflection symmetry. We place one 
He bubble at the "north"-pole, while another one 
at the "south"-pole with some time delay. In Table \ref{table:testHS}
we tabulate the configuration and initial detonation
properties of our test models. 
It can be seen that the all the cases give a positive 
response to the He detonation, regardless of being
one or two He bubbles and their delay time. This suggests that 
as long as the He envelope has exceeded the marginal
thickness then the shock compression, either by 
shock-shock collision or by shock-wall collision 
can create similar heating to the surface matter
of the CO core. 

\begin{table*}
\begin{center}
\caption{The models for the study of reflection symmetry effects
in the sub-Chandrasekhar SNe Ia. Delay time is the difference
between the two He detonation bubble in the unit of s. 
"C-det?" corresponds to whether C detonation can be 
triggered or not. If yes, $\rho_{{\rm C-det}}$ and 
$T_{{\rm C-det}}$ are the density and temperature
of the triggered grid in units of $10^7$ g cm$^{-3}$ and
$10^9$ K respectively. $t_{{\rm C-det}}$ is the ignition
time in units of s. Position is the coordinate in units
of km.}
\label{table:testHS}
\begin{tabular}{|c|c|c|c|c|c|c|}
\hline
Model & Delay time & C-det? & Position & $t_{{\rm C-det}}$ & $\rho_{{\rm C-det}}$ & $T_{{\rm C-det}}$ \\ \hline
Test-QS   & 0   & Yes & (3420,0)     & 0.94 & 1.04 & 2.55 \\
Test-HS-0 & 0   & Yes & (3380,0)     & 0.94 & 1.06 & 2.84 \\
Test-HS-1 & 0.2 & Yes & (3290,-720)  & 1.03 & 1.09 & 2.83 \\
Test-HS-2 & 0.4 & Yes & (3120,-1400) & 1.15 & 1.04 & 2.89 \\
Test-HS-3 & 1.0 & Yes & (1330,-2910) & 1.43 & 1.00 & 2.00 \\ \hline
\end{tabular}
\end{center}
\end{table*}

We carry out 5 hydrodynamics simulations to extract the effects
of reflection symmetry. Test-QS corresponds to the model with 
reflection symmetry, where we choose the same configuration as 
the Benchmark Model 110-100-2-50. This means the Model 
Test-QS is exactly the benchmark model. Models Test-HS-0 - Test-HS-3
do not assume reflection symmetry and has a He ignition delay
time from 0 - 1 s. Model Test-HS-0 acts as a control test to see
if the hemisphere extension is consistent with a quadrant sphere
modeling; while in Model Test-HS-3 we delay the second ignition
so long such that the C detonation is triggered. From Table 
\ref{table:testHS}, when the delay time becomes larger, the 
position of the C detonation moves away from the "equator",
since the upper He bubble has more time to propagate before
the shock collision. However, no significant change in the 
trigger density and temperature is observed, showing that 
the trigger of C detonation does not depend strongly on
the minor details of the He detonation. 

In Figure \ref{fig:det_HSTest0} we plot the temperature colour
plots with the He and C detonation structure at 0.5 s,
at the trigger of C detonation and 0.2 s after the trigger
of C detonation respectively for the Model Test-HS0. 
The detonation structure of both He and C demonstrates a
high degree of symmetry throughout the simulation. The 
detonation occurs at equator around the surface 
of CO core. The reflected shock leads to a clear distinction
between the pre-heated region and post-heated region.
From Table \ref{table:testHS}, 
it can be seen that when the two He detonations are placed
explicitly, the C detonation is triggered along the "equator"
of the WD at the same time as Model Test-QS. However, slight
differences in density and temperature appear between the 
grid positions in the quadrant of sphere and hemisphere. They
are different that in the Test-QS no grid is placed on
the reflection plane while explicit grid is put on 
the reflection place in the Model Test-HS-0. As a result, 
it allows an explicit compression of matter on the 
equator when the two shocks merge. 

In Figure \ref{fig:det_HSTest1} we plot similar to Figure \ref{fig:det_HSTest0}
but for Model Test-HS-1. Due to the delayed He detonation, 
the area swept by the upper He detonation wave has a larger volume 
than the lower one. As a result, the collision point is lower.
Despite that, the collision point remains to be the hottest
point which can trigger C detonation. Due to the assymetric 
expansion of the star, the detonation in the CO core 
has more features compared to the previous case. 

In Figure \ref{fig:det_HSTest2} we plot similar to Figure \ref{fig:det_HSTest0}
but for Model Test-HS-2. The further delayed second He detonation bubble
allows the collision to occur at an even lower position. The newly formed
C detonation can propagate as in previous cases. The shock reflection 
in the He envelope can be clearly seen. 

In Figure \ref{fig:det_HSTest3} we plot similar to Figure \ref{fig:det_HSTest0}
but for Model Test-HS-3. We delayed putting in the second detonation 
so long that the C detonation has been triggered beforehand. 
In this case, it is identical to the one-bubble scenario
where the shock convergence at the "south"-pole of the 
He envelope creates the desired shock compression and penetration
into the CO core, which heats up sufficiently the fuel
for spontaneous runaway. The geometrical convergence around
the "south"-pole allows the shock to be strengthened with 
an increasing post-shock temperature when it approaches
the rotation-axis. The temperature is already adequately high
to trigger the C detonation before the He shock collides
with the axis. The triggered C detonation can then propagate
inside the CO core. 

From all these four cases it suffices to demonstrate that 
the C detonation can be ignited by He detonation, the reflection 
symmetry of the $z = 0$ plane can provide the necessary 
shock collision for shock compressing the fuel in order to 
raise its temperature for spontaneous nuclear runaway. 
Even without the symmetry plane, we demonstrated that the
collision of He detonation, regardless of their ignition 
time, will provide also the necessary shock heating on the 
CO core surface. We also presented that in the one-bubble
limit, i.e. the delay time much greater than the C detonation
time, the geometric convergence in the models can also provide the
required shock compression. This suggests that as long as the 
He envelope mass is large enough for triggering C detonation
naturally, the configuration of He detonation plays a less
important role for the detonation structure. 
Since in these tests we only aim at showing the robustness of
triggering the C detonation with or without reflection 
symmetry, the complete nucleosynthesis and the effects
of shock collision on the nuclear burning will be 
left as future work. However, it remains unclear whether
the WD can be represented comprehensively by a 
sphere in hydrostatic equilibrium prior to its runaway.
The effects of a non-static atmosphere, as a result of 
He burning before its runaway, will be an interesting
future work to further test the robustness of the 
C detonation mechanism by bubbles.

\begin{figure*}
\centering
\caption{(Figures removed for reducing the pdf file size, see published version 
for full figures.) 
The He and C detonation structure and the temperature
colour plot of Model Test-HS-0 at 0.5 s, at the trigger of 
C detonation and at 0.2 s after the C detonation trigger.}
\label{fig:det_HSTest0}
\end{figure*}

\begin{figure*}
\centering
\caption{(Figures removed for reducing the pdf file size, see published version 
for full figures.) 
Similar to Figure \ref{fig:det_HSTest0}, but for Model
Test-HS1 at 0.5 s, at the trigger of 
C detonation and at 0.2 s after the C detonation trigger.}
\label{fig:det_HSTest1}
\end{figure*}

\begin{figure*}
\centering
\caption{(Figures removed for reducing the pdf file size, see published version 
for full figures.) 
Similar to Figure \ref{fig:det_HSTest0}, but for Model
Test-HS2 at 0.5 s, at the trigger of 
C detonation and at 0.1 s after the C detonation trigger. }
\label{fig:det_HSTest2}
\end{figure*}

\begin{figure*}
\centering
\caption{(Figures removed for reducing the pdf file size, see published version 
for full figures.) 
Similar to Figure \ref{fig:det_HSTest0}, but for Model
Test-HS3 at 0.5 s, at the trigger of 
C detonation and at 0.075 s after the C detonation trigger.}
\label{fig:det_HSTest3}
\end{figure*}

\newpage

\section{Test 1: Resolution Study in the Propagation of Detonation}
\label{sec:test1}

\begin{figure*}
\centering
\includegraphics*[width=5.4cm,height=4.8cm]{fig37a.eps}
\includegraphics*[width=5.4cm,height=4.8cm]{fig37b.eps}
\includegraphics*[width=5.4cm,height=4.8cm]{fig37c.eps}
\caption{(left panel) The time evolution of the 
central temperature for the Models Test1-fine $(\Delta x = 7.5$ km), Test1 $(\Delta x = 15$ km) and 
Test1-coarse $(\Delta x = 7.5$ km). (middle panel) Similar to the left panel, but 
for the total energy. (right panel) Similar to the left panel,
but for the total burnt mass.}
\label{fig:det_dx_test}
\end{figure*}

\begin{table*}
\begin{center}
\caption{The model parameters for the one-dimensional resolution study.
$M$, $M_{{\rm He}}$ are in unit of $M_{\odot}$. $\Delta x$ is grid size in unit
of km. $t_{{\rm burn}}$ is the time needed for the C-detonation wave 
to burn everything in unit of s. $E_{{\rm fin}}$ is the final
asymptotic energy given by the simulation, in unit of $10^{50}$ erg. 
$T_{{\rm max}}$ is the maximum central temperature experienced in the 
simulations. The numbers in the brackets stand for the percentage
difference between that model with the higher resolution model.}
\begin{tabular}{|c|c|c|c|c|c|c|}
\hline
Model & $\Delta x$ & $M$ & $M_{{\rm He}}$ & $t_{{\rm burn}}$ & $E_{{\rm fin}}$ & $T_{{\rm max}}$ \\ \hline
Test1-fine & 7.5 & 1.1 & 0.1 & 0.31 & 15.9 & 7.0 \\ \hline
Test1 & 15.0 & 1.1 & 0.1 & 0.34 (9.6) & 15.8 (0.63) & 6.6 (5.7) \\ \hline
Test1-coarse & 30.0 & 1.1 & 0.1 & 0.38 (11.8) & 15.3 (3.2) & 6.2 (6.1) \\ \hline
\end{tabular}
\label{table:Test1}
\end{center}
\end{table*}

In the main text we have studied extensively how each of the 
model parameter contributes to the diversity of chemical composition.
However, besides the chemical composition which should 
be compatible with solar composition, the simulation 
results should be convergent with respect to different
resolution. Here we examine in more details how our models depend on 
the choice of resolution.

The first test is done to a static CO core with He envelope
as in our benchmark model. We choose the configuration the same
as Model 110-100-2-50 except for the initial He-detonation.
We put a spherical C-detonation with a radius 100 km at the beginning 
and allow it to propagate. The spherical detonation will preserve
mostly its symmetry and propagate. Thus, it is literally 
a one-dimensional problem. But we remark that 
it is still a two-dimensional problem because in cylindrical 
coordinate the spherical structure is broken down to 
$r-$ and $z-$ component along the constant radius contour.

We put the model parameters and the explosion
energetics including thermodynamics information in Table
\ref{table:Test1}. We choose the standard resolution 
at 15 km, which is the same as in the main text. A coarser
model with a resolution of 30 km and a finer model with that
of 7.5 km are prepared in a similar manner. We can see
that when the resolution increases, the global quantities
including the explosion energy and burning time
converge, though it does not follow the exact scaling
used in the spatial discretization scheme. The local quantity,
i.e. the global maximum temperature, shows a much slower 
convergence rate. Despite that, the three models show a 
decreasing relative change, showing that the results are on 
the convergence side.

In Figure \ref{fig:det_dx_test} we plot the time evolution
of the central temperature, total energy and total burnt mass
for the three test models. All three models show an 
initial peak at $t = 0.1$ s because of the shock 
imposed by the initial detonation. The peak temperature
increases when resolution decreases. A typical change 
of 5 \% increases is observed when $\Delta x$ drops by half.
After that, the star gradually
expands and the star cools down. The models with a lower resolution 
has a lower peak temperature. Our code shows linear convergence in the temperature.
It is because the smaller the grid size it has, the closer to the $1/r$ 
divergence when the shock converges. The cooling rate of the central
grid also depends on the resolution. The model with a smaller resolution 
cools faster and the change of temperature shows a linear dependence.
A 5 \% difference can be seen at the central at $t = 0.4$ when 
resolution reduces by half.

The total energy and its energy generation are also 
dependent on the spatial resolution. The total energy includes
the kinetic, internal and gravitational energy. The energy growth
and its final energy are also weakly dependent on the spatial 
resolution. Models with a higher resolution has a faster energy
growth and higher final energy. The relative difference is
$\sim 1$ \% when resolution reduces by half. This suggests that
when $\Delta x$ decreases, the level set can capture the 
front surface with more details, which increases its surface area.
As a result, the detonation can effectively propagate faster,
and release more energy while the star has less time to expand
before it is swept by the detonation wave.

The total burnt mass shows how much mass is swept up by the 
detonation wave. It has a similar trend as the total energy
but the result is independent of the energy production algorithm.
The models shows a larger and weaker scaling relation
for different $\Delta x$. A smaller $\Delta x$ gives a lower
time for the detonation wave to complete burning the whole star.
A difference of $\sim 10 \%$ is observed.

\section{Test 2: Resolution Study of Shock Convergence}
\label{sec:test2}

\begin{figure*}
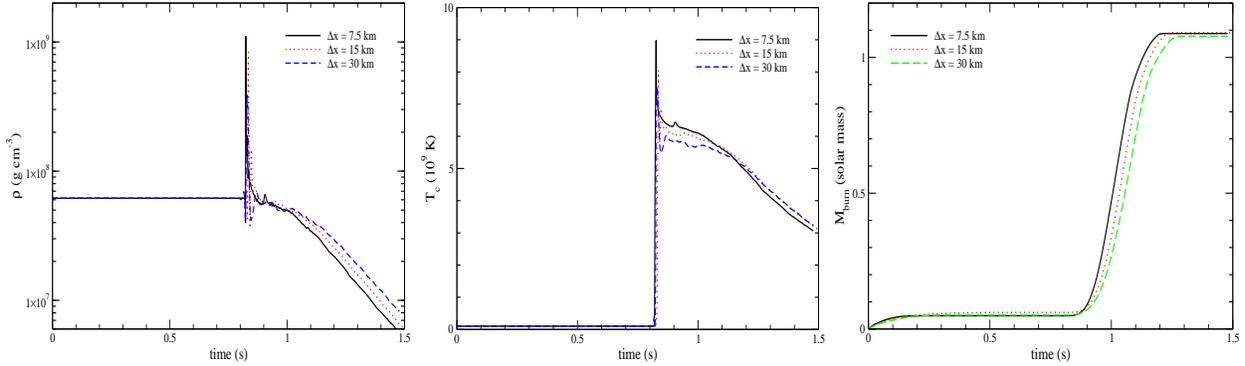

\centering
\includegraphics*[width=5.4cm,height=4.8cm]{fig38a.eps}
\includegraphics*[width=5.4cm,height=4.8cm]{fig38b.eps}
\includegraphics*[width=5.4cm,height=4.8cm]{fig38c.eps}
\caption{(left panel) The time evolution of the 
central temperature for the Models Test2-fine $(\Delta x = 7.5$ km), Test2 $(\Delta x = 15$ km) and 
Test2-coarse $(\Delta x = 30$ km). (middle panel) Similar to the left panel, but 
for the central density. (right panel) Similar to the left panel,
but for the total burnt mass.}
\label{fig:det_dx_test2}
\end{figure*}

\begin{table*}
\begin{center}
\caption{The model parameters for the one-dimensional resolution study.
$M$, $M_{{\rm He}}$ are in unit of $M_{\odot}$. $\Delta x$ is grid size in unit
of km. $t_{{\rm burn}}$ is the time needed for the C-detonation wave 
to burn 1 $M_{\odot}$ in unit of s. $\rho_{{\rm max}}$ is the maximum central
density in the simulation, in unit of $10^8$ g cm$^{-3}$. 
$T_{{\rm max}}$ is the maximum central temperature experienced in the 
simulations in units of $10^9$ K. The numbers in the brackets stand for the percentage
difference between that model with the higher resolution model.}
\begin{tabular}{|c|c|c|c|c|c|c|}
\hline
Model & $\Delta x$ & $M$ & $M_{{\rm He}}$ & $t_{{\rm burn}}$ & $\rho_{{\rm max}}$ & $T_{{\rm max}}$ \\ \hline
Test2-fine & 7.5 & 1.1 & 0.05 & 1.13 & 11.1 & 9.0 \\ \hline
Test2 & 15.0 & 1.1 & 0.05 & 1.16 (2.6) & 8.0 (2.8) & 8.0 (11.1) \\ \hline
Test2-coarse & 30.0 & 1.1 & 0.05 & 1.19 (2.6) & 3.8 (5.2) & 7.5 (6.3) \\ \hline
\end{tabular}
\label{table:Test2}
\end{center}
\end{table*}

In this test we study how the choice of spatial resolution affects the convergence of
detonation shock. Geometric convergence exists in both C- and He-detonation
in different manners. 
For C-detonation, we have showed that "X"-Type detonation
(such as Model 110-100-2-50 (X)) can result in the first C-detonation along the 
symmetry axis. This detonation later propagates to the center. But in the three-dimensional
projection, it corresponds to a C-detonation ring shrinking into a point. 
Similarly, the "S"-Type detonation (such as Model 100-050-2-S50 (S))
can result in a spherical shock propagating towards center. 
For He-detonation, similar phenomenon occurs in the "Y"-Type detonation
such as Model 110-050-2-B50 (Y). The geometric convergence occurs when the 
He-detonation propagate from the convergence. However, the 
discontinuity is described numerically in the 
discretized manner by the Eulerian meshes. As a result, 
the local thermodynamics behaviour at the point of convergence
can depend strongly on the spatial resolution.

To study how the geometric convergence of shock depends on the simulation, 
we repeat the simulations for Model 110-050-2-S50 (S) at a spatial resolution $\Delta x$
of 7.5, 15 and 30 km. We remind $\Delta x = 15$ km is the default resolution. 
We set up a WD with a $M = 1.1 ~M_{\odot}$ and $M_{\rm He}
= 0.05 ~M_{\odot}$ at $Z = Z_{\odot}$. The initial He-detonation is 
spherical at 30 km away from the CO-core.  
We use the spherical He-detonation near the CO-core interface. Then we allow the 
He-detonation to propagate and trigger the inward propagating shock. 
The shock converges at the stellar core and triggers the C-detonation,
which propagates outwards. In Table \ref{table:Test2} we tabulate the 
parameters necessary for this resolution study.

In the left panel of Figure \ref{fig:det_dx_test2} we plot the evolution
of central density for the three tests. The density is an important 
quantity not only because it is the essential part in the Euler equation,
but also the energy production frequently refers density as the 
input parameter. In the figure, the central density has its peak 
at $t \sim 0.8$ s. This corresponds to the moment when the 
spherical shock arrives at the center of the star. The peak value 
can increase from $4 \times 10^8$ up to $10^9$ g cm$^{-3}$ 
when resolution increases. Again, this suggests that the code
obtain a weakly converging result when describing the local
properties in the center.

In the middle panel of Figure \ref{fig:det_dx_test2} we plot similar
to the left panel but for the central temperature. The central temperature
can be important especially when it is related to the burnt matter 
because it controls the NSE process and the electron capture process.
The central temperature can increase from $7 \times 10^9$ to 
$9 \times 10^9$ K for the three models here. Again, smaller
$\Delta x$ allows a faster drop in the central temperature. 
The sequence does not show a convergent trend.  Despite that we
remind that the smaller the resolution we have, the smaller contribution
the divergent result to the whole system is.

In the right panel of Figure \ref{fig:det_dx_test2} we plot the burnt mass
against time. There is no significant burning at the beginning since
only He is burnt. After $t = 0.9$ s, the detonation wave begins to 
sweep across the fuel efficiently. Again, it shows a weakly
converging sequence that a smaller $\Delta x$ allows faster burning of 
material. A reduction by 5 \% by mass of the whole star to be completely
burnt is observed, when resolution drops by half. This shows that,
even the local quantities can rely on $\Delta x$, the finer $\Delta x$ is,
the smaller contribution for an individual cell to the global system, 
especially the center cell gives. As a result, the resolution-dependent
feature is averaged out in general.

\section{Test 3: Resolution Study of C-detonation Trigger}
\label{sec:test3}

\begin{figure*}
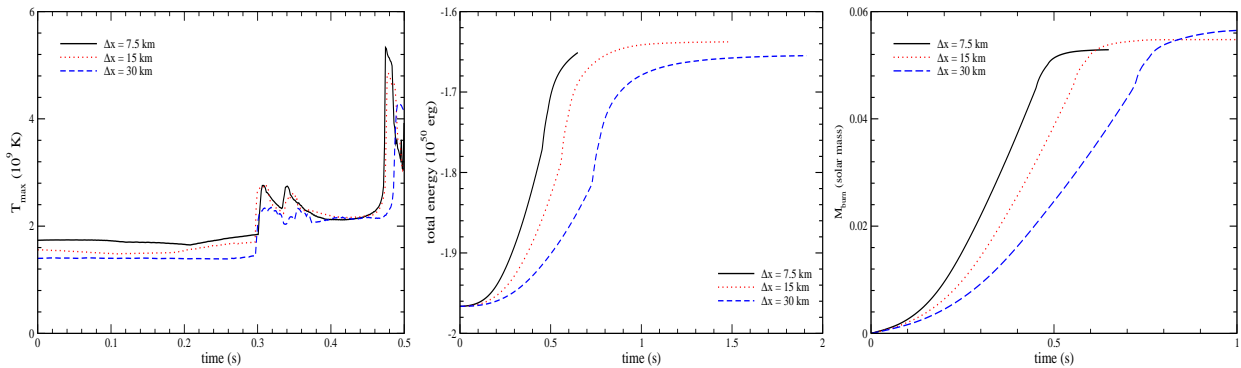

\centering
\includegraphics*[width=5.4cm,height=4.8cm]{fig39a.eps}
\includegraphics*[width=5.4cm,height=4.8cm]{energy_test3_plot.eps}
\includegraphics*[width=5.4cm,height=4.8cm]{fig39b.eps}
\caption{(left panel) The time evolution of the 
central temperature for the Models Test3-fine $(\Delta x = 7.5$ km), Test3 $(\Delta x = 15$ km) and 
Test3-coarse $(\Delta x = 30$ km). (middle panel) Similar to the left panel, but 
for the total energy. (right panel) Similar to the left panel,
but for the total burnt mass.}
\label{fig:det_dx_test3}
\end{figure*}

\begin{table*}
\begin{center}
\caption{The model parameters for the one-dimensional resolution study.
$M$, $M_{{\rm He}}$ are in unit of $M_{\odot}$. $\Delta x$ is grid size in unit
of km. $t_{{\rm burn}}$ is the time needed for the C-detonation wave 
to burn everything in unit of s. $E_{{\rm fin}}$ is the final
asymptotic energy given by the simulation, in unit of $10^{50}$ erg. 
$T_{{\rm max}}$ is the maximum central temperature experienced in the 
simulations in units of $10^9$ K. $2^{{\rm nd}}$ detonation means if
the carbon detonation is triggered throughout the simulation. 
The numbers in the brackets stand for the percentage
difference between that model with the higher resolution model.}
\begin{tabular}{|c|c|c|c|c|c|c|c|}
\hline
Model & $\Delta x$ & $M$ & $M_{{\rm He}}$ & $t_{{\rm burn}}$ & $T_{{\rm max}}$ & $E_{{\rm fin}}$ & $2^{{\rm nd}}$ detonation \\ \hline
Test3-fine & 7.5 & 1.1 & 0.05 & 0.49 & 4.3 & $\sim$-1.63 & No \\ \hline
Test3 & 15.0 & 1.1 & 0.05 & 0.58 (18.4) & 4.8 (11.6) & -1.64 (0.6) & No \\ \hline
Test3-coarse & 30.0 & 1.1 & 0.05 & 0.75 (29.3) & 5.3 (10.4) & -1.66 (1.2) & No \\ \hline
\end{tabular}
\label{table:Test3}
\end{center}
\end{table*}

In this test we study how the choice of spatial resolution affects the convergence of
shock in the trigger of second (C-) detonation. In the simulation, similar to the 
previous test, the geometric convergence plays an important role for creating 
the necessary hot spot, if the initial He-detonation possesses certain 
symmetry in space. For example, 
We choose to study Model 110-100-2-3R50 (N). It is because, by comparing 
Models 110-100-2-50 (X), 110-100-2-2R50 (Y) and 110-100-2-3R50 (N), 
they do not show a regular trend in the detonation pattern. 
Also, given the fact that Model 110-100-2-50 (X) can trigger the second
detonation, with more initial He being burnt at the beginning,
Model 110-100-2-3R50 should be more probable to be ignited. 
Therefore, it becomes interesting to question if the choice of resolution
plays a role.

To test the validity of our result, we also perform convergence study for 
the Model 110-100-2-3R50 (N) at three resolutions of 7.5, 15 and 30 km. 
Again, 15 km is the default simulation size used in our calculation. 
We set up the same initial model with a total mass of 1.1 $M_{\odot}$ 
and He mass of 0.1 $M_{\odot}$ at solar metallity. The initial detonation
is a three-bubble structure located along the rotation axis, symmetry axis 
and diagonal line. Due to the much longer computational time for the 
high resolution one, it is only computed until the reflected shock reaches
the axis of rotation symmetry (i.e. z-axis).

In the left panel of Figure \ref{fig:det_dx_test3} we plot the 
global maximum temperature against time for the three models.
Due to the multiple dimensional effects in this test, the 
time where the system reaches its maximum temperature and
the exact value are not monotonic. However, it shows a clear
sign that the difference between the two models decreases when 
$\Delta x$ drops. The peak temperature varies from $4 \times 10^9$ to 
$5 \times 10^9$ K. The relative change drop from $\sim 20 \%$
to $\sim 5 \%$ between the two sets of models.
However, we notice
that the hot spot is inside the He-envelope. So even it exceeds
the maximum threshold temperature $3 \times 10^9$ K, suitable
for matter at density below $10^7$ g cm$^{-3}$, it does not trigger
any C-detonation.

In the middle panel of Figure \ref{fig:det_dx_test3} we plot
the total energy against time. This also tests the convergence
of energy production rate in the He-envelope due to the absence
of second detonation. The maximum energy is limited 
to $E = -1.6 \times 10^{50}$ erg. It can be seen that the 
He-detonation has a stronger effect on the energy production rate. 
We observe a difference in the $\sim 1 \%$ of final energy by reducing half of
grid size but a difference of $\sim 10$ \% time for the model to reach the 
same energy.

In the right panel of Figure \ref{fig:det_dx_test3} we plot
the total burnt mass against time. The maximum burnt mass is limited 
to $M_{{\rm He}} = 0.05 M_{\odot}$. The 
He-detonation has larger but weakly converging differences in 
its propagation against different resolution. This conforms with 
the energy production rate in the middle panel. 
They all show
to burn the same amount of matter, but the amount of time
differs by $20 \%$ and is weakly converging.  

The above test demonstrates that the trigger of C-detonation by shock convergence is 
in general robust at the current resolution. However,
the necessary $\Delta x$ to determine the C-detonation trigger
can be different, which 
depends on the the chemical composition and also 
the numerical algorithm such as how nuclear reaction
scheme is implemented. 
For further discussion in how resolution affects the discrimination of 
C-detonation trigger,
we refer interested readers to some recent
resolution studies for the 
colliding WD scenario in e.g. \cite{Katz2019,Kushnir2019} and for the 
near-Chandrasekhar mass deflagration-detonation transition scenario in e.g. \cite{Fisher2019}.

\section{Comparison of Models in the Literature}
\label{sec:Comparison}
We have studied the two-dimensional SNe Ia model using the sub-Chandrasekhar mass
WD with the C detonation induced by surface He detonation. In this
work, we compared effects of different detonation structure. Here, we consider
the realizability of the detonation structure and compare
with previous works in
the literature. 

\subsection{\cite{Shigeyama1992}}

The spherical detonation is the same as the classical DD Model \citep{Shigeyama1992}.
The model is adopted for SN1990N, which contains clear Si and Ca signatures with high velocities.
The Model 105-050-2-S50 is comparable with their Model CDT5 but with two qualitative
differences. The two models share a similar CO core mass with the same
metallicity at $Z_{\odot}$. Furthermore, the spherical He-detonation setting in Model
105-050-2-S50 ensures the evolution is spherical, which is compatible to their 
one-dimensional simulation with spherical symmetry.
In their work, the detonation is triggered by hand, assuming the He
detonation on the surface has finished and sent an inward-going shock wave.
Thus, there is no direct He burning considered. Second, that model assumes
a direct CO detonation at the center, which comes from the assumed symmetry in the 
detonation wave. In our model, the He detonation is the "X"-Type detonation. 
They find a yield  of $0.56 ~M_{\odot}$ and $ 1.3 \times 10^{51}$ erg for the 
$^{56}$Ni production and total energy. Our model has a stronger detonation that we 
find 0.60 $M_{\odot}$ and $ 1.07 \times 10^{51}$ erg respectively.
The spherical detonation model is one of the higher viable shapes of detonation
when the convection in the He layer is weak. In that case, the layer closest
to the CO boundary has always the highest and uniform temperature. The whole layer will be the 
first site to trigger explosive He burning. 
 
\subsection{\cite{Fink2007}} 
 
We compare our one-bubble model with the models
in \cite{Fink2007} in the detonation structure. They consider
an isothermal WD model of total mass $0.9 - 1.0$ $M_{\odot}$.
They also explored different detonation pattern, including 
spherical, one-, two-, and five-bubble detonation 
structure. Their model $z4.24A\_$2dq$\_$256 has a similar
model configuration as our Model 105-050-2-2R50.

They observe the second detonation starts 
at 1.08 s after the He detonation. Our model shows a very close results
of 1.07 s. However, they find a yield of 0.01 $M_{\odot}$ unburned 
fuel, 0.40 $M_{\odot}$ $^{56}$Ni and 0.51 $M_{\odot}$ $^{28}$Si.
Our model shows more $^{56}$Ni production of mass 0.49 $M_{\odot}$ but
a lower IME at 0.18 $M_{\odot}$. There is more $^{16}$O fuel of
mass 0.11 $M_{\odot}$. The differences between the two models come from
the burning scheme. An instantaneous input of energy is provided
in the model of detonation wave, while our scheme applies the three-step
burning scheme. The burning of $^{16}$O is suppressed when 
the ash temperature is not sufficiently high, especially around 
$10^7$ g cm$^{-3}$, so that the estimated NQSE and NSE
timescales become very long for all the burning to take place. 
For WD models where convection and turbulence are important, the fluid
motion always disturbs the heat-generating He layer. As a result, 
local hot spot is possible to form. When temperature is close to the 
explosive burning of $^{4}$He, the formation of a hot spot is likely to 
be the first location of He detonation. 

\subsection{\cite{Shen2018}}

%In \cite{Shen2017} the sub-Chandrasekhar mass WD detonation 
%is also computed.
In \cite{Shen2018} the sub-Chandrasekhar mass WD detonation
model is also modeled in the framework of dynamically
driven double degenerate double detonation (DDDDDD) model. 
In this framework, when the two WDs pass by 
each other, the tidal force of the secondary WD triggers
the C detonation of the primary WD, while the secondary WD
later leaves the system without disrupting itself.
The major difference of this physical picture from the 
other one is that the companion WD remains intact after
the SN Ia, unlike the standard white dwarf violent merger.
This provides a smaller total mass in the system, where 
the ejecta may explode more easily with a higher velocity.
In that work, SN Ia model with a mass range of 0.8 - 1.1 $M_{\odot}$
with a metallicity from 0 - 2 $Z_{\odot}$ and
C/O mass fraction ratio from 0.3 - 1 are computed
in the one-dimensional limit. Here we compare one of the 
most similar models, the Model 100-005-1-S50, with their
1 $M_{\odot}$, solar metallicity, C + O = 1 Model. 
We choose this model because the initial detonation
and the C detonation are spherically symmetric,
also the final $^{56}$Ni mass is similar. We have
0.6 $M_{\odot}$ while their model has 0.53 $M_{\odot}$.

\begin{figure}
\centering
\includegraphics*[width=8cm,height=5.7cm]{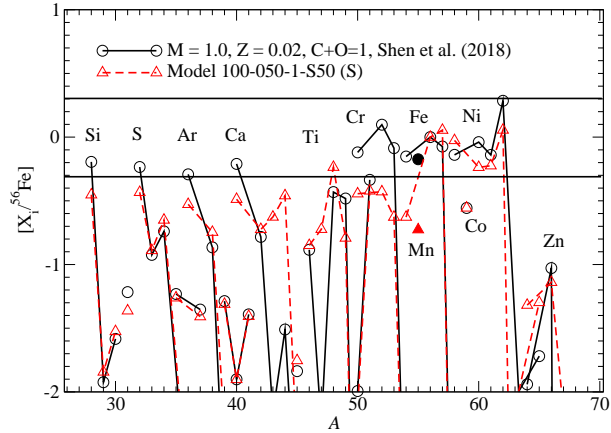}
\caption{$[X_i/^{56}$Fe] for
the model from \cite{Shen2018} (1 $M_{\odot}$, solar metallicity
and C + O = 1) and our Model 100-005-1-S50.}
\label{fig:final_compShen}
\end{figure}

In Figure \ref{fig:final_compShen} we plot the scaled
mass fraction of the stable isotopes of the two models. 
We can see that in general the two models agree well
qualitatively. Both models share the similar relative 
mass fractions of the same elements. Some minor elements,
including P, Cl, Na and Sc are surprisingly close to 
each other, despite their relatively small amounts (subject
to larger systematic uncertainty) and the very different
treatments in the explosion scenario, initial configuration,
explosion treatment, and in particular, the hydrodynamics. 
Major elements, Si, S, Ar, Ca, Fe, Ni and Zn, are 
still close to each other. However, their model shows a 
systematic higher mass fraction for the high-$Y_{\rm e}$
end isotopes (i.e. close to 0.5), e.g. $^{28}$Si, $^{32}$S, $^{36}$Ar, 
$^{40}$Ca, $^{52}$Cr and $^{54}$Fe. This shows that 
they have more incomplete burning such that more
IMEs and light Fe-peak elements
are formed. However, there are also some differences
in Ti and Fe. Their model obtains a higher 
abundance ratios of $^{49}$Ti and $^{50,52,53}$Cr than our model, 
but a lower ratio of $^{46-48}$Ti.
We note that this 
feature is prominent in asymmetric detonation but not 
in symmetric detonation. Also, their $^{55}$Mn production
is a few times higher, despite the low density matter 
in the star. A more detailed study of how He detonation
and C detonation are affected by the numerical treatment
will be an interesting future work. 

\subsection{\cite{Polin2018}}

In \cite{Polin2018} the one-dimensional sub-Chandrasekhar mass
models are also calculated for a wide range of WD masses from 0.6 to 1.2 $M_{\odot}$
and He envelope masses from 0.01 to 0.08 $M_{\odot}$
using the CASTRO code. Their work
studies the observational influences from the He envelope mass.
It is found that two subclasses of light curves emerge. 
For a lighter He envelope, the light curve resembles with some features in SNe Ia,
including the correlation between mass, brightness
and velocity in the spectra. 
For a more massive He envelope, the light curve contains early
UV flux and appears to be red due to iron-peak elements on
the surface and later it turns blue. 

Their methodology and initial models are different from ours. In their work, the 
nuclear reaction is directly solved by introducing the 13-isotope
network containing $^{4}$He, $^{12}$C, $^{16}O$, $^{20}$Ne, 
$^{24}$Mg, $^{28}$Si, $^{32}$S, $^{36}$Ar, $^{40}$Ca, 
$^{44}$Ti, $^{48}$Cr, $^{52}$Fe and $^{56}$Fe. The nuclear 
reaction zone is specifically refined using the AMR option in
CASTRO. A mixed transition between the CO-core and He-envelope
is introduced. Also, at the beginning a width of $\sim 1$ km
spherical He-detonation is put in by hand. On the other hand,
we use a simplified 7-isotope network patched with the 
three-step burning scheme. The nuclear reaction is not directly
resolved but relied on the level-set, which assumes the front of 
the most rapid reaction is directly represented by lines,
where slower nuclear reactions take place assuming a given 
timescale. No mixing between CO-core and He-envelope core
is introduced in our initial model. Also, due to the two-dimensional
nature, our models include initial He-detonations from
spherical to different aspherical structure, but the typical size is larger
$(\sim 30$ km).

Since their work does not aim for nucleosynthesis, here we
only compare with their global chemical yields, 
in particular their models with 0.05 $M_{\odot}$ He
in the envelope. Their models show a different growth rate
in the $M_{{\rm Ni}}$ as a function of $M$. 
They obtain a $^{56}$Ni mass from $\sim$ 0.2, to 
0.5 and then 0.8 $M_{\odot}$ $^{56}$Ni in the 0.9, 1.0
and 1.1 $M_{\odot}$. On the other hand, we have 0.02, 0.6 and 0.8 $M_{\odot}$
$^{56}$Ni respectively from our Models 090-050-2-S50 (S), 100-050-2-S50 (S)
and 110-050-2-S50 (S). Large differences appear at low mass models. One major reason
could be the nuclear reaction at the low density for the CO-detonation.
In our model, we have used a three-step nuclear burning reaction,
with the timescale dependent on the local density. On the other
hand, they solve the nuclear reaction directly using the
13-isotope network in the hydrodynamics. And they also 
use the adaptive mesh refinement for resolving the nuclear burning
at small scales. Both procedures can capture in greater details
how the low-density matter achieves complete burning, 
which may enhance the IMEs
and $^{56}$Ni production. 
On the other hand, for a more massive WD model, our results agree with 
theirs well. 

\subsection{\cite{Jacobs2016}}

An extension of the comparison includes the pre-supernova
models evolved from multi-D hydrodynamics model. However, the exact
site of nuclear runaway in our work is an model parameter. 
In fact, the detailed position and its runaway time can be modeled
by following the exact hydrodynamics evolution over a few of
convective turnover timescale. 
For example, in \cite{Jacobs2016}
the three-dimensional sub-Chandrasekhar mass WDs of masses from 0.85 to 
1.23 $M_{\odot}$ are studied using the low-Mach number code MAESTRO.
The evolution path prior to its nuclear runaway is studied. 

Three nuclear runaway types are observed: localized runaway,
quasi-equilibrium and convective runaway. Localized runaway
corresponds to the runaway taking place by a unique hot spot. 
This occurs when the convection fails to transport heat away
generated from the nuclear reaction near the interface efficiently.
Quasi-equilibrium stands for the opposite of the localized runaway.
The convection can remove the heat efficiently so that no particular
hot spot can grow. However, it is unclear finally the runaway is
localized or collective. Convective runaway means the collective 
runaway in the form of helium nova.

In their study, there is not a clear trend in how 
they observed that localized runaway
takes place in models with a mass 0.8, 1.1 and 1.2 $M_{\odot}$.
Models with a mass of 1.0 $M_{\odot}$ tends to have quasi-equilibrium.
Convective runaway takes place in the low mass model with a 
low $M_{{\rm He}}$. From this it shows that for massive star model
the single spot runaway, e.g. 120-050-2-50 (X) is a more realistic model
than collective runaway. Models with a mass $M = 1.1 ~M_{\odot}$ tend
to occur in a single spot runaway, and hence benchmark model 
including 110-100-2-50 (X) is the most likely initial configuration.
There is no clear conclusion for our benchmark
models due to the quasi-equilibrium outcome for $M = 1.0 ~M_{\odot}$.
There is no models of mass 0.9 $M_{\odot}$ presented in their work
to compare with ours.

\newpage

\bibliographystyle{apj}
\pagestyle{plain}
\bibliography{biblio} 
 
\newpage

\begin{table*}

\caption{The nucleosynthesis yields for the stable isotopes of the benchmark model 
at different metallicity. The model at solar metallicity is Model 110-100-2-50.
Masses are in units of solar mass.}
\begin{center}
\begin{tabular}{|c|c|c|c|c|c|c|c|}

\hline
 Isotope & $Z = 0$ & $Z = 0.1 ~Z_{\odot}$ & $Z = 0.5 ~Z_{\odot}$ & 
 $Z = Z_{\odot}$ & $Z = 2 ~Z_{\odot}$ & $Z = 3 ~Z_{\odot}$ & $Z = 5 ~Z_{\odot}$ \\ \hline

 $^{12}$C & $3.39 \times 10^{-3}$ & $3.38 \times 10^{-3}$ & $3.38 \times 10^{-3}$ & $3.35 \times 10^{-3}$ & $3.29 \times 10^{-3}$ & $2.62 \times 10^{-3}$ & $2.20 \times 10^{-3}$ \\
 $^{13}$C & $3.33 \times 10^{-10}$ & $1.22 \times 10^{-10}$ & $3.41 \times 10^{-10}$ & $1.25 \times 10^{-9}$ & $4.59 \times 10^{-9}$ & $2.9 \times 10^{-8}$ & $1.20 \times 10^{-8}$ \\
 $^{14}$N & $1.16 \times 10^{-8}$ & $4.53 \times 10^{-9}$ & $1.37 \times 10^{-8}$ & $3.80 \times 10^{-8}$ & $9.63 \times 10^{-8}$ & $6.40 \times 10^{-7}$ & $9.43 \times 10^{-8}$ \\
 $^{15}$N & $2.44 \times 10^{-6}$ & $3.97 \times 10^{-8}$ & $3.2 \times 10^{-9}$ & $1.8 \times 10^{-9}$ & $4.42 \times 10^{-10}$ & $1.86 \times 10^{-9}$ & $6.88 \times 10^{-10}$ \\
 $^{16}$O & $1.14 \times 10^{-1}$ & $1.14 \times 10^{-1}$ & $1.16 \times 10^{-1}$ & $1.17 \times 10^{-1}$ & $1.19 \times 10^{-1}$ & $1.12 \times 10^{-1}$ & $1.4 \times 10^{-1}$ \\
 $^{17}$O & $1.86 \times 10^{-10}$ & $5.92 \times 10^{-10}$ & $5.31 \times 10^{-9}$ & $2.8 \times 10^{-8}$ & $6.32 \times 10^{-8}$ & $1.44 \times 10^{-7}$ & $5.44 \times 10^{-8}$ \\
 $^{18}$O & $7.64 \times 10^{-12}$ & $2.4 \times 10^{-11}$ & $1.37 \times 10^{-10}$ & $3.4 \times 10^{-10}$ & $4.87 \times 10^{-10}$ & $2.67 \times 10^{-9}$ & $1.5 \times 10^{-9}$ \\
\end{tabular}
\label{table:stable_yield}
\end{center}
\end{table*} 
 
\begin{table*}

\caption{$cont'd$ of Table \ref{table:stable_yield}.}
\begin{center}
\begin{tabular}{|c|c|c|c|c|c|c|c|}

\hline
 Isotope & $Z = 0$ & $Z = 0.1 $ & $Z = 0.5 $ & 
 $Z = 1$ & $Z = 2$ & $Z = 3 $ & $Z = 5$ \\ \hline

 $^{50}$Cr & $1.0 \times 10^{-4}$ & $1.7 \times 10^{-4}$ & $1.50 \times 10^{-4}$ & $2.18 \times 10^{-4}$ & $3.50 \times 10^{-4}$ & $4.46 \times 10^{-4}$ & $5.18 \times 10^{-4}$ \\
 $^{52}$Cr & $1.78 \times 10^{-2}$ & $1.76 \times 10^{-2}$ & $1.59 \times 10^{-2}$ & $1.60 \times 10^{-2}$ & $1.56 \times 10^{-2}$ & $1.57 \times 10^{-2}$ & $1.73 \times 10^{-2}$ \\
 $^{53}$Cr & $4.25 \times 10^{-4}$ & $4.74 \times 10^{-4}$ & $4.39 \times 10^{-4}$ & $4.87 \times 10^{-4}$ & $5.20 \times 10^{-4}$ & $5.96 \times 10^{-4}$ & $7.64 \times 10^{-4}$ \\
 $^{54}$Cr & $8.73 \times 10^{-6}$ & $8.73 \times 10^{-6}$ & $1.28 \times 10^{-5}$ & $1.37 \times 10^{-4}$ & $1.30 \times 10^{-5}$ & $1.72 \times 10^{-5}$ & $6.57 \times 10^{-6}$ \\
 $^{55}$Mn & $1.79 \times 10^{-3}$ & $1.95 \times 10^{-3}$ & $1.89 \times 10^{-3}$ & $2.28 \times 10^{-3}$ & $2.60 \times 10^{-3}$ & $3.85 \times 10^{-3}$ & $7.26 \times 10^{-3}$ \\
 $^{54}$Fe & $8.76 \times 10^{-4}$ & $1.36 \times 10^{-3}$ & $3.59 \times 10^{-3}$ & $7.80 \times 10^{-3}$ & $1.23 \times 10^{-2}$ & $1.94 \times 10^{-2}$ & $3.96 \times 10^{-2}$ \\
 $^{56}$Fe & $6.73 \times 10^{-1}$ & $6.69 \times 10^{-1}$ & $6.34 \times 10^{-1}$ & $6.10 \times 10^{-1}$ & $5.83 \times 10^{-1}$ & $5.57 \times 10^{-1}$ & $5.15 \times 10^{-1}$ \\
\hline
 
\hline
\end{tabular}
\label{table:stable_yieldb}
\end{center}
\end{table*}

\begin{table*}

\caption{Similar to Table \ref{table:stable_yield} but for 
the radioactive isotopes of the benchmark model.
Masses are in units of solar mass. }
\begin{center}
\begin{tabular}{|c|c|c|c|c|c|c|c|}
 \hline
 $Z$ $(Z_{odot})$ & $Z = 0$ & $Z = 0.1$ & $Z = 0.5$& 
 $Z = 1$ & $Z = 2$ & $Z = 3$ & $Z = 5 $ \\ \hline
$^{22}$Na & $7.32 \times 10^{-9}$ & $6.59 \times 10^{-9}$ & $1.12 \times 10^{-8}$ & $1.25 \times 10^{-8}$ & $1.1 \times 10^{-8}$ & $7.67 \times 10^{-9}$ & $5.26 \times 10^{-9}$ \\
 $^{26}$Al & $1.91 \times 10^{-6}$ & $4.30 \times 10^{-6}$ & $7.4 \times 10^{-6}$ & $6.98 \times 10^{-6}$ & $4.45 \times 10^{-6}$ & $2.40 \times 10^{-6}$ & $1.15 \times 10^{-6}$ \\
 $^{39}$Ar & $5.90 \times 10^{-9}$ & $6.5 \times 10^{-9}$ & $1.28 \times 10^{-8}$ & $1.99 \times 10^{-8}$ & $5.54 \times 10^{-8}$ & $1.5 \times 10^{-7}$ & $1.1 \times 10^{-7}$ \\
 $^{40}$K & $4.99 \times 10^{-8}$ & $5.16 \times 10^{-8}$ & $8.37 \times 10^{-8}$ & $1.23 \times 10^{-7}$ & $2.26 \times 10^{-7}$ & $2.25 \times 10^{-7}$ & $1.15 \times 10^{-7}$ \\
 $^{41}$Ca & $2.4 \times 10^{-6}$ & $3.63 \times 10^{-6}$ & $6.65 \times 10^{-6}$ & $8.49 \times 10^{-6}$ & $9.99 \times 10^{-6}$ & $9.51 \times 10^{-6}$ & $7.23 \times 10^{-6}$ \\
 $^{44}$Ti & $5.52 \times 10^{-4}$ & $5.50 \times 10^{-4}$ & $5.15 \times 10^{-4}$ & $5.14 \times 10^{-4}$ & $5.12 \times 10^{-4}$ & $5.15 \times 10^{-4}$ & $4.72 \times 10^{-4}$ \\
 $^{48}$V & $2.93 \times 10^{-6}$ & $2.94 \times 10^{-6}$ & $3.22 \times 10^{-6}$ & $3.26 \times 10^{-6}$ & $3.30 \times 10^{-6}$ & $2.81 \times 10^{-6}$ & $2.9 \times 10^{-6}$ \\
\hline

\end{tabular}
\label{table:radioactive_yield}
\end{center}
\end{table*}

\begin{table*}

\caption{The nucleosynthesis yields for the stable isotopes of the benchmark model with
a spherical He detonation as a trigger based on Model 110-050-2-B50 at solar metallicity. 
Masses are in units of solar mass.}
\begin{center}
\begin{tabular}{|c|c|c|c|c|c|c|c|}
\hline
 $Z$ ($Z_{\odot}$) & 0 & 0.1 & 0.5 & 1 & 2 & 3 & 5 \\ \hline

$^{12}$C & $4.10 \times 10^{-3}$ & $4.9 \times 10^{-3}$ & $4.5 \times 10^{-3}$ & $4.2 \times 10^{-3}$ & $3.95 \times 10^{-3}$ & $3.89 \times 10^{-3}$ & $3.77 \times 10^{-3}$ \\
 $^{13}$C & $3.66 \times 10^{-10}$ & $2.36 \times 10^{-10}$ & $2.13 \times 10^{-9}$ & $8.97 \times 10^{-9}$ & $5.42 \times 10^{-8}$ & $8.85 \times 10^{-8}$ & $2.8 \times 10^{-7}$ \\
 $^{14}$N & $1.24 \times 10^{-8}$ & $8.32 \times 10^{-9}$ & $3.57 \times 10^{-8}$ & $1.17 \times 10^{-7}$ & $5.17 \times 10^{-7}$ & $8.47 \times 10^{-7}$ & $1.95 \times 10^{-6}$ \\
 $^{15}$N & $1.94 \times 10^{-6}$ & $4.25 \times 10^{-8}$ & $3.71 \times 10^{-9}$ & $1.38 \times 10^{-9}$ & $1.49 \times 10^{-9}$ & $2.65 \times 10^{-9}$ & $8.42 \times 10^{-9}$ \\
 $^{16}$O & $9.80 \times 10^{-2}$ & $9.93 \times 10^{-2}$ & $1.1 \times 10^{-1}$ & $1.2 \times 10^{-1}$ & $1.4 \times 10^{-1}$ & $1.4 \times 10^{-1}$ & $1.2 \times 10^{-1}$ \\
 $^{17}$O & $2.98 \times 10^{-10}$ & $1.50 \times 10^{-9}$ & $1.52 \times 10^{-8}$ & $4.32 \times 10^{-8}$ & $8.5 \times 10^{-8}$ & $1.6 \times 10^{-7}$ & $1.52 \times 10^{-7}$ \\
 $^{18}$O & $1.13 \times 10^{-11}$ & $4.34 \times 10^{-11}$ & $1.94 \times 10^{-10}$ & $5.10 \times 10^{-10}$ & $1.96 \times 10^{-9}$ & $4.5 \times 10^{-9}$ & $1.31 \times 10^{-8}$ \\
\end{tabular}
\label{table:stable_yield2}
\end{center}
\end{table*} 
 
\begin{table*}

\caption{$cont'd$ of Table \ref{table:stable_yield2}.}
\begin{center}
\begin{tabular}{|c|c|c|c|c|c|c|c|}
\hline
  $Z$ ($Z_{\odot}$) & 0 & 0.1 & 0.5 & 1 & 2 & 3 & 5 \\ \hline 
 
 $^{50}$Cr & $1.69 \times 10^{-5}$ & $2.32 \times 10^{-5}$ & $5.98 \times 10^{-5}$ & $1.9 \times 10^{-4}$ & $2.10 \times 10^{-4}$ & $2.80 \times 10^{-4}$ & $3.48 \times 10^{-4}$ \\
 $^{52}$Cr & $3.34 \times 10^{-3}$ & $3.19 \times 10^{-3}$ & $2.95 \times 10^{-3}$ & $2.80 \times 10^{-3}$ & $2.73 \times 10^{-3}$ & $2.93 \times 10^{-3}$ & $3.83 \times 10^{-3}$ \\
 $^{53}$Cr & $7.22 \times 10^{-5}$ & $1.6 \times 10^{-4}$ & $1.34 \times 10^{-4}$ & $1.56 \times 10^{-4}$ & $2.3 \times 10^{-4}$ & $2.78 \times 10^{-4}$ & $4.48 \times 10^{-4}$ \\
 $^{54}$Cr & $2.68 \times 10^{-8}$ & $2.70 \times 10^{-8}$ & $3.58 \times 10^{-8}$ & $6.70 \times 10^{-8}$ & $2.54 \times 10^{-7}$ & $7.23 \times 10^{-7}$ & $2.23 \times 10^{-6}$ \\
 $^{55}$Mn & $3.56 \times 10^{-4}$ & $4.89 \times 10^{-4}$ & $7.29 \times 10^{-4}$ & $9.15 \times 10^{-4}$ & $1.39 \times 10^{-3}$ & $2.41 \times 10^{-3}$ & $6.72 \times 10^{-3}$ \\
 $^{54}$Fe & $1.97 \times 10^{-4}$ & $6.46 \times 10^{-4}$ & $2.70 \times 10^{-3}$ & $5.20 \times 10^{-3}$ & $9.87 \times 10^{-3}$ & $1.49 \times 10^{-2}$ & $3.68 \times 10^{-2}$ \\
 $^{56}$Fe & $7.25 \times 10^{-1}$ & $7.21 \times 10^{-1}$ & $7.4 \times 10^{-1}$ & $6.84 \times 10^{-1}$ & $6.45 \times 10^{-1}$ & $6.8 \times 10^{-1}$ & $5.32 \times 10^{-1}$ \\
\hline

\end{tabular}
\label{table:stable_yield2b}
\end{center}
\end{table*}

\begin{table*}

\caption{The nucleosynthesis yields for radioactive isotopes of the benchmark models. 
Masses are in units of solar mass. }
\begin{center}
\begin{tabular}{|c|c|c|c|c|c|c|c|}
 \hline
  $Z$ ($Z_{\odot}$) & 0 & 0.1 & 0.5 & 1 & 2 & 3 & 5 \\ \hline 
$^{22}$Na & $9.63 \times 10^{-9}$ & $8.92 \times 10^{-9}$ & $1.40 \times 10^{-8}$ & $1.56 \times 10^{-8}$ & $1.24 \times 10^{-8}$ & $9.89 \times 10^{-9}$ & $7.6 \times 10^{-9}$ \\
 $^{26}$Al & $2.54 \times 10^{-6}$ & $5.47 \times 10^{-6}$ & $8.71 \times 10^{-6}$ & $8.12 \times 10^{-6}$ & $5.7 \times 10^{-6}$ & $3.16 \times 10^{-6}$ & $1.44 \times 10^{-6}$ \\
 $^{39}$Ar & $9.85 \times 10^{-12}$ & $1.14 \times 10^{-10}$ & $1.75 \times 10^{-9}$ & $8.69 \times 10^{-9}$ & $4.57 \times 10^{-8}$ & $9.15 \times 10^{-8}$ & $1.3 \times 10^{-7}$ \\
 $^{40}$K & $5.92 \times 10^{-10}$ & $1.91 \times 10^{-9}$ & $1.42 \times 10^{-8}$ & $5.44 \times 10^{-8}$ & $1.59 \times 10^{-7}$ & $1.84 \times 10^{-7}$ & $9.9 \times 10^{-8}$ \\
 $^{41}$Ca & $1.5 \times 10^{-6}$ & $2.59 \times 10^{-6}$ & $4.44 \times 10^{-6}$ & $5.71 \times 10^{-6}$ & $6.75 \times 10^{-6}$ & $6.65 \times 10^{-6}$ & $5.31 \times 10^{-6}$ \\
 $^{44}$Ti & $6.3 \times 10^{-4}$ & $6.1 \times 10^{-4}$ & $6.0 \times 10^{-4}$ & $5.99 \times 10^{-4}$ & $5.97 \times 10^{-4}$ & $5.97 \times 10^{-4}$ & $5.95 \times 10^{-4}$ \\
 $^{48}$V & $1.11 \times 10^{-6}$ & $1.12 \times 10^{-6}$ & $1.13 \times 10^{-6}$ & $1.15 \times 10^{-6}$ & $1.19 \times 10^{-6}$ & $1.21 \times 10^{-6}$ & $1.20 \times 10^{-6}$ \\
\hline

\end{tabular}
\label{table:radioactive_yield2}
\end{center}
\end{table*} 
 
\begin{table*}

\caption{The nucleosynthesis yields for the stable isotopes of the benchmark model with
a spherical He detonation as a trigger based on Model 100-050-2-S50 at solar metallicity. 
Masses are in units of solar mass.}
\begin{center}
\begin{tabular}{|c|c|c|c|c|c|c|c|}
\hline
 $Z$ ($Z_{\odot}$) & 0 & 0.1 & 0.5 & 1 & 2 & 3 & 5 \\ \hline

$^{12}$C & $1.18 \times 10^{-3}$ & $1.17 \times 10^{-3}$ & $1.16 \times 10^{-3}$ & $1.15 \times 10^{-3}$ & $1.13 \times 10^{-3}$ & $1.13 \times 10^{-3}$ & $1.11 \times 10^{-3}$ \\
 $^{13}$C & $1.4 \times 10^{-10}$ & $8.4 \times 10^{-11}$ & $8.87 \times 10^{-10}$ & $3.2 \times 10^{-9}$ & $1.21 \times 10^{-8}$ & $3.70 \times 10^{-8}$ & $1.9 \times 10^{-7}$ \\
 $^{14}$N & $3.94 \times 10^{-9}$ & $1.76 \times 10^{-9}$ & $9.2 \times 10^{-9}$ & $1.81 \times 10^{-8}$ & $7.5 \times 10^{-8}$ & $2.52 \times 10^{-7}$ & $6.87 \times 10^{-7}$ \\
 $^{15}$N & $1.0 \times 10^{-6}$ & $4.93 \times 10^{-9}$ & $5.68 \times 10^{-10}$ & $2.13 \times 10^{-10}$ & $3.20 \times 10^{-10}$ & $1.42 \times 10^{-9}$ & $5.54 \times 10^{-9}$ \\
 $^{16}$O & $5.90 \times 10^{-2}$ & $6.34 \times 10^{-2}$ & $6.55 \times 10^{-2}$ & $6.64 \times 10^{-2}$ & $6.68 \times 10^{-2}$ & $6.64 \times 10^{-2}$ & $6.48 \times 10^{-2}$ \\
 $^{17}$O & $9.61 \times 10^{-11}$ & $2.70 \times 10^{-10}$ & $4.18 \times 10^{-9}$ & $1.8 \times 10^{-8}$ & $2.3 \times 10^{-8}$ & $2.43 \times 10^{-8}$ & $2.52 \times 10^{-8}$ \\
 $^{18}$O & $3.29 \times 10^{-12}$ & $8.33 \times 10^{-12}$ & $6.40 \times 10^{-11}$ & $9.43 \times 10^{-11}$ & $4.35 \times 10^{-10}$ & $2.21 \times 10^{-9}$ & $8.97 \times 10^{-9}$ \\
\end{tabular}
\label{table:stable_yield3}
\end{center}
\end{table*} 
 
\begin{table*}

\caption{$cont'd$ of Table \ref{table:stable_yield3}.}
\begin{center}
\begin{tabular}{|c|c|c|c|c|c|c|c|}
\hline
 $Z$ ($Z_{\odot}$) & 0 & 0.1 & 0.5 & 1 & 2 & 3 & 5 \\ \hline
 
 $^{50}$Cr & $5.56 \times 10^{-6}$ & $1.34 \times 10^{-5}$ & $6.64 \times 10^{-5}$ & $1.41 \times 10^{-4}$ & $3.5 \times 10^{-4}$ & $4.30 \times 10^{-4}$ & $5.91 \times 10^{-4}$ \\
 $^{52}$Cr & $4.11 \times 10^{-3}$ & $3.86 \times 10^{-3}$ & $3.30 \times 10^{-3}$ & $2.94 \times 10^{-3}$ & $2.65 \times 10^{-3}$ & $2.77 \times 10^{-3}$ & $3.68 \times 10^{-3}$ \\
 $^{53}$Cr & $4.72 \times 10^{-5}$ & $1.21 \times 10^{-4}$ & $1.78 \times 10^{-4}$ & $2.14 \times 10^{-4}$ & $2.77 \times 10^{-4}$ & $3.74 \times 10^{-4}$ & $5.67 \times 10^{-4}$ \\
 $^{54}$Cr & $9.21 \times 10^{-10}$ & $1.18 \times 10^{-9}$ & $1.14 \times 10^{-8}$ & $4.40 \times 10^{-8}$ & $2.43 \times 10^{-7}$ & $8.6 \times 10^{-7}$ & $2.68 \times 10^{-6}$ \\
 $^{55}$Mn & $3.9 \times 10^{-4}$ & $5.59 \times 10^{-4}$ & $1.3 \times 10^{-3}$ & $1.32 \times 10^{-3}$ & $1.87 \times 10^{-3}$ & $3.6 \times 10^{-3}$ & $6.30 \times 10^{-3}$ \\
 $^{54}$Fe & $1.10 \times 10^{-4}$ & $8.14 \times 10^{-4}$ & $4.37 \times 10^{-3}$ & $8.69 \times 10^{-3}$ & $1.66 \times 10^{-2}$ & $2.48 \times 10^{-2}$ & $4.79 \times 10^{-2}$ \\
 $^{56}$Fe & $6.43 \times 10^{-1}$ & $6.38 \times 10^{-1}$ & $6.20 \times 10^{-1}$ & $6.0 \times 10^{-1}$ & $5.65 \times 10^{-1}$ & $5.32 \times 10^{-1}$ & $4.67 \times 10^{-1}$ \\
\hline

\end{tabular}
\label{table:stable_yield3b}
\end{center}
\end{table*}

\begin{table*}

\caption{Similar to \ref{table:stable_yield3} but for the radioactive isotopes. 
Masses are in units of solar mass. }
\begin{center}
\begin{tabular}{|c|c|c|c|c|c|c|c|}
 \hline
$Z$ ($Z_{\odot}$) & 0 & 0.1 & 0.5 & 1 & 2 & 3 & 5 \\ \hline
  $^{22}$Na & $2.93 \times 10^{-9}$ & $3.22 \times 10^{-9}$ & $4.6 \times 10^{-9}$ & $3.74 \times 10^{-9}$ & $2.87 \times 10^{-9}$ & $2.42 \times 10^{-9}$ & $2.0 \times 10^{-9}$ \\
 $^{26}$Al & $8.57 \times 10^{-7}$ & $1.74 \times 10^{-6}$ & $2.12 \times 10^{-6}$ & $1.70 \times 10^{-6}$ & $1.4 \times 10^{-6}$ & $7.15 \times 10^{-7}$ & $4.1 \times 10^{-7}$ \\
 $^{39}$Ar & $4.11 \times 10^{-13}$ & $1.11 \times 10^{-10}$ & $1.66 \times 10^{-9}$ & $7.27 \times 10^{-9}$ & $3.3 \times 10^{-8}$ & $5.13 \times 10^{-8}$ & $5.39 \times 10^{-8}$ \\
 $^{40}$K & $4.83 \times 10^{-11}$ & $1.59 \times 10^{-9}$ & $1.22 \times 10^{-8}$ & $3.86 \times 10^{-8}$ & $8.64 \times 10^{-8}$ & $9.6 \times 10^{-8}$ & $5.63 \times 10^{-8}$ \\
 $^{41}$Ca & $6.17 \times 10^{-7}$ & $2.38 \times 10^{-6}$ & $4.36 \times 10^{-6}$ & $5.66 \times 10^{-6}$ & $6.62 \times 10^{-6}$ & $6.34 \times 10^{-6}$ & $4.76 \times 10^{-6}$ \\
 $^{44}$Ti & $2.70 \times 10^{-4}$ & $2.68 \times 10^{-4}$ & $2.66 \times 10^{-4}$ & $2.64 \times 10^{-4}$ & $2.62 \times 10^{-4}$ & $2.61 \times 10^{-4}$ & $2.59 \times 10^{-4}$ \\
 $^{48}$V & $1.45 \times 10^{-7}$ & $1.47 \times 10^{-7}$ & $1.70 \times 10^{-7}$ & $1.96 \times 10^{-7}$ & $2.34 \times 10^{-7}$ & $2.49 \times 10^{-7}$ & $2.36 \times 10^{-7}$ \\
\hline

\end{tabular}
\label{table:radioactive_yield3}
\end{center}
\end{table*}

\begin{table*}

\caption{The nucleosynthesis yields for stable isotopes from the 
models using one He detonation bubble as the initial configuration.
The minimum He envelope mass is used for each choice of 
progenitor mass. All models are of solar metallicity and
masses are in units of solar mass.}

\begin{center}
\begin{tabular}{|c|c|c|c|c|c|c|c|}
 \hline

$M$ & 0.90 & 0.95 & 1.00 & 1.05 & 1.10 & 1.15 & 1.20 \\ \hline
$M_{{\rm He}}$ & 0.15 & 0.15 & 0.10 & 0.10 & 0.10 & 0.10 & 0.05 \\ \hline

  $^{12}$C & $1.78 \times 10^{-2}$ & $7.5 \times 10^{-3}$ & $8.85 \times 10^{-3}$ & $1.26 \times 10^{-2}$ & $3.35 \times 10^{-3}$ & $1.6 \times 10^{-4}$ & $2.10 \times 10^{-3}$ \\
 $^{13}$C & $6.99 \times 10^{-8}$ & $2.73 \times 10^{-9}$ & $3.87 \times 10^{-9}$ & $1.84 \times 10^{-8}$ & $1.25 \times 10^{-9}$ & $4.83 \times 10^{-11}$ & $5.8 \times 10^{-10}$ \\
 $^{14}$N & $1.50 \times 10^{-6}$ & $7.88 \times 10^{-8}$ & $9.0 \times 10^{-8}$ & $4.92 \times 10^{-7}$ & $3.80 \times 10^{-8}$ & $8.42 \times 10^{-10}$ & $1.41 \times 10^{-8}$ \\
 $^{15}$N & $8.65 \times 10^{-9}$ & $1.83 \times 10^{-9}$ & $2.69 \times 10^{-9}$ & $3.45 \times 10^{-9}$ & $1.8 \times 10^{-9}$ & $1.13 \times 10^{-10}$ & $7.37 \times 10^{-10}$ \\
 $^{16}$O & $2.63 \times 10^{-1}$ & $2.19 \times 10^{-1}$ & $2.35 \times 10^{-1}$ & $1.85 \times 10^{-1}$ & $1.17 \times 10^{-1}$ & $5.23 \times 10^{-2}$ & $6.54 \times 10^{-2}$ \\
 $^{17}$O & $3.23 \times 10^{-7}$ & $4.35 \times 10^{-8}$ & $4.80 \times 10^{-8}$ & $1.68 \times 10^{-7}$ & $2.8 \times 10^{-8}$ & $2.95 \times 10^{-10}$ & $7.14 \times 10^{-9}$ \\
\end{tabular}
\label{table:stable_yield4}
\end{center}
\end{table*} 
 
\begin{table*} 
\caption{$cont'd$ of Table \ref{table:stable_yield4}.}

\begin{center}
\begin{tabular}{|c|c|c|c|c|c|c|c|}
 \hline

$M$ & 0.90 & 0.95 & 1.00 & 1.05 & 1.10 & 1.15 & 1.20 \\ \hline
$M_{{\rm He}}$ & 0.15 & 0.15 & 0.10 & 0.10 & 0.10 & 0.10 & 0.05 \\ \hline 
 
 $^{50}$Cr & $2.8 \times 10^{-4}$ & $1.29 \times 10^{-4}$ & $1.35 \times 10^{-4}$ & $1.25 \times 10^{-4}$ & $2.18 \times 10^{-4}$ & $1.83 \times 10^{-4}$ & $1.20 \times 10^{-4}$ \\
 $^{52}$Cr & $1.36 \times 10^{-2}$ & $1.74 \times 10^{-2}$ & $1.6 \times 10^{-2}$ & $1.67 \times 10^{-2}$ & $1.60 \times 10^{-2}$ & $1.24 \times 10^{-2}$ & $8.18 \times 10^{-3}$ \\
 $^{53}$Cr & $4.36 \times 10^{-4}$ & $3.84 \times 10^{-4}$ & $3.40 \times 10^{-4}$ & $3.76 \times 10^{-4}$ & $4.87 \times 10^{-4}$ & $4.52 \times 10^{-4}$ & $2.42 \times 10^{-4}$ \\
 $^{54}$Cr & $5.24 \times 10^{-8}$ & $9.91 \times 10^{-8}$ & $4.6 \times 10^{-8}$ & $2.79 \times 10^{-8}$ & $1.37 \times 10^{-4}$ & $5.88 \times 10^{-7}$ & $3.12 \times 10^{-8}$ \\
 $^{55}$Mn & $1.96 \times 10^{-3}$ & $1.78 \times 10^{-3}$ & $1.59 \times 10^{-3}$ & $2.11 \times 10^{-3}$ & $2.28 \times 10^{-3}$ & $2.16 \times 10^{-3}$ & $2.85 \times 10^{-3}$ \\
 $^{54}$Fe & $6.59 \times 10^{-3}$ & $4.18 \times 10^{-3}$ & $4.55 \times 10^{-3}$ & $4.65 \times 10^{-3}$ & $7.80 \times 10^{-3}$ & $6.52 \times 10^{-3}$ & $1.17 \times 10^{-2}$ \\
\hline

\end{tabular}
\label{table:stable_yield4b}
\end{center}
\end{table*}

\begin{table*}

\caption{Similar to Table \ref{table:stable_yield4} but for 
the radioactive isotopes of the selected models after explosion. 
Masses are in units of solar mass. }

\begin{center}
\begin{tabular}{|c|c|c|c|c|c|c|c|}
 \hline
$M$ & 0.90 & 0.95 & 1.00 & 1.05 & 1.10 & 1.15 & 1.20 \\ \hline
$M_{{\rm He}}$ & 0.15 & 0.15 & 0.10 & 0.10 & 0.10 & 0.10 & 0.05 \\ \hline
$^{22}$Na & $8.72 \times 10^{-8}$ & $2.71 \times 10^{-8}$ & $3.50 \times 10^{-8}$ & $3.2 \times 10^{-8}$ & $1.25 \times 10^{-8}$ & $2.0 \times 10^{-9}$ & $9.56 \times 10^{-9}$ \\
 $^{26}$Al & $4.50 \times 10^{-5}$ & $1.46 \times 10^{-5}$ & $1.72 \times 10^{-5}$ & $1.49 \times 10^{-5}$ & $6.98 \times 10^{-6}$ & $1.50 \times 10^{-6}$ & $4.78 \times 10^{-6}$ \\
 $^{39}$Ar & $2.90 \times 10^{-8}$ & $1.20 \times 10^{-8}$ & $1.35 \times 10^{-8}$ & $9.97 \times 10^{-9}$ & $1.99 \times 10^{-8}$ & $2.96 \times 10^{-9}$ & $4.99 \times 10^{-9}$ \\
 $^{40}$K & $1.73 \times 10^{-7}$ & $6.72 \times 10^{-8}$ & $9.15 \times 10^{-8}$ & $5.46 \times 10^{-8}$ & $1.23 \times 10^{-7}$ & $2.31 \times 10^{-8}$ & $2.94 \times 10^{-8}$ \\
 $^{41}$Ca & $9.96 \times 10^{-6}$ & $6.19 \times 10^{-6}$ & $7.5 \times 10^{-6}$ & $5.27 \times 10^{-6}$ & $8.49 \times 10^{-6}$ & $5.46 \times 10^{-6}$ & $5.8 \times 10^{-6}$ \\
 $^{44}$Ti & $1.0 \times 10^{-3}$ & $1.4 \times 10^{-3}$ & $1.14 \times 10^{-3}$ & $6.58 \times 10^{-4}$ & $5.14 \times 10^{-4}$ & $3.69 \times 10^{-4}$ & $5.11 \times 10^{-4}$ \\
\hline

\end{tabular}
\label{table:radioactive_yield4}
\end{center}
\end{table*} 
 
\begin{table*}

\caption{The nucleosynthesis yields for stable isotopes from the selected models
using a He detonation ring as the initial configuration.
based on the benchmark model 110-050-2-B50. 
$M_{{\rm He}} = 0.05$ $M_{\odot}$ for all models in this table.
Masses are in units of solar mass. }

\begin{center}
\begin{tabular}{|c|c|c|c|c|c|c|}
 \hline

$M (M_{\odot})$ & 0.90 & 0.95 & 1.00 & 1.05 & 1.10 & 1.20 \\ \hline

 $^{12}$C & $6.40 \times 10^{-2}$ & $6.0 \times 10^{-2}$ & $1.15 \times 10^{-3}$
  & $6.86 \times 10^{-3}$ & $4.2 \times 10^{-3}$ & $7.84 \times 10^{-6}$ \\
 $^{13}$C & $2.2 \times 10^{-8}$ & $1.60 \times 10^{-8}$ & $3.2 \times 10^{-9}$ 
 & $7.8 \times 10^{-9}$ & $8.97 \times 10^{-9}$ & $1.6 \times 10^{-11}$ \\
 $^{14}$N & $5.8 \times 10^{-7}$ & $4.40 \times 10^{-7}$ & $1.63 \times 10^{-8}$
  & $1.14 \times 10^{-7}$ & $7.62 \times 10^{-8}$ & $6.96 \times 10^{-12}$ \\
 $^{15}$N & $1.2 \times 10^{-8}$ & $1.14 \times 10^{-8}$ & $2.13 \times 10^{-10}
 $ & $1.71 \times 10^{-9}$ & $1.38 \times 10^{-9}$ & $9.11 \times 10^{-12}$ \\
 $^{16}$O & $4.28 \times 10^{-1}$ & $3.86 \times 10^{-1}$ & $6.64 \times 10^{-2}
 $ & $1.35 \times 10^{-1}$ & $1.2 \times 10^{-1}$ & $6.89 \times 10^{-3}$ \\
 $^{17}$O & $2.96 \times 10^{-7}$ & $2.56 \times 10^{-7}$ & $1.8 \times 10^{-8}$
  & $6.97 \times 10^{-8}$ & $4.32 \times 10^{-8}$ & $1.23 \times 10^{-13}$ \\
 $^{18}$O & $3.82 \times 10^{-9}$ & $3.52 \times 10^{-9}$ & $9.43 \times 10^{-11
 }$ & $6.91 \times 10^{-10}$ & $5.10 \times 10^{-10}$ & $4.95 \times 10^{-16}$ \\
\end{tabular}
\label{table:stable_yield5}
\end{center}
\end{table*} 
 
\begin{table*} 
\caption{$cont'd$ of Table \ref{table:stable_yield5}.}

\begin{center}
\begin{tabular}{|c|c|c|c|c|c|c|}
 \hline

$M (M_{\odot})$ & 0.90 & 0.95 & 1.00 & 1.05 & 1.10 & 1.20 \\ \hline

 $^{50}$Cr & $4.32 \times 10^{-5}$ & $8.96 \times 10^{-5}$ & $2.29 \times 10^{-4}$ & $2.1 \times 10^{-4}$ & $1.9 \times 10^{-4}$ & $7.79 \times 10^{-5}$ \\
 $^{52}$Cr & $1.27 \times 10^{-3}$ & $1.72 \times 10^{-3}$ & $2.87 \times 10^{-3}$ & $2.96 \times 10^{-3}$ & $2.80 \times 10^{-3}$ & $9.80 \times 10^{-3}$ \\
 $^{53}$Cr & $4.95 \times 10^{-5}$ & $1.28 \times 10^{-4}$ & $2.2 \times 10^{-4}$ & $2.1 \times 10^{-4}$ & $1.56 \times 10^{-4}$ & $2.71 \times 10^{-4}$ \\
 $^{54}$Cr & $1.79 \times 10^{-8}$ & $8.10 \times 10^{-8}$ & $8.83 \times 10^{-8}$ & $5.64 \times 10^{-8}$ & $6.70 \times 10^{-8}$ & $1.59 \times 10^{-8}$ \\
 $^{55}$Mn & $3.18 \times 10^{-4}$ & $7.12 \times 10^{-4}$ & $1.8 \times 10^{-3}$ & $1.16 \times 10^{-3}$ & $9.15 \times 10^{-4}$ & $1.99 \times 10^{-3}$ \\
 $^{54}$Fe & $1.71 \times 10^{-3}$ & $4.12 \times 10^{-3}$ & $1.11 \times 10^{-2}$ & $1.7 \times 10^{-2}$ & $5.20 \times 10^{-3}$ & $5.59 \times 10^{-3}$ \\
 $^{56}$Fe & $3.58 \times 10^{-2}$ & $1.7 \times 10^{-1}$ & $3.12 \times 10^{-1}$ & $4.68 \times 10^{-1}$ & $6.84 \times 10^{-1}$ & $9.21 \times 10^{-1}$ \\
\hline

\end{tabular}
\label{table:stable_yield5b}
\end{center}
\end{table*}

\begin{table*}

\caption{Similar to Table \ref{table:stable_yield5}, but for the radioactive isotopes. 
Masses are in units of solar mass. }

\begin{center}
\begin{tabular}{|c|c|c|c|c|c|c|}
 \hline

$M (M_{\odot})$ & 0.90 & 0.95 & 1.00 & 1.05 & 1.10 & 1.20 \\ \hline

$^{22}$Na & $1.12 \times 10^{-7}$ & $9.86 \times 10^{-8}$ & $3.74 \times 10^{-9
 }$ & $2.49 \times 10^{-8}$ & $1.56 \times 10^{-8}$ & $4.95 \times 10^{-11}$ \\
 $^{26}$Al & $7.39 \times 10^{-5}$ & $5.12 \times 10^{-5}$ & $1.70 \times 10^{-6
 }$ & $1.29 \times 10^{-5}$ & $8.12 \times 10^{-6}$ & $4.10 \times 10^{-9}$ \\
 $^{39}$Ar & $3.52 \times 10^{-8}$ & $2.39 \times 10^{-8}$ & $7.27 \times 10^{-9
 }$ & $9.79 \times 10^{-9}$ & $8.69 \times 10^{-9}$ & $5.15 \times 10^{-10}$ \\
 $^{40}$K & $2.77 \times 10^{-7}$ & $1.69 \times 10^{-7}$ & $3.86 \times 10^{-8}
 $ & $5.74 \times 10^{-8}$ & $5.44 \times 10^{-8}$ & $3.13 \times 10^{-9}$ \\
 $^{41}$Ca & $2.77 \times 10^{-6}$ & $4.13 \times 10^{-6}$ & $5.66 \times 10^{-6
 }$ & $6.52 \times 10^{-6}$ & $5.71 \times 10^{-6}$ & $1.22 \times 10^{-6}$ \\
 $^{44}$Ti & $2.33 \times 10^{-4}$ & $4.37 \times 10^{-5}$ & $2.64 \times 10^{-4
 }$ & $1.2 \times 10^{-4}$ & $5.99 \times 10^{-4}$ & $5.65 \times 10^{-4}$ \\
 $^{48}$V & $2.67 \times 10^{-7}$ & $1.43 \times 10^{-7}$ & $1.96 \times 10^{-7}
 $ & $1.59 \times 10^{-7}$ & $1.15 \times 10^{-6}$ & $6.95 \times 10^{-7}$ \\
\hline

\end{tabular}
\label{table:radioactive_yield5}
\end{center}
\end{table*} 
 
\begin{table*}

\caption{The nucleosynthesis yields for stable isotopes from the selected models
using a spherical He detonation as the initial configuration.
based on the benchmark model 110-050-2-S50. 
$M_{{\rm He}} = 0.05$ $M_{\odot}$ for all models in this table.
Masses are in units of solar mass. }
\label{table:stable_yield6}
\begin{center}
\begin{tabular}{|c|c|c|c|c|c|c|}
 \hline

$M (M_{\odot})$ & 0.90 & 0.95 & 1.00 & 1.05 & 1.10 & 1.20 \\ \hline

  $^{12}$C & $5.27 \times 10^{-3}$ & $4.1 \times 10^{-3}$ & $1.15 \times 10^{-3}$
  & $2.97 \times 10^{-4}$ & $7.65 \times 10^{-4}$ & $7.84 \times 10^{-6}$ \\
 $^{13}$C & $1.15 \times 10^{-7}$ & $3.32 \times 10^{-7}$ & $3.2 \times 10^{-9}$
  & $4.6 \times 10^{-11}$ & $3.70 \times 10^{-10}$ & $1.6 \times 10^{-11}$ \\
 $^{14}$N & $4.51 \times 10^{-7}$ & $4.93 \times 10^{-7}$ & $1.63 \times 10^{-8}
 $ & $2.22 \times 10^{-10}$ & $1.54 \times 10^{-9}$ & $6.96 \times 10^{-12}$ \\
 $^{15}$N & $3.24 \times 10^{-9}$ & $2.61 \times 10^{-9}$ & $2.13 \times 10^{-10
 }$ & $6.54 \times 10^{-11}$ & $1.27 \times 10^{-10}$ & $9.11 \times 10^{-12}$ \\
 $^{16}$O & $2.67 \times 10^{-1}$ & $1.14 \times 10^{-1}$ & $6.64 \times 10^{-2}
 $ & $3.90 \times 10^{-2}$ & $2.30 \times 10^{-2}$ & $6.89 \times 10^{-3}$ \\
 $^{17}$O & $1.88 \times 10^{-7}$ & $1.62 \times 10^{-7}$ & $1.8 \times 10^{-8}$
  & $8.12 \times 10^{-11}$ & $8.10 \times 10^{-10}$ & $1.23 \times 10^{-13}$ \\
 $^{18}$O & $3.51 \times 10^{-9}$ & $3.29 \times 10^{-9}$ & $9.43 \times 10^{-11
 }$ & $1.99 \times 10^{-12}$ & $1.46 \times 10^{-11}$ & $4.95 \times 10^{-16}$ \\
\end{tabular}
\end{center}
\end{table*} 
 
\begin{table*} 
\caption{$cont'd$ of Table \ref{table:stable_yield6}.}
\label{table:stable_yield6b}
\begin{center}
\begin{tabular}{|c|c|c|c|c|c|c|}
 \hline

$M (M_{\odot})$ & 0.90 & 0.95 & 1.00 & 1.05 & 1.10 & 1.20 \\ \hline

$^{50}$Cr & $3.1 \times 10^{-4}$ & $1.63 \times 10^{-4}$ & $1.41 \times 10^{-4}$ & $1.22 \times 10^{-4}$ & $1.13 \times 10^{-4}$ & $2.93 \times 10^{-5}$ \\
 $^{52}$Cr & $5.29 \times 10^{-4}$ & $2.99 \times 10^{-3}$ & $2.94 \times 10^{-3}$ & $1.73 \times 10^{-3}$ & $2.14 \times 10^{-3}$ & $7.38 \times 10^{-3}$ \\
 $^{53}$Cr & $8.54 \times 10^{-5}$ & $2.63 \times 10^{-4}$ & $2.14 \times 10^{-4}$ & $1.38 \times 10^{-4}$ & $1.44 \times 10^{-4}$ & $2.6 \times 10^{-4}$ \\
 $^{54}$Cr & $1.3 \times 10^{-7}$ & $5.22 \times 10^{-8}$ & $4.40 \times 10^{-8}$ & $4.64 \times 10^{-8}$ & $3.17 \times 10^{-8}$ & $1.45 \times 10^{-10}$ \\
 $^{55}$Mn & $4.60 \times 10^{-4}$ & $1.58 \times 10^{-3}$ & $1.32 \times 10^{-3}$ & $7.88 \times 10^{-4}$ & $8.56 \times 10^{-4}$ & $6.32 \times 10^{-4}$ \\
 $^{54}$Fe & $1.17 \times 10^{-2}$ & $1.2 \times 10^{-2}$ & $8.69 \times 10^{-3}$ & $7.1 \times 10^{-3}$ & $5.94 \times 10^{-3}$ & $6.38 \times 10^{-4}$ \\
 $^{56}$Fe & $1.55 \times 10^{-2}$ & $4.52 \times 10^{-1}$ & $6.0 \times 10^{-1}$ & $7.4 \times 10^{-1}$ & $8.17 \times 10^{-1}$ & $10.48 \times 10^{-1}$ \\
\hline

\end{tabular}
\end{center}
\end{table*}

\begin{table*}

\caption{Similar to Table \ref{table:stable_yield6}, but for the radioactive isotopes. 
Masses are in units of solar mass. }
\label{table:radioactive_yield6}
\begin{center}
\begin{tabular}{|c|c|c|c|c|c|c|}
 \hline

$M (M_{\odot})$ & 0.90 & 0.95 & 1.00 & 1.05 & 1.10 & 1.20 \\ \hline

 $^{22}$Na & $1.94 \times 10^{-8}$ & $5.10 \times 10^{-9}$ & $3.74 \times 10^{-9
 }$ & $1.31 \times 10^{-9}$ & $2.46 \times 10^{-9}$ & $4.95 \times 10^{-11}$ \\
 $^{26}$Al & $1.39 \times 10^{-5}$ & $2.16 \times 10^{-6}$ & $1.70 \times 10^{-6
 }$ & $8.41 \times 10^{-7}$ & $1.3 \times 10^{-6}$ & $4.10 \times 10^{-9}$ \\
 $^{39}$Ar & $2.33 \times 10^{-8}$ & $1.1 \times 10^{-8}$ & $7.27 \times 10^{-9}
 $ & $5.89 \times 10^{-9}$ & $3.48 \times 10^{-9}$ & $5.15 \times 10^{-10}$ \\
 $^{40}$K & $1.59 \times 10^{-7}$ & $5.28 \times 10^{-8}$ & $3.86 \times 10^{-8}
 $ & $3.31 \times 10^{-8}$ & $1.73 \times 10^{-8}$ & $3.13 \times 10^{-9}$ \\
 $^{41}$Ca & $1.56 \times 10^{-5}$ & $6.89 \times 10^{-6}$ & $5.66 \times 10^{-6
 }$ & $4.75 \times 10^{-6}$ & $3.43 \times 10^{-6}$ & $1.22 \times 10^{-6}$ \\
 $^{44}$Ti & $3.27 \times 10^{-5}$ & $6.5 \times 10^{-5}$ & $2.64 \times 10^{-4}
 $ & $5.75 \times 10^{-5}$ & $1.93 \times 10^{-4}$ & $5.65 \times 10^{-4}$ \\
 $^{48}$V & $2.8 \times 10^{-7}$ & $1.81 \times 10^{-7}$ & $1.96 \times 10^{-7}$
  & $5.81 \times 10^{-8}$ & $3.20 \times 10^{-7}$ & $6.95 \times 10^{-7}$ \\
\hline

\end{tabular}
\end{center}
\end{table*}

\end{document}